\documentclass{aastex63}

\newcommand\rmi{\mathrm{i}}
\newcommand{\comment}[1]{}

\usepackage{graphicx}
\shortauthors{Gu et al.}

\begin{document}

\title{Revealing the drag instability in one-fluid non-ideal MHD simulations of a 1D isothermal C-shock}

\correspondingauthor{Pin-Gao Gu}
\email{gu@asiaa.sinica.edu.tw}

\author{Pin-Gao Gu}
\affiliation{Institute of Astronomy \& Astrophysics, Academia Sinica,
Taipei 10617, Taiwan}

\author{Che-Yu Chen}
\affiliation{Lawrence Livermore National Laboratory, Livermore, CA 94550, USA}
\affiliation{Department of Astronomy, University of Virginia, Charlottesville, VA 22904, USA}

\author{Emma Shen}
\affiliation{Institute of Astronomy \& Astrophysics, Academia Sinica, Taipei 10617, Taiwan}
\affiliation{Cavendish Laboratory, University of Cambridge, Cambridge CB3 0HE, UK}

\author{Chien-Chang Yen}
\affiliation{Institute of Astronomy \& Astrophysics, Academia Sinica,
Taipei 10617, Taiwan}
\affiliation{Department of Mathematics, Fu-Jen Catholic University,
New Taipei City 24205, Taiwan}


\author{Min-Kai Lin}
\affiliation{Institute of Astronomy \& Astrophysics, Academia Sinica, Taipei 10617, Taiwan}
\affiliation{Physics Division, National Center for Theoretical Sciences, Taipei 10617, Taiwan}




\begin{abstract}
C-type shocks are believed to be ubiquitous in turbulent molecular clouds thanks to ambipolar diffusion. We investigate whether the drag instability in 1D isothermal C-shocks, inferred from the local linear theory of Gu \& Chen, can appear in non-ideal magnetohydrodynamic simulations. Two C-shock models (with narrow and broad steady-state shock widths) are considered to represent the typical environment of star-forming clouds. The ionization-recombination equilibrium is adopted for the one-fluid approach.
In the 1D simulation, the inflow gas is continuously perturbed by a sinusoidal density fluctuation with a constant frequency. The perturbations clearly grow after entering the C-shock region until they start being damped at the transition to the postshock region. We show that the profiles of a predominant Fourier mode extracted locally from the simulated growing perturbation match those of the growing mode derived from the linear analysis. Moreover, the local growth rate and wave frequency derived from the predominant mode generally agree with those from the linear theory. Therefore, we confirm the presence of the drag instability in simulated 1D isothermal C-shocks. We also explore the nonlinear behavior of the instability by imposing larger-amplitude perturbations to the simulation. We find that the drag instability is subject to wave-steepening, leading to saturated perturbation growth.
Issues concerning local analysis, nonlinear effects, one-fluid approach, and astrophysical applications are discussed.
\end{abstract}



\section{Introduction} \label{sec:intro}
Stars form in the cold and dense cores of molecular clouds in the interstellar medium \citep[e.g.,][]{KE12,HI19,Girichidis20}. Observations and numerical simulations suggest that molecular clouds exhibit supersonic and magnetized turbulence  \citep[e.g.,][]{ES04,BP07,HF12}. The resulting shocks can lead to gas compression and energy dissipation, which have critical impacts on the environment of star-forming clouds \citep[e.g.,][]{Smith00}. Additionally, the star-forming clouds are weakly ionized by cosmic rays with a typical ionization fraction $\lesssim 10^{-6}$ \citep[e.g.,][]{Draine,Tielens05,Dalgarno06,Indriolo12}. The ions, coupled with the stressed magnetic fields, strive to drift against the ion-neutral drag in a cloud.  This non-ideal magnetohydrodynamic (MHD) phenomenon, known as ambipolar diffusion, pervades the weakly ionized plasma in star-forming regions \citep{Spitzer1956,MS1956,Shu,Zweibel}. It also influences the structure of interstellar shocks. 
When a supersonic shock travels slower than the magnetosonic speed of the ions, a jump-type shock (J-shock) is preceded by a magnetic precursor due to the ambipolar diffusion \citep{Mullan,Draine80}. The extent of the steady magnetic precursor depends on the interplay between the magnetic field and thermodynamics. When the field strength is strong enough or when the radiative cooling is efficient, the J-shock following behind disappears and the neutral flow remains supersonic throughout, thereby forming a continuous-type shock (C-shock) with a smooth transition in physical quantities between pre- and postshock regions \citep{HCM1989,RD1990,DM93}.\footnote{For example, C-shocks form when the shock speed is not higher than $\sim$ 40-50 km/s to dissociate H$_2$ and thus quench strong H$_2$ line emissions in the environment of the BN-KL region of Orion molecular clouds \citep{Chernoff1982}.}

With turbulence-enhanced ambipolar diffusion,
C-shocks 
have been posited to play a key role in core and star formation \citep[e.g.,][]{LN04,CO12,CO14}. Observationally, various efforts have been made
 to detect such features in turbulent molecular clouds \citep[e.g.,][]{LiHoude08,Hezareh10,Hezareh14,XuLi16,Tang18}, though most of them are indirect measurements and highly dependent on the adopted dynamical and chemical models \citep[e.g.,][please see also the Introduction of Gu \& Chen 2020, hereafter \citet{GC20}]{Flower98,Flower10,Gusdorf08,LehmannWardle16,Valdivia17}. More elaborate models and observational applications for a C-shock substructure can include a J-shock tail when the dynamical timescale of the astrophysical flow, such as outflows from young stellar objects or supernovae, is short enough that the shock structure of multifluid as a whole has not reached a steady state \citep[e.g.,][]{Chieze,Lesaffre,Karska,Anderi}. In this work, we restrict ourselves to stationary (i.e. steady-state) isothermal C-shocks in the typical environment of star-forming clouds \citep{CO12},  where the cloud lifetime is sufficiently long, typically about tens of millions of years \citep{Engargiola,Blitz,Kawamura,Murray,Miura,Meidt,JK18}.
 The simplicity enables us to set up an equilibrium C-shock background and investigate the shock substructure induced by dynamical instabilities alone.

C-shocks in a steady state are subject to dynamical instabilities. It is well-known that C-shocks are susceptible to the Wardle instability, provided that the cosmic ionization and ion-electron recombination processes are not efficient \citep{Wardle90,Wardle91,SmithML97,Stone,Falle}. In the typical environment of star-forming clouds, the timescale of the ionization-recombination process is short vis-\'a-vis other dynamical timescales of interest, and therefore, the ionization-recombination equilibrium is nearly attained. Under this condition for the ambipolar diffusion, \citet{GLV} first showed that a fast drift between the ions and neutrals can provide free energy to facilitate an unstable mode known as the drag instability, which  is a local linear overstability phenomenon associated with an exponentially growing mode of a propagating wave due to the ion-neutral drag. The authors postulated that the instability could occur in a C-shock where the magnetic field is compressed and thus the ion-neutral drift velocity is greatly enhanced, namely, supersonic. 
\citet{GC20}  performed a local Wentzel-Kramers-Brillouin-Jeffreys (WKBJ) analysis to study local perturbations in a 1D isothermal C-shock. The wavenumber and growth/damping rate associated with a local mode of a given wave frequency can be found at each location. Their study 
confirmed the postulation of  \citet{GLV}.
As the unstable mode is advected with the fast shock flow downstream, it
travels a few wavelengths over one growth timescale in the shock frame. 
The total growth of the drag instability is thus limited by the C-shock width. Consequently, the instability requires a finite amplitude to grow to a nonlinear phase. \citet{Gu21} extended the analysis to 2D perpendicular and oblique C-shocks and found that the instability finds its way to avoid the threat from magnetic tension on small scales by exciting the slow mode. For most of the oblique shock angles, the most unstable modes in a 2D C-shock are transversely (i.e. normal to the shock flow) large mode, and thus their growth rate is in general comparable to that derived in the 1D case.

Consequent to the ongoing progress in the linear theory of drag instability, we continue the effort to investigate the instability in numerical simulations. In the present work, we focus on the instability in 1D isothermal C-shocks, aimed at revealing the instability in non-ideal MHD simulations under the guidance of linear theory. Numerical simulations also help in exploring the drag instability beyond the local linear theory -- to calculate the global growth of the instability across the entire shock width and to investigate the nonlinear outcome of the instability with an initial large perturbation. The contents of this paper are structured as follows. In Section~\ref{sec:eqns}, we introduce basic equations for a 1D isothermal C-shock, including the local WKBJ analysis for the perturbations in the one-fluid approach for the neutrals.  In Section~\ref{sec:verify}, we describe the process of setting up a non-ideal MHD simulation and explain the method based on the linear theory, to identify the drag instability in numerical simulations. The simulated results are presented in comparison with the linear results in Section~\ref{sec:result}, followed by the nonlinear results from simulations in Section~\ref{sec:nonlinear}. A couple of points for the drag instability in C-shocks are discussed in Section~\ref{sec:discussion}, concerning the local linear analysis, nonlinear effects, one-fluid approach, and astrophysical applications.
Finally, the method and results are summarized in Section~\ref{sec:sum}.

\section{Basic equations}
\label{sec:eqns}
\citet{CO12} simulated C-shocks produced by converging flows under the strong-coupling
approximation, i.e., the ion-neutral drag force is balanced by the Lorentz force acting on the ion: ${\bf f_d}={\bf f_L}=(1/4\pi)(\nabla \times {\bf B}) \times {\bf B}$, where ${\bf B}$ is the magnetic field. As the timescale for the ionization-recombination equilibrium is short in the typical environment of star forming clouds, the two-fluid equations governing ambipolar diffusion can be
reduced to the following one-fluid equations for the density $\rho_n$, pressure $p_n$, and velocity $\bf v_n$ of the neutrals as well as $\bf B$ \citep[e.g.,][]{Mac95}:
\begin{eqnarray}
\frac{\partial \rho_n}{\partial t} + \nabla \cdot (\rho_n {\bf v_n} ) =0, \label{eq:mhd1}\\
\rho_n \left[ {\partial {\bf v_n} \over \partial t} + ({\bf v_n} \cdot \nabla){\bf v_n} \right] + \nabla p_n = {1\over 4\pi} (\nabla \times {\bf B}) \times {\bf B}, \\
{\partial {\bf B} \over \partial t} + \nabla \times ({\bf B} \times {\bf v_n} ) = \nabla \times \left\{ {\bf B} \times \left[ {1\over 4 \pi \gamma \rho_i \rho_n} {\bf B} \times ( \nabla \times {\bf B}  \right] \right\},
\label{eq:mhd3}
\end{eqnarray}
where the ion density $\rho_i$ is related to the neutral density by $\rho_i=\sqrt{(\xi_{CR}/\beta )\rho_n}\equiv 10^{-6} \chi_{i,0} m_i \sqrt{\rho_n/m_n}$ in the ionization-recombination equilibrium, $\xi_{CR}$ is the cosmic-ray ionization rate, $\beta$ is the recombination rate coefficient, $m_i=30m_H$ is the ion mass, $m_n=2.3m_H$ is the neutral mass, and $\gamma=3.5\times 10^{13}$ cm$^3$ s$^{-1}$ g$^{-1}$ is the drag force coefficient \citep{Draine}. 
With the basic state for a 1D steady-state isothermal C-shock given by \citet{CO12}, 
the local perturbations $U(\omega,k)=(\delta B, \delta \rho_n, \delta v_n)^T$ multiplied by $\exp(\rmi kx+ \rmi \omega t)$ are governed by the following linearized equation under the WKBJ approximation:
\begin{equation}
C_n U = \rmi \omega U, 
\label{eq:linear_eq}
\end{equation}
where
\begin{eqnarray}
C_n=\left[\arraycolsep=6pt
\begin{array}{ccc}
-\rmi k V_n-D_{ambi} k^2  & {3\over 2} \rmi kV_d {B\over \rho_n} & -\rmi k B \\
0 & -\rmi k V_n & -\rmi k \rho_n \\
-\rmi k {V_{A,n}^2 \over B} & -\rmi k {c_s^2 \over \rho_n} -{\gamma \rho_i V_d \over \rho_n}& -\rmi k V_n 
\end{array}
\right],
\label{eq:Cn}
\end{eqnarray}
and a few terms associated with the background gradients in $C_n$ have been ignored due to the spatial scale of local perturbations $1/k \ll$ the local gradient length scale of background states. Moreover,
the constraint of ionization equilibrium, $\delta \rho_i/\rho_i = (1/2) \delta \rho_n/\rho_n$, has been applied to the perturbation of Equation(\ref{eq:mhd3}) in deriving Equation(\ref{eq:linear_eq}). 
In Equation(\ref{eq:Cn}),  $V_{A,n}=B/\sqrt{4\pi \rho_n}=B/\sqrt{4\pi n_n m_n}$ is the Alfv\'en speed of the neutrals with the number density $n_n$, $D_{ambi} \equiv V_{A,n}^2/\gamma \rho_i$ is the ambipolar diffusion coefficient, $V_d=- D_{ambi} d\ln B/dx$ is the drift velocity of the ion relative to the neutral due to ambipolar diffusion,  and $c_s= 0.2$
km/s is the isothermal sound speed at a temperature equal to 10 K. 

\citet{GC20} focused on the drag instability within a C-shock in the linear theory including both ion and neutral species, i.e., a two-fluid linear approach without assuming ionization equilibrium and strong coupling for perturbations. Nevertheless, the authors have also derived a few key terms in
Equation(\ref{eq:linear_eq}) to strongly suggest that
the drag instability should remain in the one-fluid approach for a 1D steady-state isothermal C-shock. In the following with Equation(\ref{eq:linear_eq}), we provide a formal and complete derivation for the confirmation. 
Equation(\ref{eq:linear_eq}) admits non-trivial solutions when the determinant of $C_n$ is zero. It then follows that
\begin{equation}
    k^2 V_{A,n}^2 \left( {3\over 2} \rmi k V_d + \Gamma \right)=(\Gamma + D_{ambi} k^2) (-\Gamma^2-k^2 c_s^2 + \rmi k V_d \gamma \rho_i ),\label{eq:complex_dispersion}
\end{equation}
where $\Gamma=\rmi \omega+ \rmi k V_n$ is the the growth/damping rate and wave frequency in the frame of the neutrals.
Following \citet{GLV} and \citet{GC20}, we consider a mode with $\Gamma$ 
to be smaller than the recombination rate $2\beta \rho_i$ and the ambipolar drift rate across the mode wavelength $k|V_d|$, but still larger than the neutral collision rate with the ions $\gamma \rho_i$ and the sound-crossing time over one mode wavelength $kc_s$. Besides, $D_{ambi} k^2 \gg D_{ambi}k/(d\ln B/dx)^{-1} = k|V_d|$ because $k(d\ln B/dx)^{-1} \gg 1$.
Therefore, Equation(\ref{eq:complex_dispersion}) is reduced to
\begin{equation}
    k^2 V_{A,n}^2 \left( {3\over 2} \rmi k V_d  \right) \sim D_{ambi} k^2 (-\Gamma^2+\rmi k V_d \gamma \rho_i ),\label{eq:dp2}
\end{equation}
which yields
\begin{equation}
\Gamma \sim \pm {1+\rmi \over 2} \sqrt{k|V_d|\gamma \rho_i}.
\label{eq:Gamma}
\end{equation}

The above derivation can be physically understood as follows. $-D_{ambi}k^2 \delta B/B \sim (3/2)\rmi k V_d \delta \rho_n/\rho_n$ from the first row of Equation(\ref{eq:linear_eq}) is a result of the balance between perturbed magnetic pressure and drag force; namely, the perturbed version of the strong-coupling approximation. In addition, $\Gamma \delta v_n \sim -\rmi k V_{A,n}^2 \delta B/B -\gamma \rho_i V_d \delta \rho_n/\rho_n$ from the third row of Equation(\ref{eq:linear_eq}) arising from the perturbed magnetic pressure acting on the neutrals. Together with the second row of Equation(\ref{eq:linear_eq}) (i.e., perturbed continuity equation), these simplified equations yield Equation(\ref{eq:Gamma}),
which is the same as the dispersion relation for this particular pair of modes in the two-fluid linear approach. The positive sign corresponds to the growth rate and wave frequency of the drag instability. The phase velocity of the unstable mode in the shock frame is given by $v_{ph}=$Re$[-\omega]/k$, which is slightly larger than the background velocity $V_n$ because the Doppler-shift frequency due to the background flow $kV_n$ is much larger than the intrinsic wave frequency $\sqrt{k|V_d|\gamma \rho_i}/2$ in the frame of the neutrals. In other words, the unstable mode is essentially advected by the shock flow downstream \citep{GC20}.

It should be noted that the simplified dispersion relation described by Equation(\ref{eq:Gamma}) is meant to capture the basic physics underlying the drag instability. In reality, the ionization-recombination equilibrium is not perfectly maintained, and thus there can exist a phase difference between the ion and neutral density perturbations. This subtle detail is incapable of being studied in the one-fluid approach. We shall discuss this issue in Section~\ref{subsec:1fluid}.

\section{Verify the drag instability in a C-shock MHD simulation}
\label{sec:verify}
\subsection{Set up a non-ideal MHD simulation for the stability analysis of a C-shock}
We employ the \textsc{Athena} code to study the presence of the drag instability in a 1D isothermal C-shock. \textsc{Athena} is a grid-based Godunov MHD code designed for astrophysical problems \citep{Stone08}, which has been widely used for studies of galaxy dynamics, star formation, and accretion disks. 
In particular, the ambipolar diffusion effect in \textsc{Athena} was developed and tested by \cite{Bai_Stone_AD}, and has been adopted in star formation simulations studied by \cite{CO14}.
\textsc{Athena} is therefore an ideal numerical tool for evolving Equations(\ref{eq:mhd1})-(\ref{eq:mhd3}) for the environments of star-forming clouds in our study. 
To study the problem of the drag instability in a C-shock, we make use of the sample test problem provided in \textsc{Athena} which generates the steady-state C shock profile following Equations(\ref{eq:mhd1})-(\ref{eq:mhd3}) in \citet{Mac95}, and implement the ionization-recombination equilibrium via a power-law dependence of $\rho_i$ on $\rho_n$, i.e., $\rho_i \propto \rho_n^{0.5}$.

\citet{CO12} analytically derived the steady-state profile of 1D isothermal C-shocks 
under the typical environment of star-forming clouds.
We use such an equilibrium solution to construct the base-state C-shock profile as the initial condition of our 1D simulations. 
Since the drag instability is dominated by density perturbation \citep{GC20},
we introduce density fluctuations on top of the background state 
in the incoming inflow of the preshock region,
and let them propagate downstream at the inflow velocity $v_0$.
We set the density perturbation to be in the form of sinusoidal waves
with given amplitude ($\ll 1$) and wavenumber $k_0$.
To maintain the amplitude and wavenumber of the perturbation in the preshock region, we continuously drive the sinusoidal waves in density with angular frequency $v_0 k_0$ at the inflow boundary during the simulation.
That is, we set the density in the upstream ghost zones to follow the same sinusoidal wave pattern based on the simulation time and the base velocity.
The simulation is then evolved with Equations(\ref{eq:mhd1})-(\ref{eq:mhd3}) but without self-gravity, to focus on the instability growth within the C-shock. 
Note that as the perturbation travels downstream through the C-shock, these sinusoidal waves 
retain their initial $\omega_{wave}$ but change $k$ due to the background gradient.

\subsection{Calculation of the growth rate from an MHD simulation}
\label{sec:method}
It can be difficult to compute the growth rate of the drag instability directly from a non-ideal MHD simulation
because the unstable mode is expected to travel 
a few wavelengths over one growth timescale in the shock frame
and because its growth rate 
and the wavenumber vary with $x$ due to the background gradient \citep{GC20}.
However, we can attempt to extract the local growth rate indirectly from a simulation by using the linear equation. Specifically,
the linearized continuity equation for the neutrals reads (i.e., the second row of Equation(\ref{eq:linear_eq}))
\begin{equation}
\rmi \omega \delta \rho_n = -\rmi k  (\delta \rho_n V_n  + \rho_n \delta v_n).
\label{eq:linear_cont}
\end{equation}

We now substitute the identity $\delta v_n/V_n = C(x) \delta \rho_n/\rho_n \exp(\rmi \Delta \phi)$ in the above equation, where $C(x)=(|\delta v_n|/V_n)/(|\delta \rho_n|/\rho_n)$ and $\Delta \phi$ is the phase difference between $\delta v_n$ and $\delta \rho_n$. Thus,  Equation(\ref{eq:linear_cont}) becomes
\begin{equation}
\omega = -  k V_n ( 1+ C\exp(\rmi \Delta \phi) ),
\end{equation}
which in turn gives the instantaneous growth rate, which is the negative of the imaginary part of $\omega$:
\begin{equation}
\Gamma_{grow}=-{\rm Im}[\omega] =  C k V_n \sin(\Delta \phi),\label{eq:growthrate}
\end{equation}
with the wave frequency given by 
\begin{equation}
\omega_{wave}={\rm Re}[\omega]=-kV_n[1+C\cos(\Delta \phi)].
\label{eq:omega_wave}
\end{equation}
Similarly, we can also compute the growth rate from the linearized momentum equation for the neutrals (i.e. the third row of Equation(\ref{eq:linear_eq})):
\begin{equation}
\Gamma_{grow}=-\omega_I={kV_{A,n}^2 C_B \sin(\Delta \phi_B-\Delta \phi) - k c_s^2 \sin(\Delta \phi)-\gamma \rho_i V_d \cos(\Delta \phi) \over V_n C},
\label{eq:growthrate2}
\end{equation}
where we have used the identity $\delta B/B= C_B (\delta \rho_n/\rho_n) \exp(\rmi \Delta \phi_B)$ with $C_B \equiv (|\delta B|/B)/(|\delta \rho_n|/\rho_n)$ and $\Delta \phi_B \equiv \phi_{\delta B}-\phi_{\delta \rho_n}$.

Due to the different eigenvalues in Equation(\ref{eq:linear_eq}), any initial sinusoidal perturbation is the linear combination of the three linearly independent 
modes from the eigenvalue problem described by Equation(\ref{eq:linear_eq}).
After the damping timescales of the decaying modes inferred from the linear theory,  the initial sinusoidal perturbation has propagated into the C-shock and settled to the unstable growing mode in a simulation. At this point, $C$, $C_B$, $\Delta \phi$, and $\Delta \phi_B$ of the unstable mode can be directly extracted from the simulation at each $x$.  The background states are also known at each $x$. Therefore, we can compare the analytical and simulated profiles of the perturbations $\delta \rho/\rho$, $\delta v_n/V_n$ and $\delta B/B$ in terms of their relative amplitudes and phase to examine whether they match.  If they match reasonably, we can estimate the instantaneous growth rate of the drag instability as a function of $x$ from the simulation at a particular time, using either Equation(\ref{eq:growthrate}) or (\ref{eq:growthrate2}). The growth rates of density, velocity, and magnetic-field perturbations should be the same during the exponential growth phase, except when the initial perturbations have not yet settled to the {\bf pure} growing mode driven by the drag instability or when the perturbations have grown significantly to be subject to nonlinear saturation.

In reality, the situation is expected to be more complex because
the perturbations are not exactly sinusoidal even over a distance of 2-3 wavelengths (e.g., see Figure~\ref{fig:pertD_inflow} later in this paper).
A range of wavenumber $k$, amplitudes ($C$ and $C_B$), and phases ($\Delta \phi$ and $\Delta \phi_B$) could be involved in simulated data, say, due to the truncation of the sinusoidal waves and due to $k$, amplitudes and phases being a continuous function of $x$ (see Figures 6 and 8 in \citet{GC20}, and Figures~\ref{fig:linear_fig3_trend} and \ref{fig:linear_V06_trend} later in this paper). 
To compare to sinusoidal modes at a particular $x$ from the linear analysis (see Figure~\ref{fig:perb_comp} later in this paper),
a Fourier transform is performed for the local perturbation from a non-ideal MHD simulation in a spatial range covering two wavelengths to obtain the predominant mode in the $k$ space. This $k$ value of the predominant mode is then applied to the linear theory for the further investigation of its relevance to the drag instability.

\subsection{Case studies}
\citet{GC20} performed a local linear analysis and found the drag instability in C-shocks. The authors adopted  the shock profile described by model Fig3CO12  as a fiducial case, which was presented in Figure 3 in \citet{CO12} with the pre-shock conditions $n_0=500$ cm$^{-3}$, $v_0=5$ km/s, $B_0=10$ $\mu$G, and $\chi_{i,0}=10$.  The shock width is about 0.4 pc and the total growth is significantly limited by the short time span of the unstable mode within the shock in their fiducial model. Following \citet{GC20}, we consider this case in this study. \citet{GC20} also conducted a parameter study for a range of pre-shock conditions 
($v_0 \sim$ 1-6 km/s, $n_0 \sim$ 100-1000 1/cm$^3$, $B_0 \sim 5-10$ $\mu$G, and 
$T = 10$\,K;
see Table 1 in \citet{CO12} and Table 2 in \citet{GC20}),
which are typical for star-forming regions in molecular clouds \citep[see e.g.,][]{McKee2010}.
They found that a stronger shock with a broader shock width promotes the total growth of the drag instability across the shock width. Following the discussion of \citet{GC20}, we also consider model V06 as a representative case for a wider C-shock, of which the pre-shock conditions are given by $n_0=200$ cm$^{-3}$, $v_0=6$ km/s, $B_0=10$ $\mu$G, and $\chi_{i,0}=5$.  In this model, the instability has more total growth in the two-fluid linear analysis because the shock width is about five times larger than that in model Fig3CO12.

\begin{figure}
\plottwo{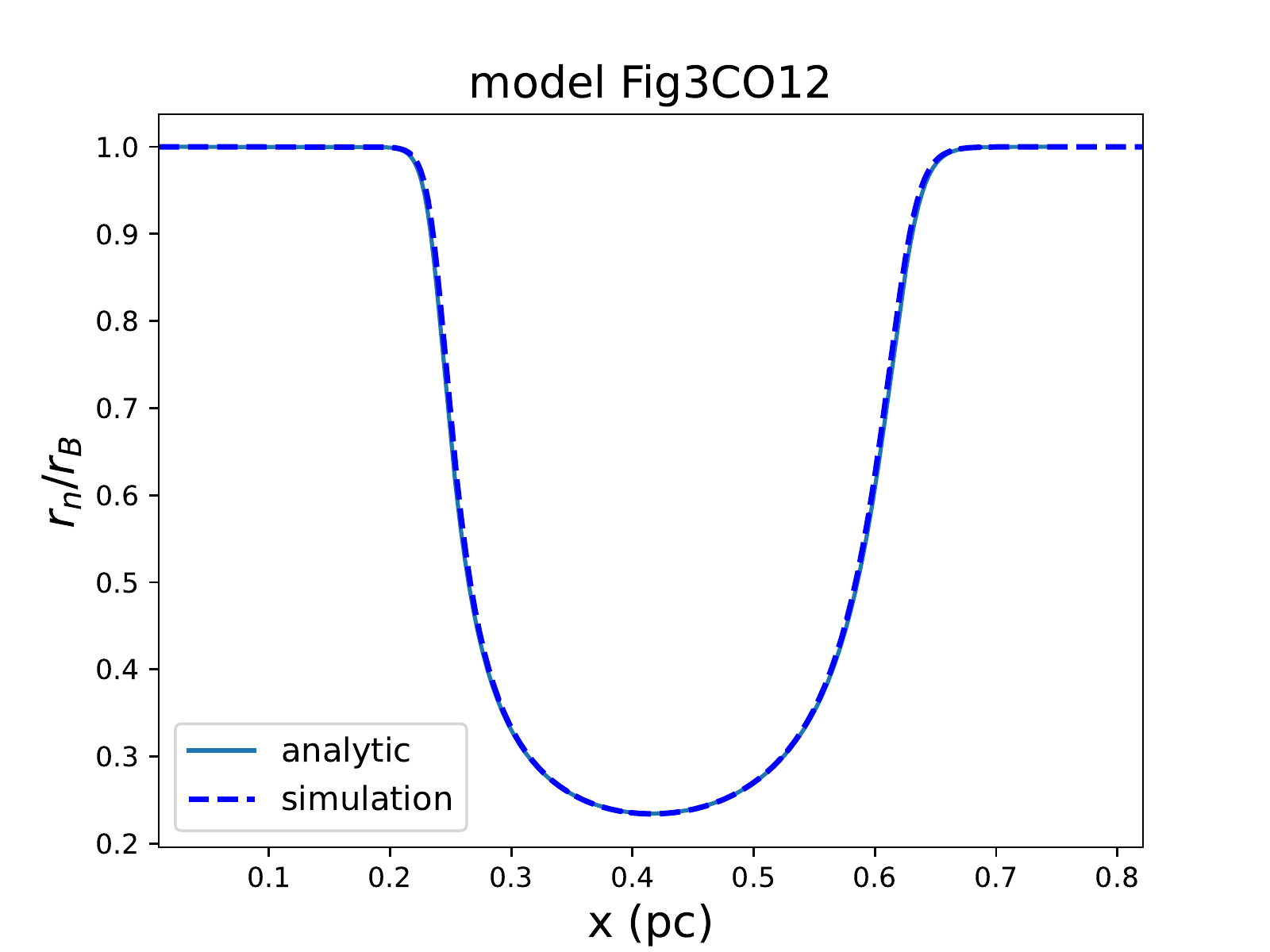}{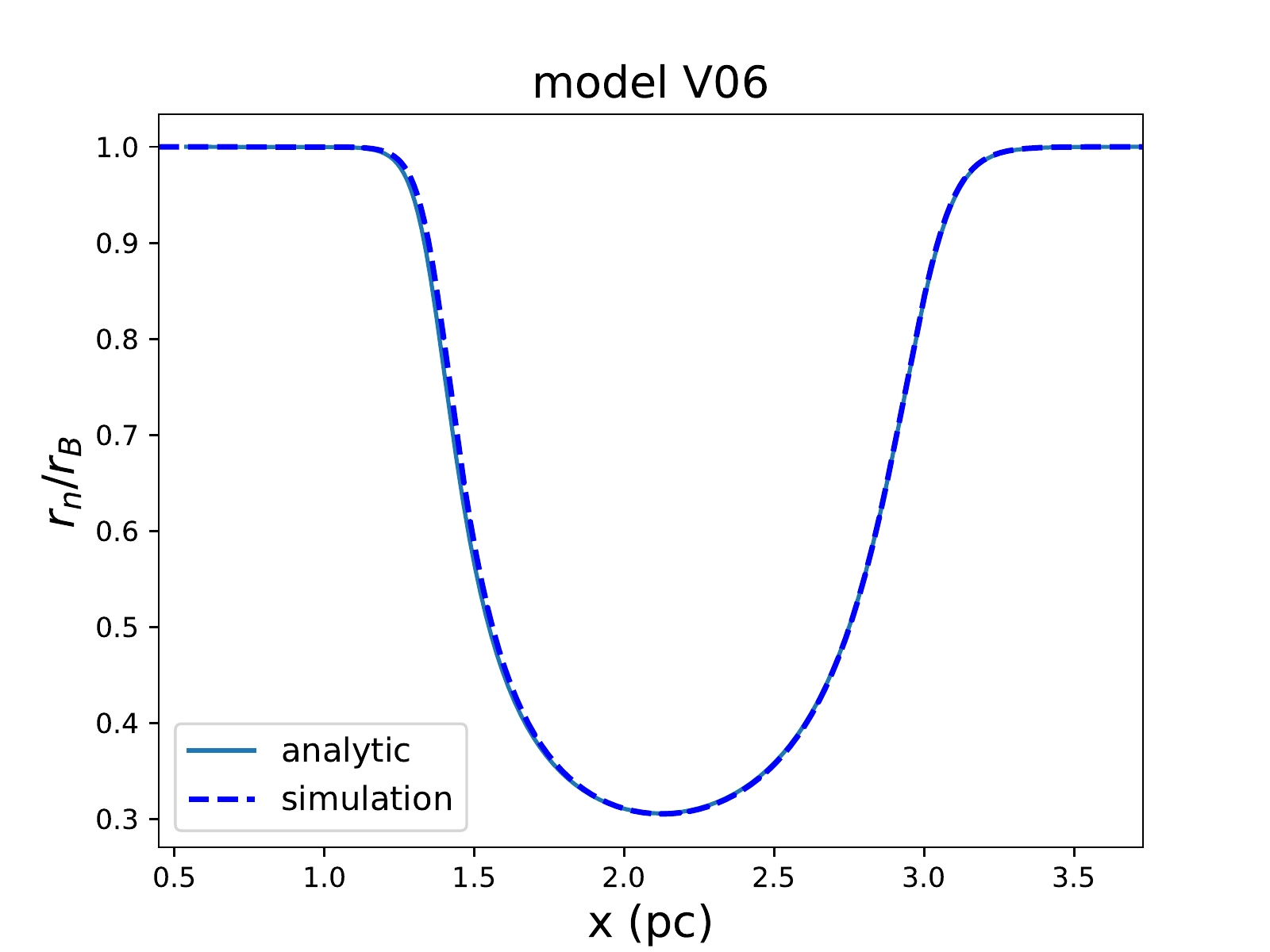}
\caption{Background state presented in terms of $r_n/r_B$ in model Fig3CO12 (left panel) and model V06 (right panel): analytical (solid line) vs. simulated (dashed) results.}
\label{fig:bg_comp}
\end{figure}

\section{Results}
\label{sec:result}
\subsection{Simulated background and perturbation profiles}
We investigate the drag instabilities in C-shocks described by models Fig3CO12 and V06. 
Figure~\ref{fig:bg_comp} shows the background states of these two C-shock models, presented in terms of the profile of $r_n/r_B$, where $r_n$ and $r_B$ are the compression ratio of neutral density and magnetic field, respectively. The simulated C-shock structures generated by \textsc{Athena} following Equations(\ref{eq:mhd1})-(\ref{eq:mhd3}) are highly consistent  with those integrated from the ordinary differential equation derived by \citet{CO12}, thus confirming the valid setup of the background states in these simulations.

\begin{figure}
\centering
\renewcommand{\arraystretch}{0}
\begin{tabular}{cc}
\includegraphics[width=0.35\linewidth]{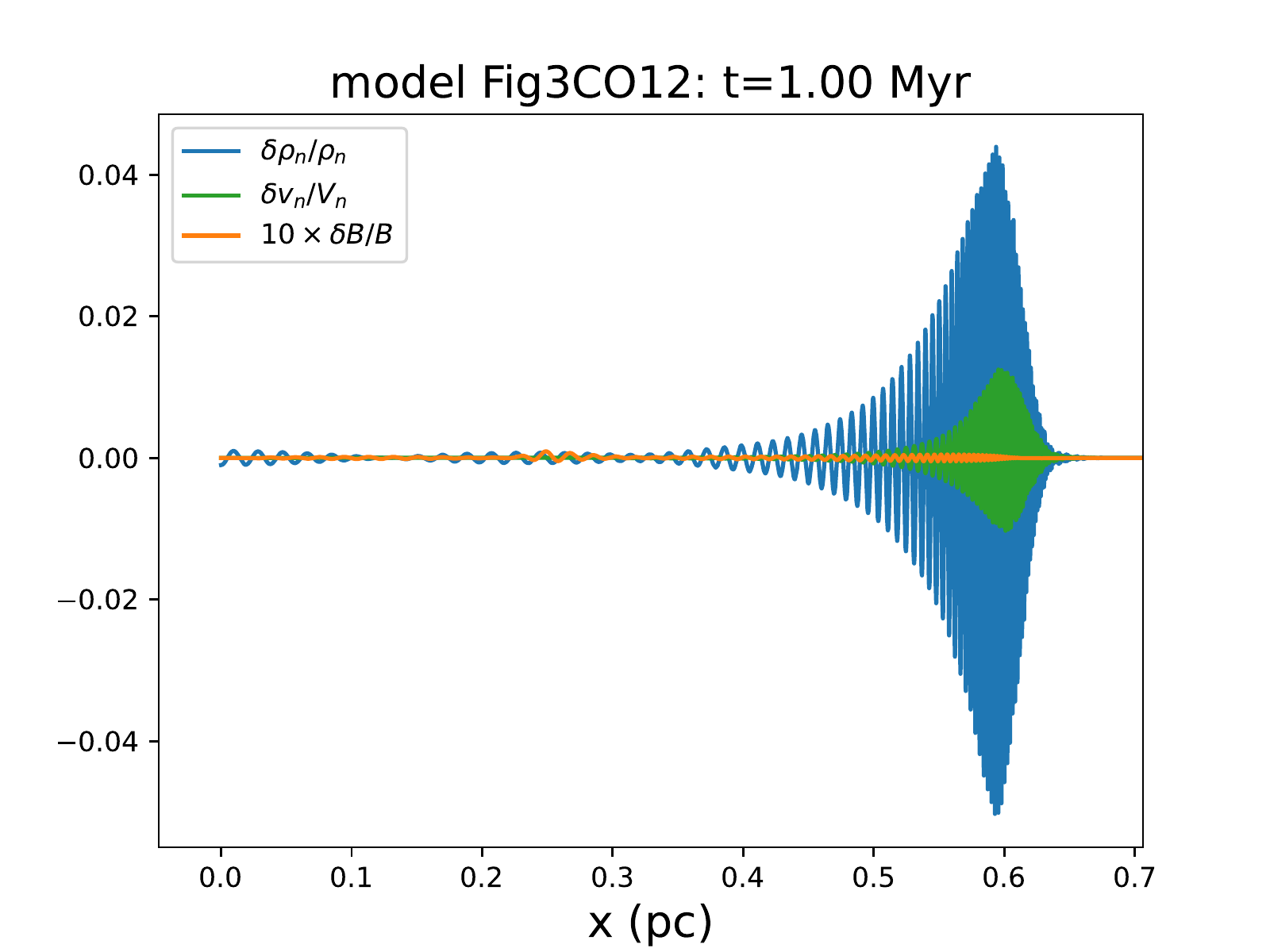} &
\includegraphics[width=0.35\linewidth]{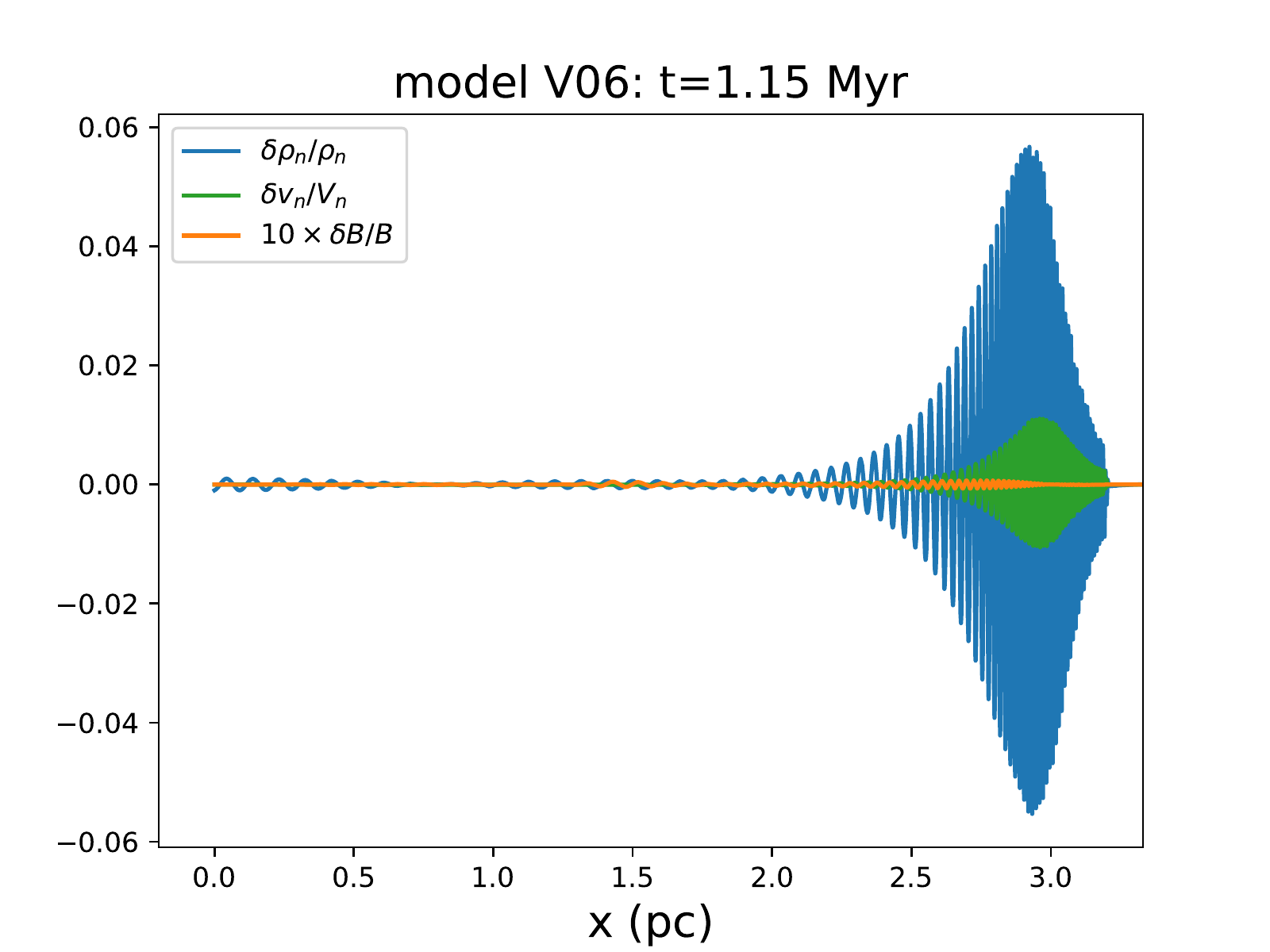} \\
\includegraphics[width=0.35\linewidth]{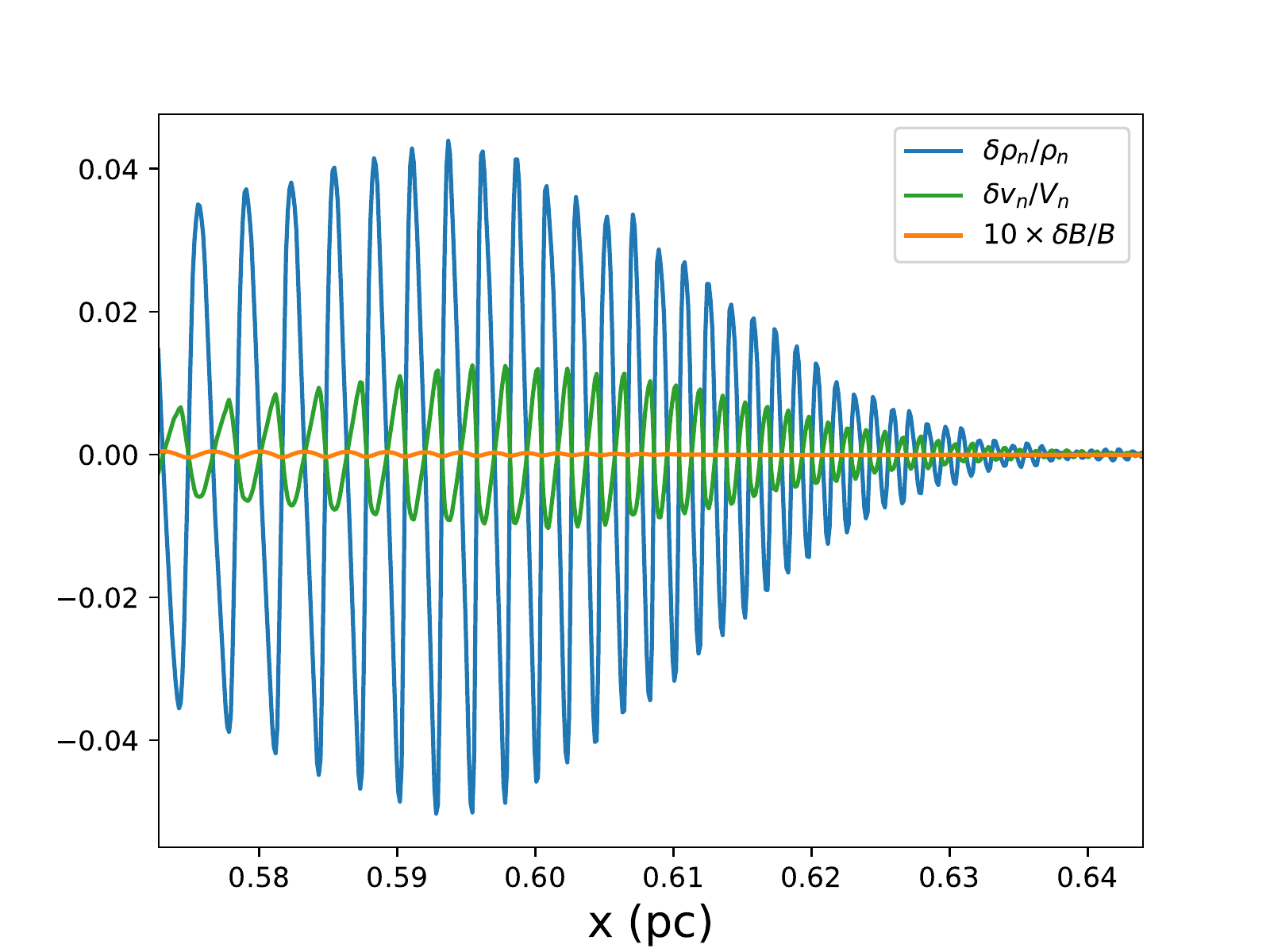}  &
\includegraphics[width=0.35\linewidth]{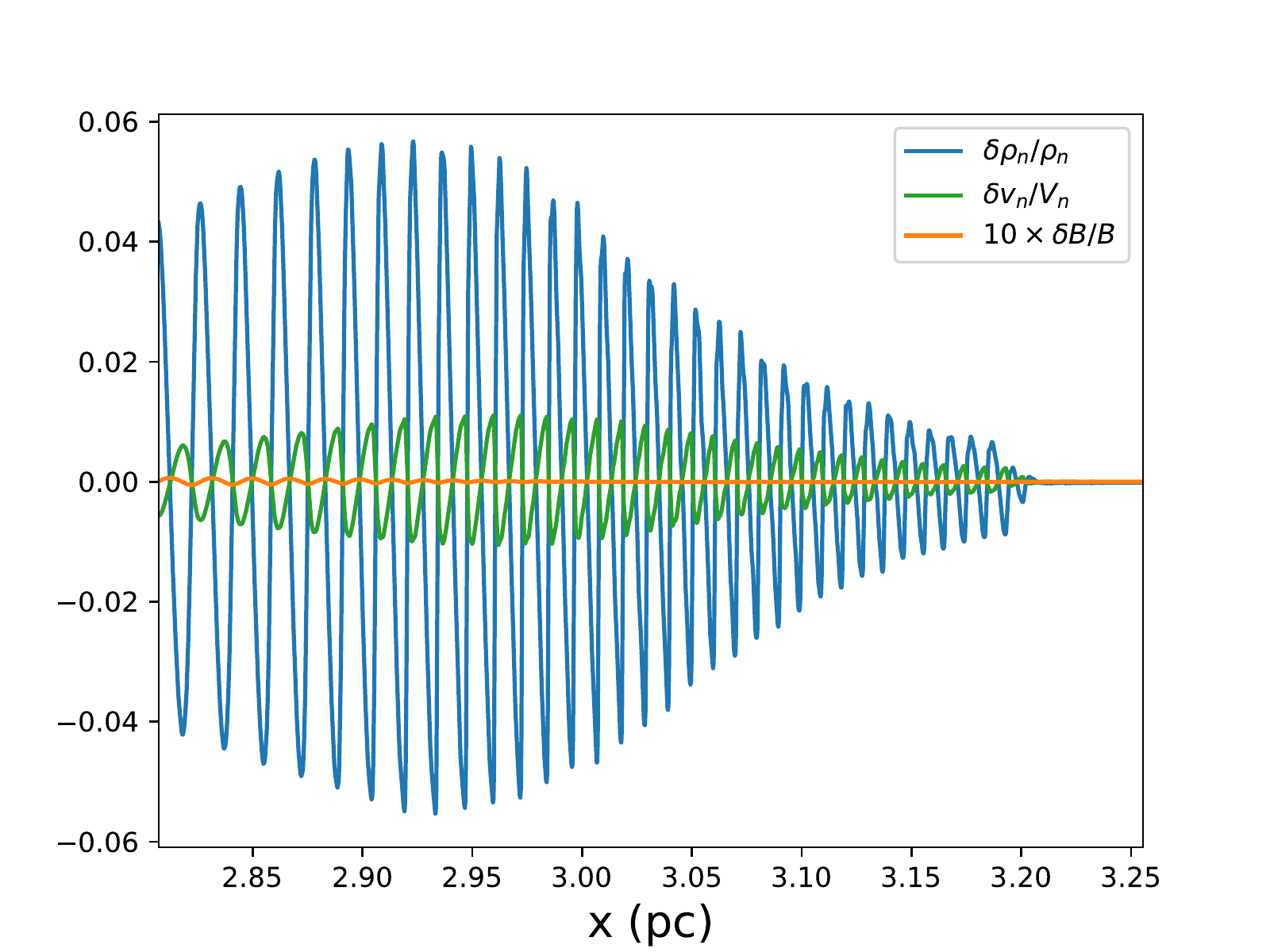} \\
\includegraphics[width=0.35\linewidth]{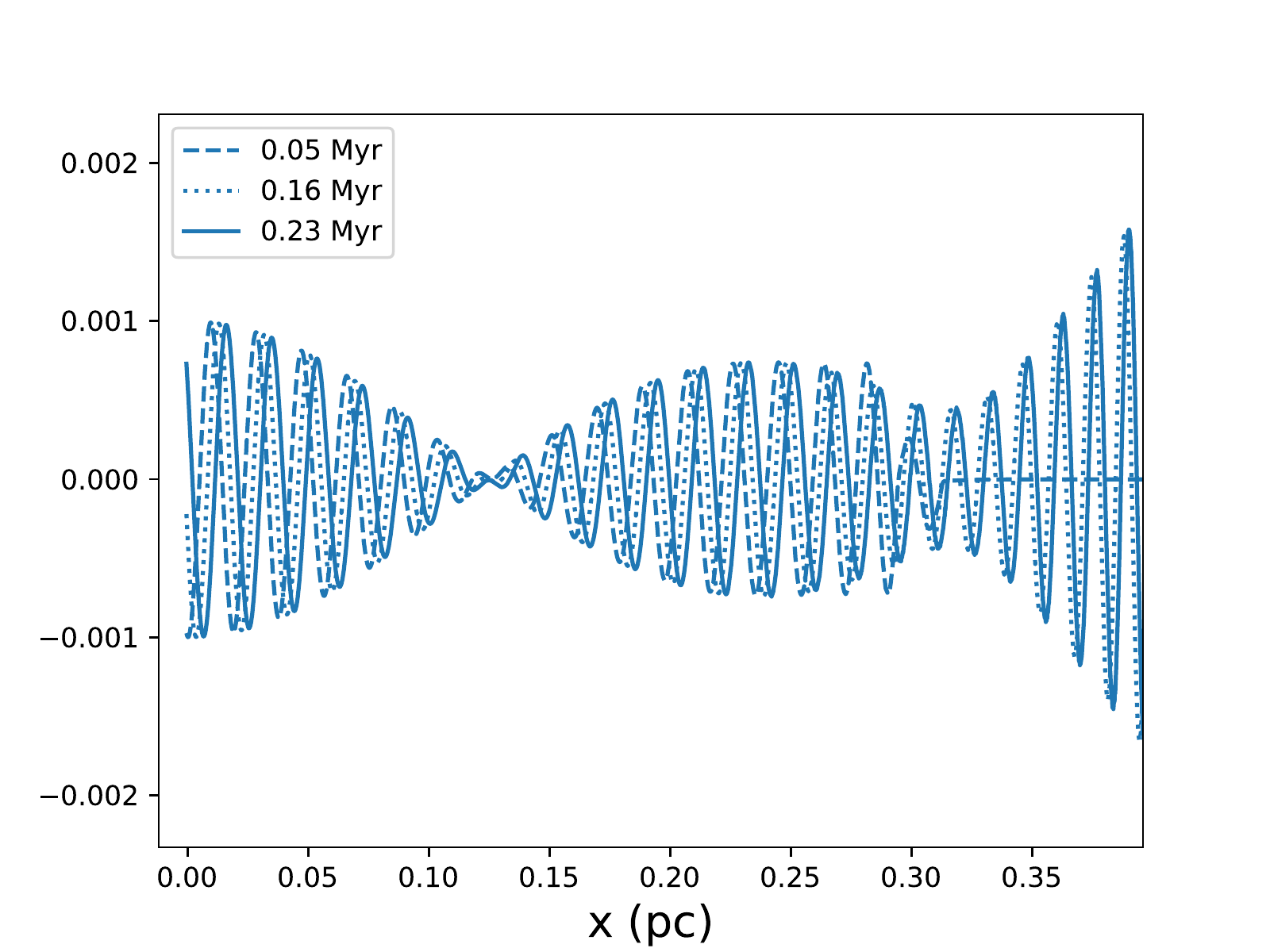}  &
\includegraphics[width=0.35\linewidth]{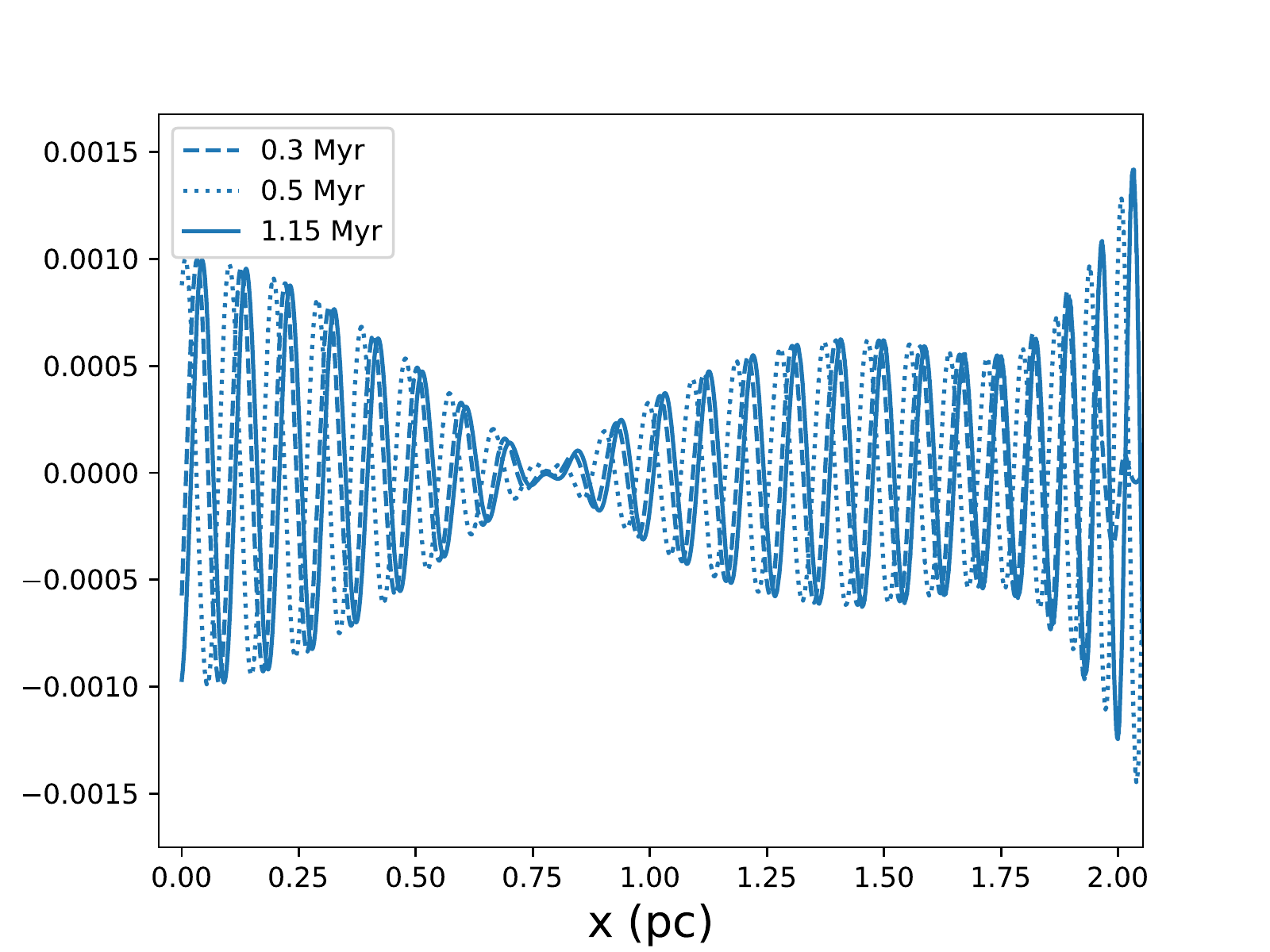} \\
\end{tabular}
\caption{The profile of perturbations from the simulations in models Fig3CO12 (left panels) and V06 (right panels). The magnetic field perturbation is multiplied by ten for viewing clarity. Top row: propagating perturbations advected by the shock flow at $t=1$ Myr for model Fig3CO12 and at $t=1.15$ Myr for model V06.  Middle row: the blow-up of the $x$ range around the beginning of the postshock region from the top panels to better inspect the individual crest and trough of the perturbations downstream. Bottom row:  the evolution of the density perturbation in the preshock and the beginning part of the shock regions to show $|\delta \rho_n/\rho_n|_{init}=0.001$ at $x=0$ and the wave beats.}
\label{fig:pertD_inflow}
\end{figure}

The simulated C-shocks with density perturbations are presented in Figure~\ref{fig:pertD_inflow}. The top panel of Figure~\ref{fig:pertD_inflow} shows snapshots of the sinusoidal propagation of the perturbations with the gas flow along the $+x$ direction, at simulation time $t=1$ Myr for model Fig3CO12 (left) and at $t=1.15$ Myr for model V06 (right), respectively. At these moments in both models, the perturbations have propagated across the entire shock width (see Figure~\ref{fig:bg_comp}), where the drag instability is to be revealed.
The perturbations (initiated at $t=0$ Myr) are maintained by continuously imposing sinusoidal waves to gas density at the inflow boundary (i.e., $x=0$) with a relative amplitude $|\delta \rho_n/\rho_n|_{init} = 0.001$.
Since the drag instability is a local instability, the numerical setup is more 
consistent with the WKBJ approximation for
perturbations with large wavenumbers $k_0$, especially those that satisfy $k_0 \gg 1/L_{\rm shock}$ where $L_{\rm shock}$ is the C-shock width. We thus choose $k_0=1/0.003$ pc$^{-1}$ in Model Fig3CO12 ($L_{\rm shock}\approx 0.4$\,pc) and $k_0=1/0.015$ pc$^{-1}$ in Model V06 ($L_{\rm shock}\approx 1.5$\,pc). 
The grid size of the computation domain is $1/8000$\,pc$^{-1}$ and $4/16000$\,pc$^{-1}$ for models Fig3CO12 and V06, respectively.\footnote{We note that the resolution of simulations presented in this manuscript differs from model to model, which is the consequence of our numerical convergence test; i.e.,~we choose to show the dataset with the highest resolution whether or not such resolution is necessary to resolve the features we aim to discuss. }

As expected from the drag instability in the linear theory, the simulated perturbation grows quickly inside the C-shock (see the top panels of Figure~\ref{fig:pertD_inflow}) with decreasing wavelength with $x$ due to shock compression (see the middle panels of Figure~\ref{fig:pertD_inflow}). 
The amplified perturbation then starts to damp at the beginning of the post-shock region ($x\approx 0.6$ pc in model Fig3CO12 and $\approx 2.9$ pc in model V06; see Figure~\ref{fig:bg_comp} and the middle panels of Figure~\ref{fig:pertD_inflow}). 
The middle panels of Figure~\ref{fig:pertD_inflow} also indicate that the profiles of the velocity perturbation appear to be tilted away from the sinusoidal shape near the end of the shock width in the simulations. It is to be recalled that the phase velocity of the perturbations is close to the background velocity $V_n$. As the velocity perturbation $\delta v_n/V_n \approx \delta v_n/v_{ph}$ increases with $x$, the weak nonlinear effect becomes non-negligible. The velocity crest travels even faster than the velocity trough, which slightly steepens the sinusoidal profiles downstream.

\subsection{Wave beats in space}
\label{sec:beat}
\begin{figure}
\centering
\setlength{\tabcolsep}{0pt}
\begin{tabular}{ccc}
\includegraphics[width=0.35\linewidth]{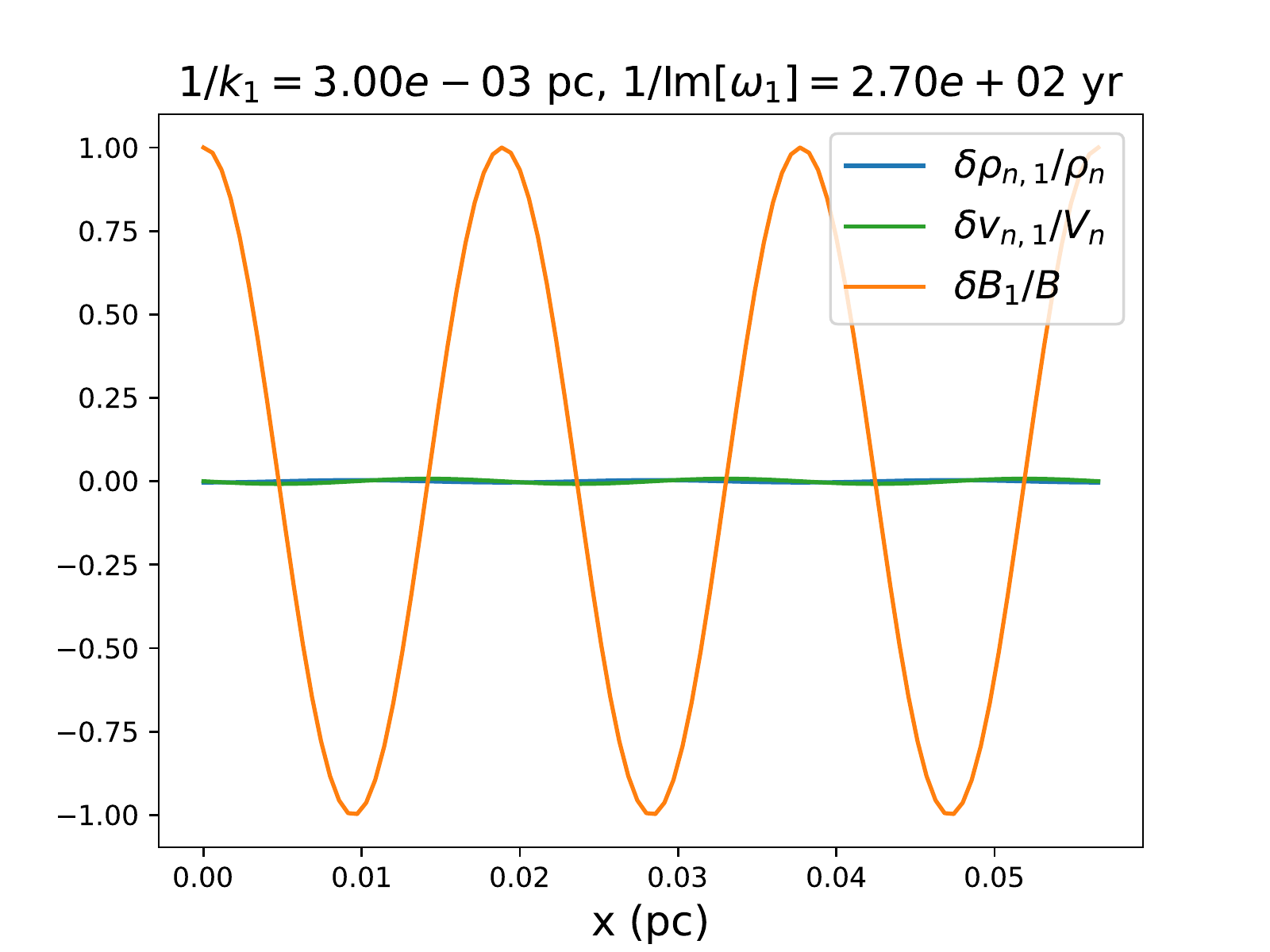} &
\includegraphics[width=0.35\linewidth]{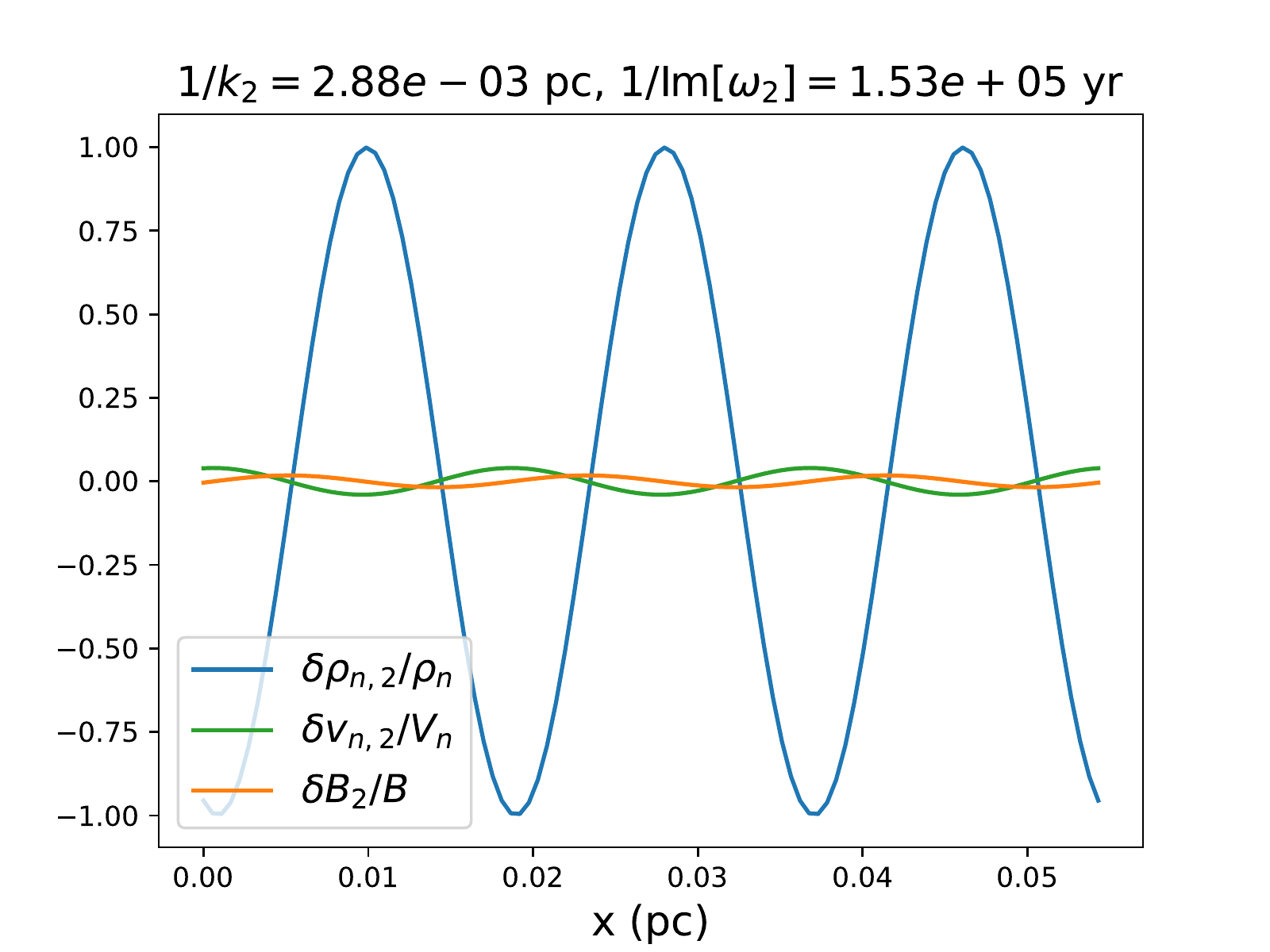}  &
\includegraphics[width=0.35\linewidth]{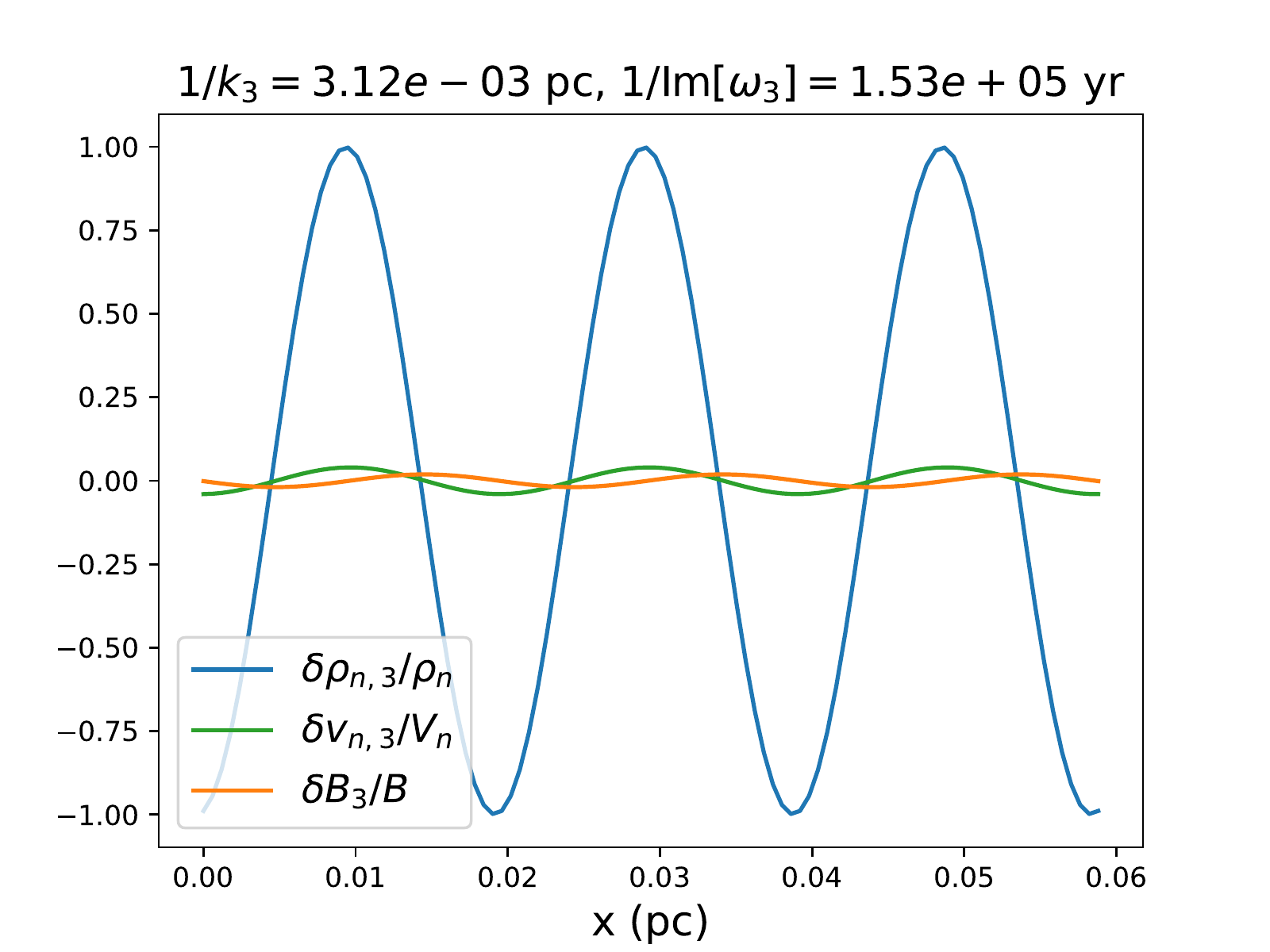}  
\end{tabular}
\caption{The three modes, denoted by the subscripts 1 (left), 2 (middle), and 3 (right panel), are derived from the linear theory with the wave angular frequency $\omega_{wave}=-v_0k_0$ in the preshock region for model Fig3CO12. The profiles of the three modes in the preshock region in model V06 are similar to those in model Fig3CO12 and hence not shown. In model V06, $1/k_1=0.015$ pc, $1/k_2=0.0145$ pc, and $1/k_3=0.0155$ pc.} 
\label{fig:Fig3_3modes}
\end{figure}

Further, the lower panels of Figure~\ref{fig:pertD_inflow} show that before the perturbation is amplified within the shock, the wave amplitude goes to zero at particular locations independent of time, i.e., $x\approx 0.125$ pc in model Fig3CO12 and $x\approx 0.78$ pc in model V06.
This simulated result for the nearly zero amplitude can be explained by the destructive interference of two modes, as the initial density perturbation disperses in the linear theory. Solving Equation(\ref{eq:linear_eq}) in the preshock region with Re$[\omega]=\omega_{wave}=-v_0k_0$, we obtain the three decaying modes (i.e., Im$[\omega]>0$) as illustrated in Figure~\ref{fig:Fig3_3modes}.
The first mode, denoted by subscript 1, is predominated by magnetic field perturbation and is thus subject to magnetic diffusion. This mode is short-lived with the damping timescale 
given by  the ambipolar diffusion timescale $1/(k^2 D_{ambi})\approx$ 270 and 850 years in the linear theory for model Fig3CO12 and V06, respectively.
In contrast, the second and the third modes, denoted by the subscripts 2 and 3 in Figure~\ref{fig:Fig3_3modes}, form a pair of modes predominated by density perturbation and are hence subject to the ion-neutral drag,  thus being long-lived with the damping timescale $\approx (\gamma \rho_i/2)^{-1}$ \citep{GC20},  which are 0.15 and 0.48 Myrs for model Fig3CO12 and V06, respectively. 

The density-only sinusoidal perturbation excited at the left boundary can be expressed by the linear combination of the three aforementioned modes. As explained in detail in Appendix~\ref{app}, the pair of the modes, i.e., the second and third modes, are strongly coupled with the initial density perturbation and are therefore predominantly excited with equal amplitude in contrast to the weakly excited first mode. Further, the second and third modes are slowly damped, compared to the short-lived first mode.
Consequently, the second and third modes are crucial for the subsequent evolution. This is the reason why a density-only perturbation is excited in the simulations. As has been demonstrated in \citet{GC20},
one of these two primary modes becomes the growth mode due to the drag instability with the other as its counterpart, yet decaying mode, once the pair of modes travel into the shock, which can be realized
in Equation(\ref{eq:Gamma}) also.\footnote{Given a constant $\omega_{wave}$, it can be shown from Equation(\ref{eq:Gamma}) that the mode with $k_2$, i.e., the mode with a slightly large wavenumber in the mode pairs, will grow in the shock and the mode  with $k_3$, i.e., with a slightly small wavenumber, will decay in the shock.}
Owing to the slight difference in the wavelength, the second and third modes produce ``wave beats in space" until one of the pair damps and the other grows significantly inside the shock width. In other words, the envelope of the wave beats does not travel with time and its spatial interval $\lambda_{beat}= 2\pi/(k_2-k_3)$ between the locations of complete destructive interference.  Using the values of $k_2$ and $k_3$ derived from the linear theory as shown in Figure~\ref{fig:Fig3_3modes}, we obtain the beat wavelength $\lambda_{beat} \approx 0.24$ and 1.43 pc for models Fig3CO12 and V06, respectively. The linear results are consistent with the simulated locations of complete destructive interference in the preshock region at $x\sim 0.12$ pc in model Fig3CO12 and $x\sim 0.71$ pc in model V06 (i.e. $\approx \lambda_{beat}/2$ from the left boundary), as shown in the bottom panels of Figure~\ref{fig:pertD_inflow}. In Appendix~\ref{app}, we elaborate on the exact locations of zero amplitude of the beat envelope, which match perfectly with the simulation results by also considering the initial phase difference between the mode pair when they are excited at the left boundary. Besides the first zero of the beat envelope, it can be also observed from the bottom panels of  Figure~\ref{fig:pertD_inflow} that the next minimal amplitude of the wave envelope due to beats occurs inside the shock, where $x\approx 0.32$ pc in model Fig3CO2 and $1.75$ pc in model V06. The amplitudes are not nil at these locations, as one of the mode pairs has been growing and the other has been decaying within the C-shock.

 \subsection{Local mode profiles and growth rates}
\begin{figure}
\centering
\renewcommand{\arraystretch}{0}
\begin{tabular}{cc} 
\includegraphics[width=0.5\linewidth]{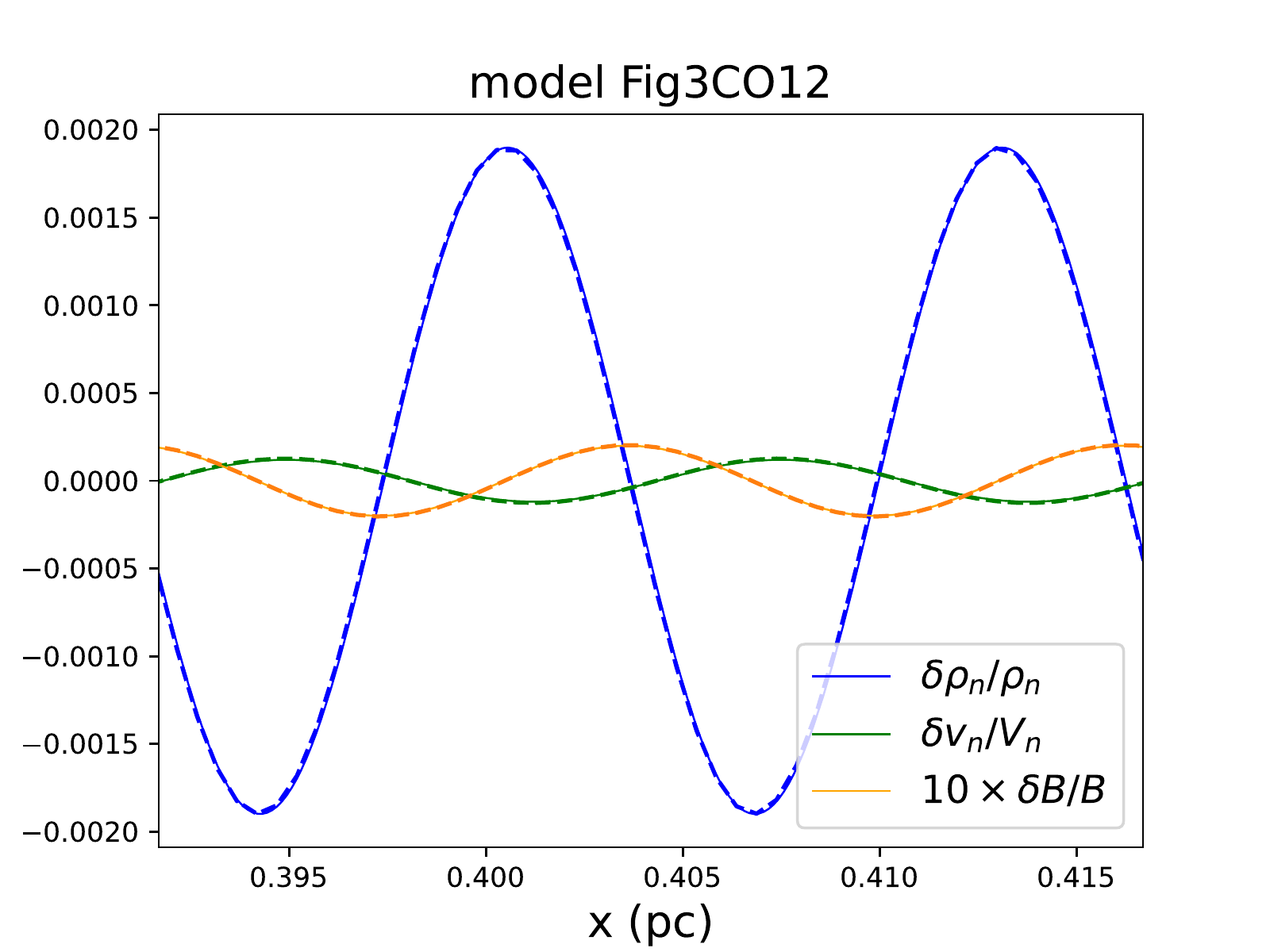} &
\includegraphics[width=0.5\linewidth]{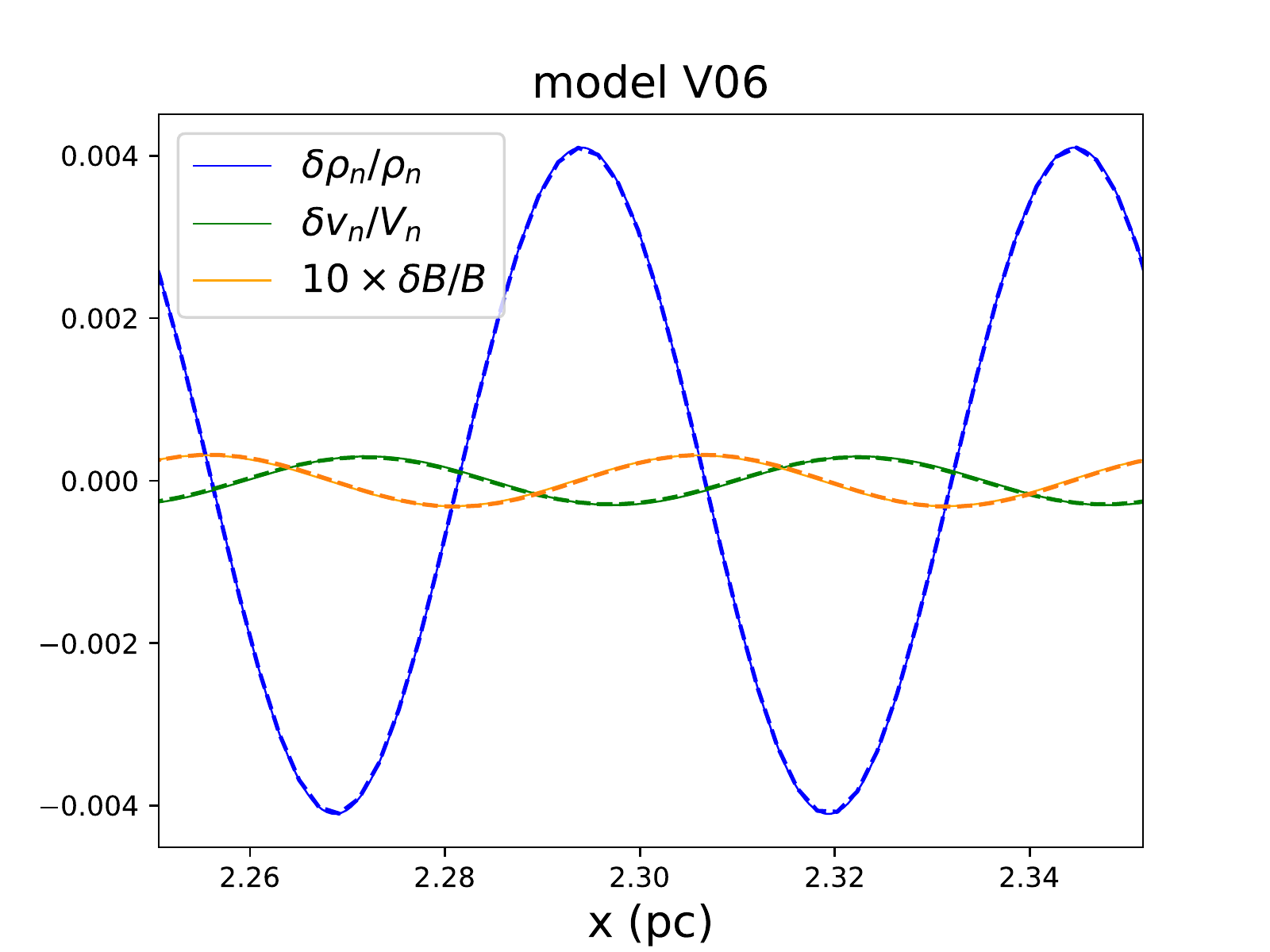}\\
\includegraphics[width=0.5\linewidth]{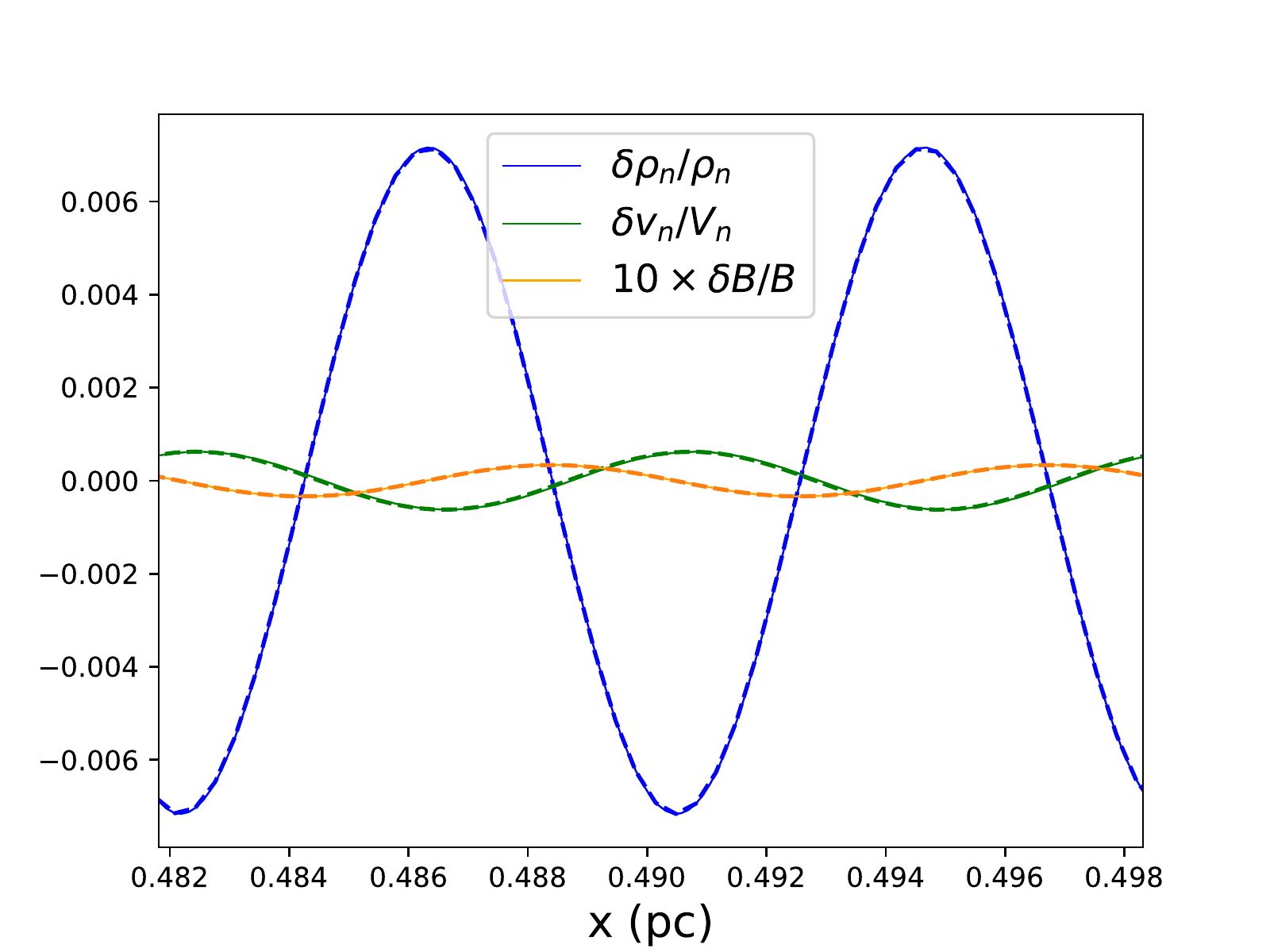} &
\includegraphics[width=0.5\linewidth]{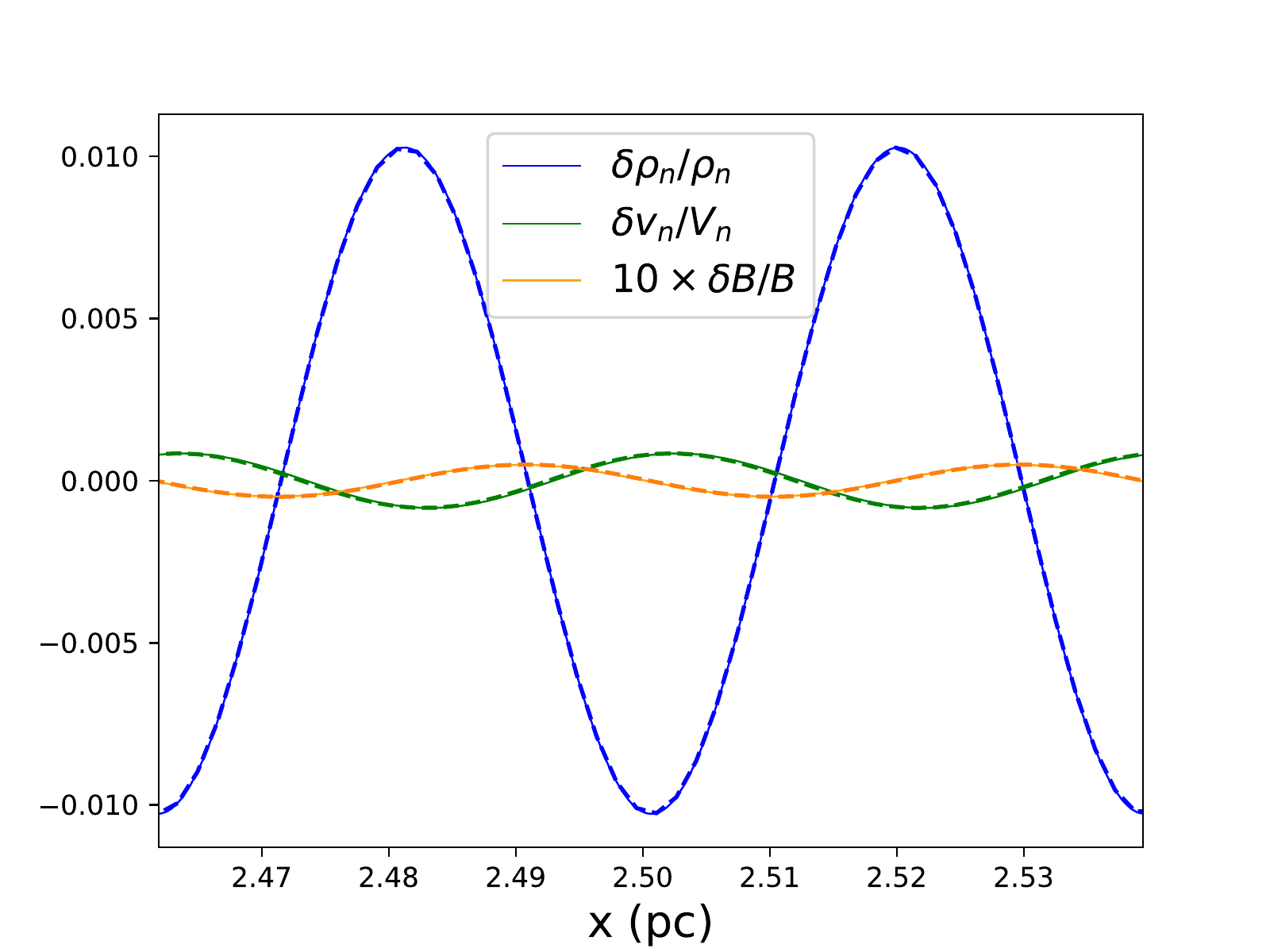} \\
\includegraphics[width=0.5\linewidth]{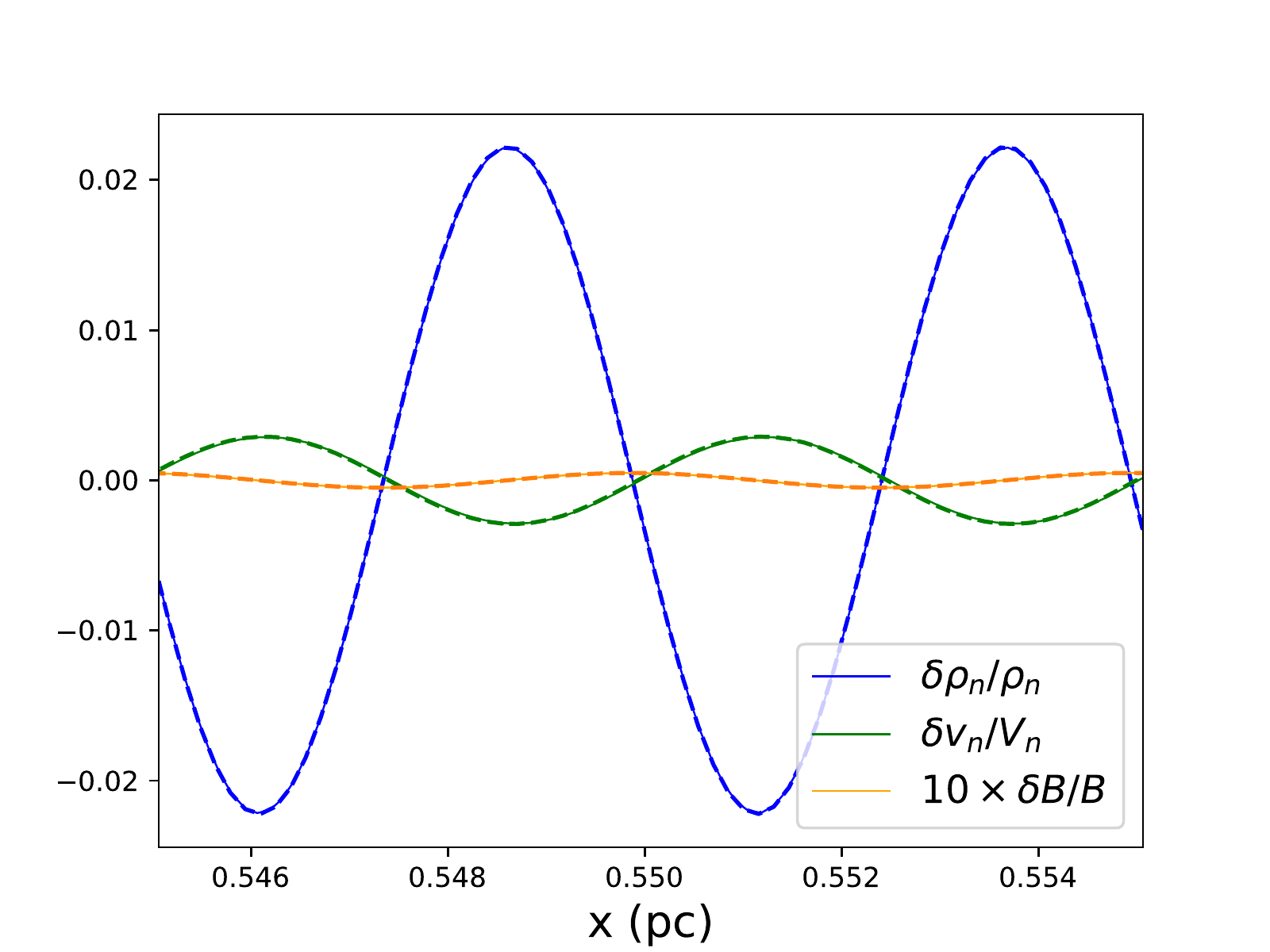} &
\includegraphics[width=0.5\linewidth]{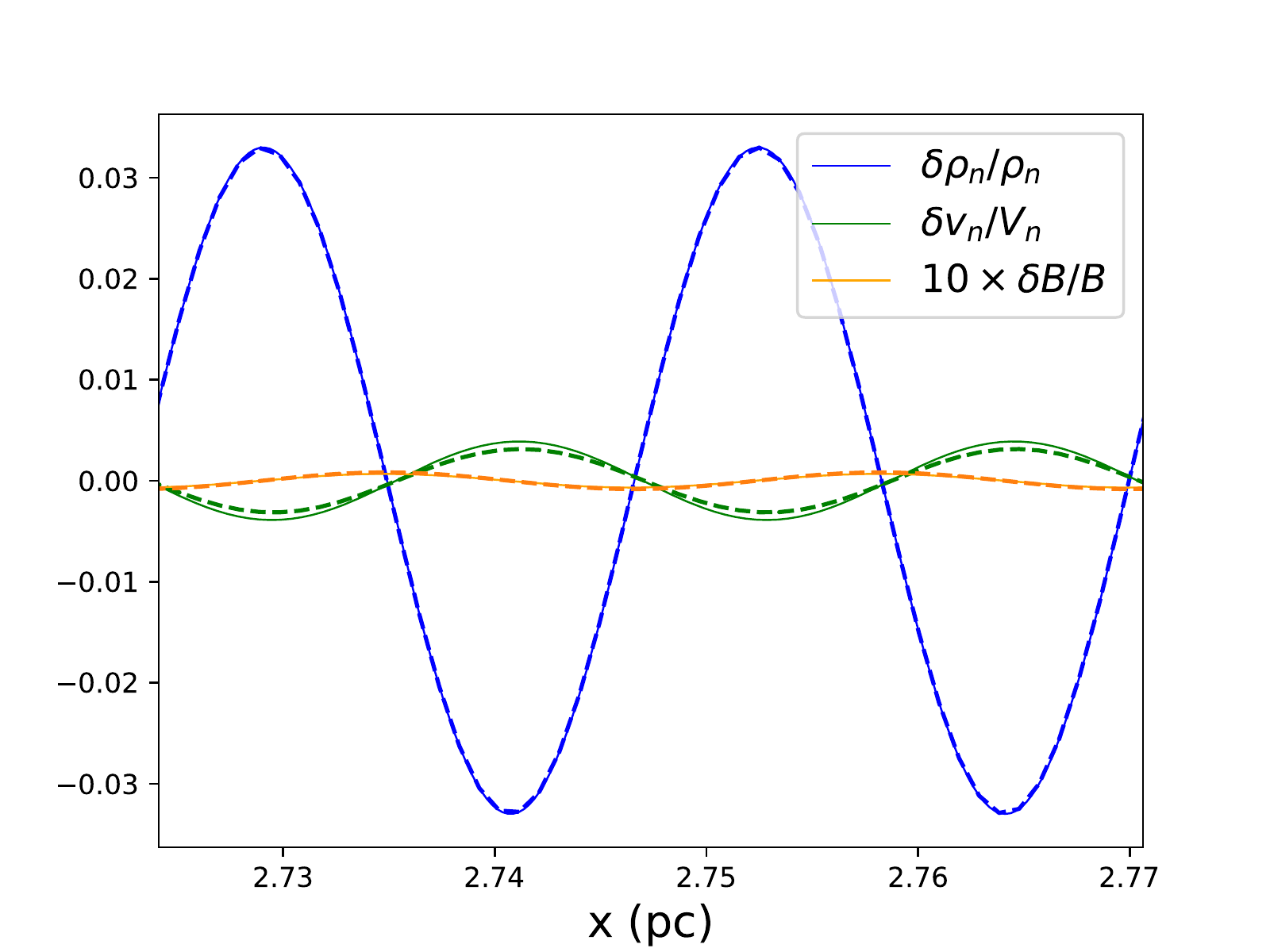} 
\end{tabular}
\caption{Comparison of the perturbation profiles between the dominant Fourier mode of the simulated perturbations (solid line) and
the analytical growing mode (dashed line, scaled to overlap with the simulated density perturbation)  for model Fig3CO12 (left panels) and model V06 (right panels). The results are presented at three locations along the two C-shocks. In model Fig3CO12, the locations correspond to $V_n\approx 3.5$ (top left panel), 2.41 (middle left), and 1.54 km/s (bottom left). In model V06, the locations correspond to
$V_n\approx 3.44$ (top right panel), 2.68 (middle right), and 1.70 km/s (bottom right). The magnetic field perturbation is multiplied by ten for viewing clarity.
}
\label{fig:perb_comp}
\end{figure}

Figure~\ref{fig:perb_comp} compares the perturbation profiles of the simulated (solid line) and analytical results (dashed). The comparison is made at the neighborhood with the range of about two wavelengths of three locations within the C-shock in each model, Fig3CO12 (left panels) and V06 (right panels).  The three locations are selected in the $x$ range of the simulation, where the perturbations have exhibited clear growth within the C-shock (see Figure~\ref{fig:pertD_inflow}). In other words, to identify the instability easily, we are looking at the region where the decaying mode has become much weaker than the growing one  in the pair of density-dominated modes and thus the total perturbation should be primarily contributed by the growing mode associated with the drag instability. 
As described earlier, the simulated profile for perturbations in Figure~\ref{fig:perb_comp}  is the predominant Fourier mode extracted from the simulated result. The wavenumber $k$ of the predominant mode obtained from the Fourier transform is applied to the linear theory to find the unstable mode driven by the drag instability.
Subsequently, we scale the amplitude of $\delta \rho_n/\rho_n$ of the growing mode from the linear theory to overlap that of the predominant mode from the simulation, while keeping the relative amplitude and phase among the perturbations in the linear theory. The procedure yields the comparison plots in Figure~\ref{fig:perb_comp}.

\begin{figure}
\centering
\setlength{\tabcolsep}{0pt}
\renewcommand{\arraystretch}{0}
\begin{tabular}{cc}
\includegraphics[width=0.35\linewidth]{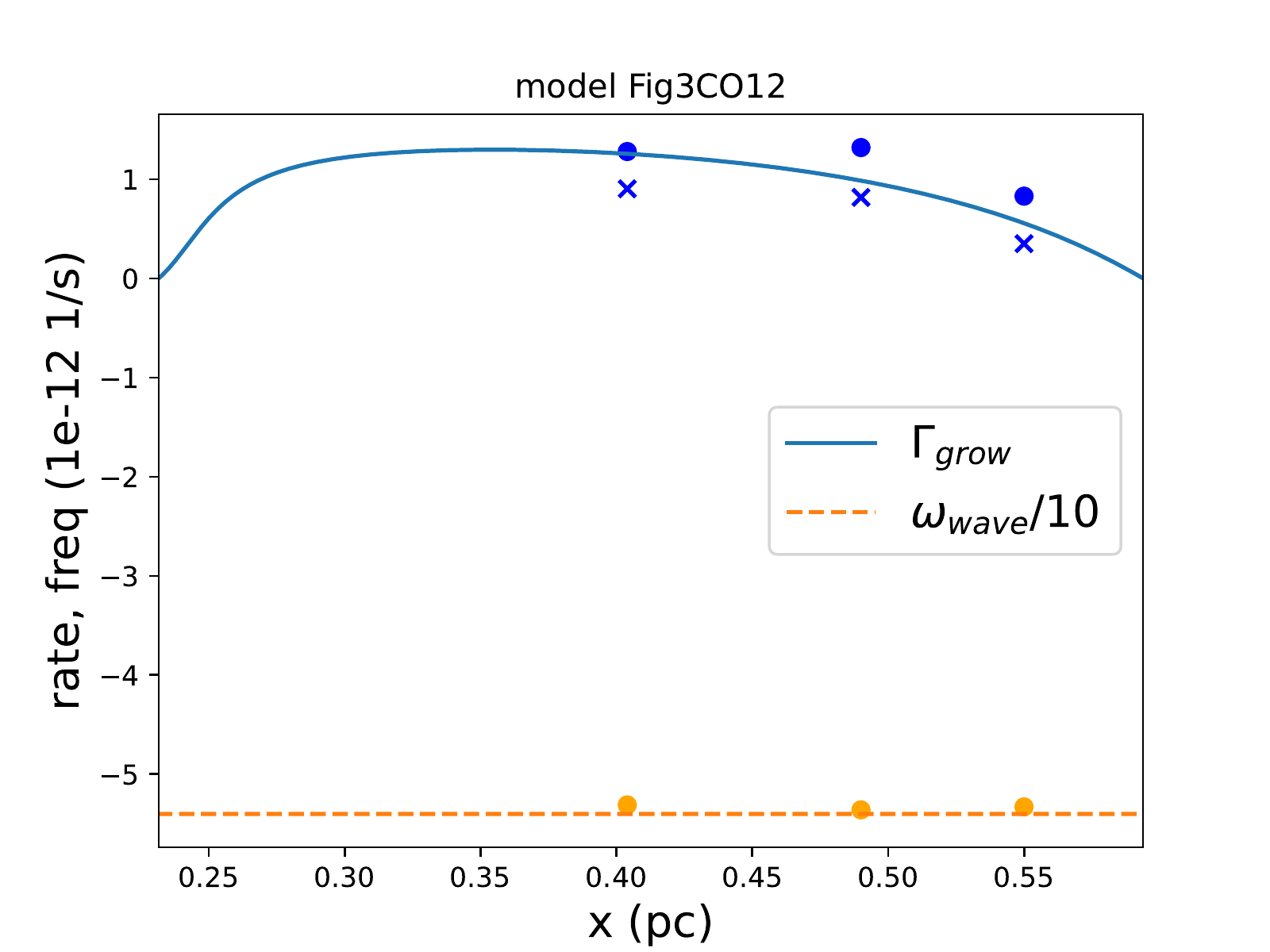} &
\includegraphics[width=0.35\linewidth]{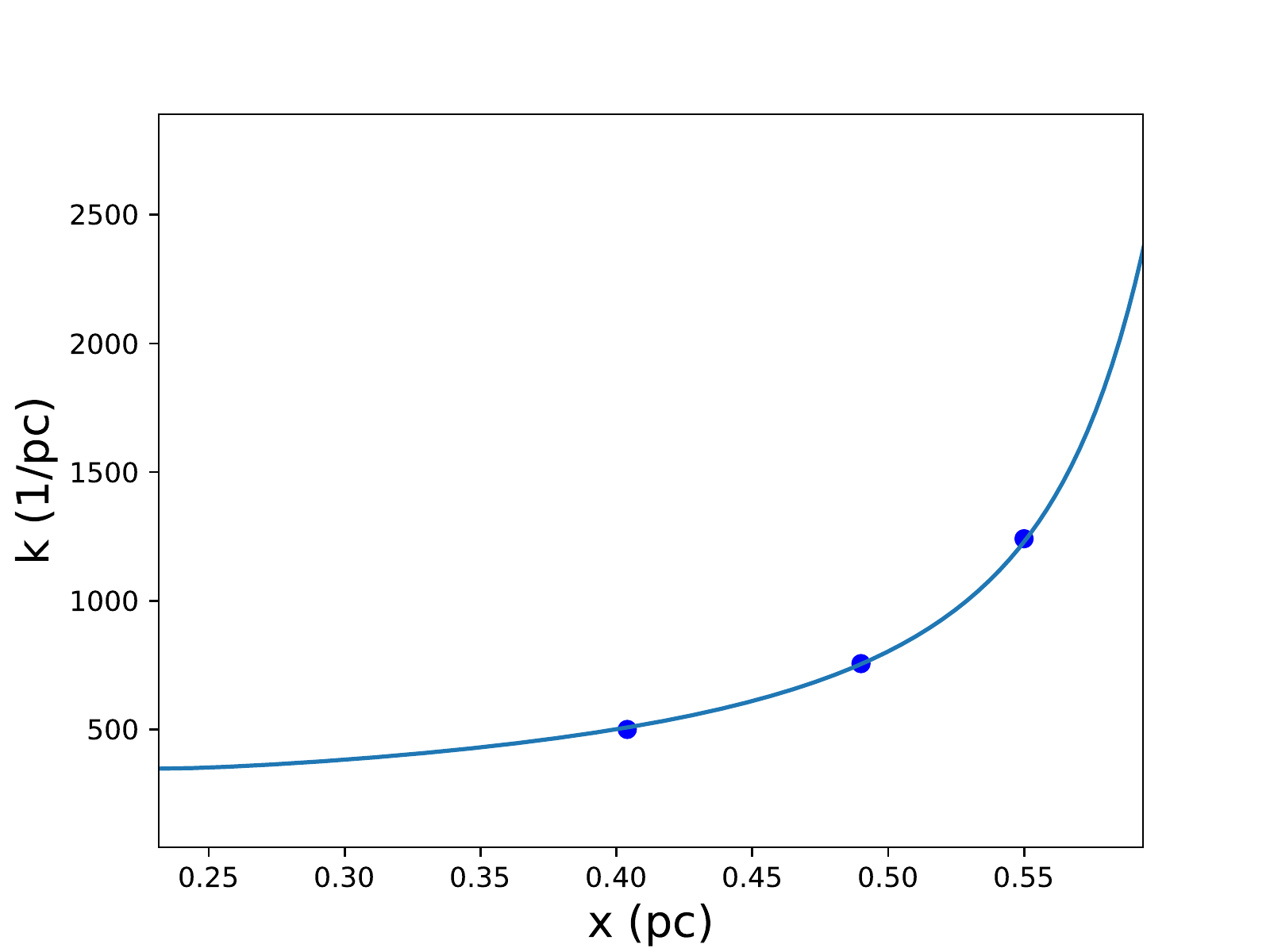} \\
\includegraphics[width=0.35\linewidth]{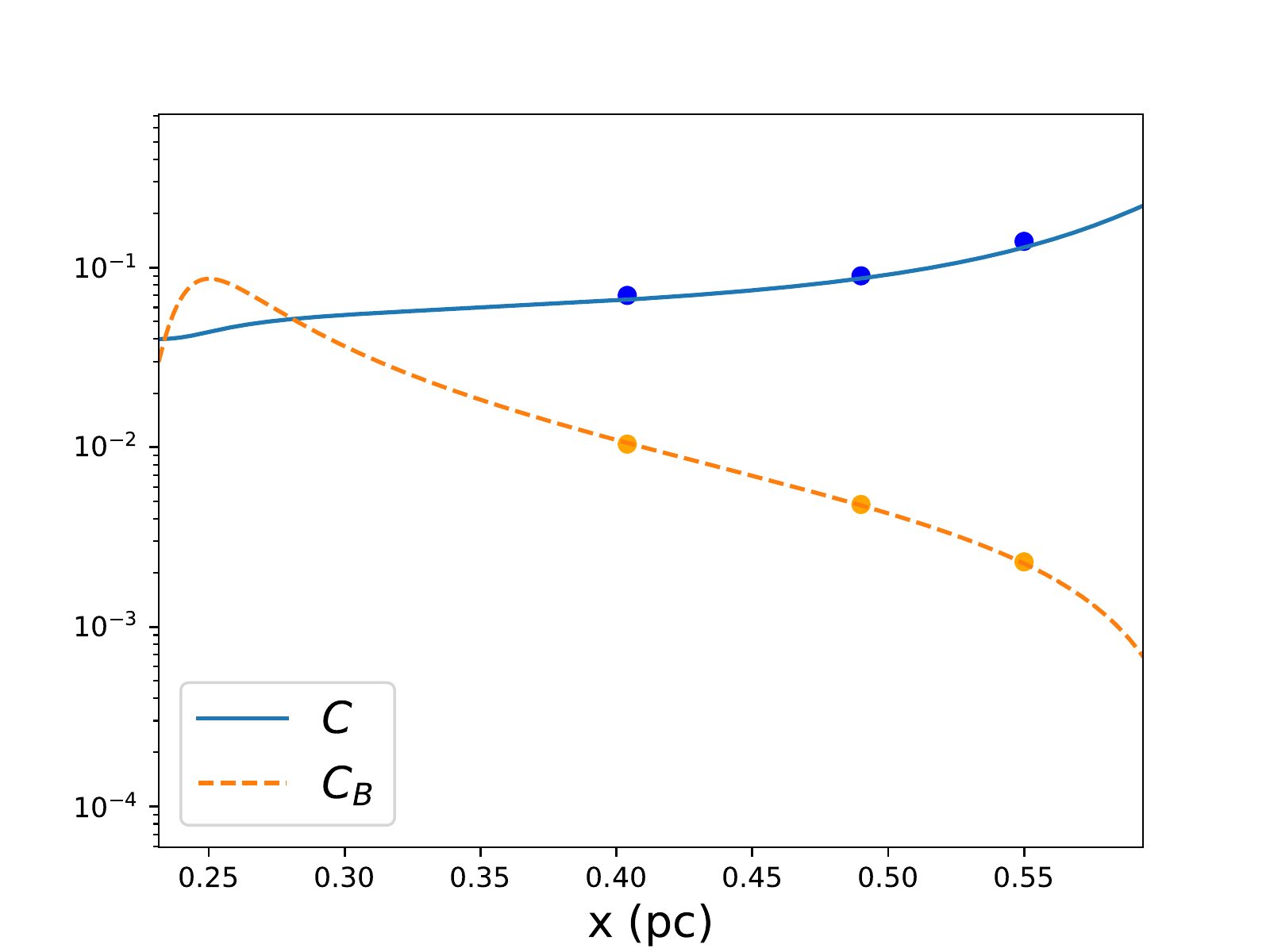}  &
\includegraphics[width=0.35\linewidth]{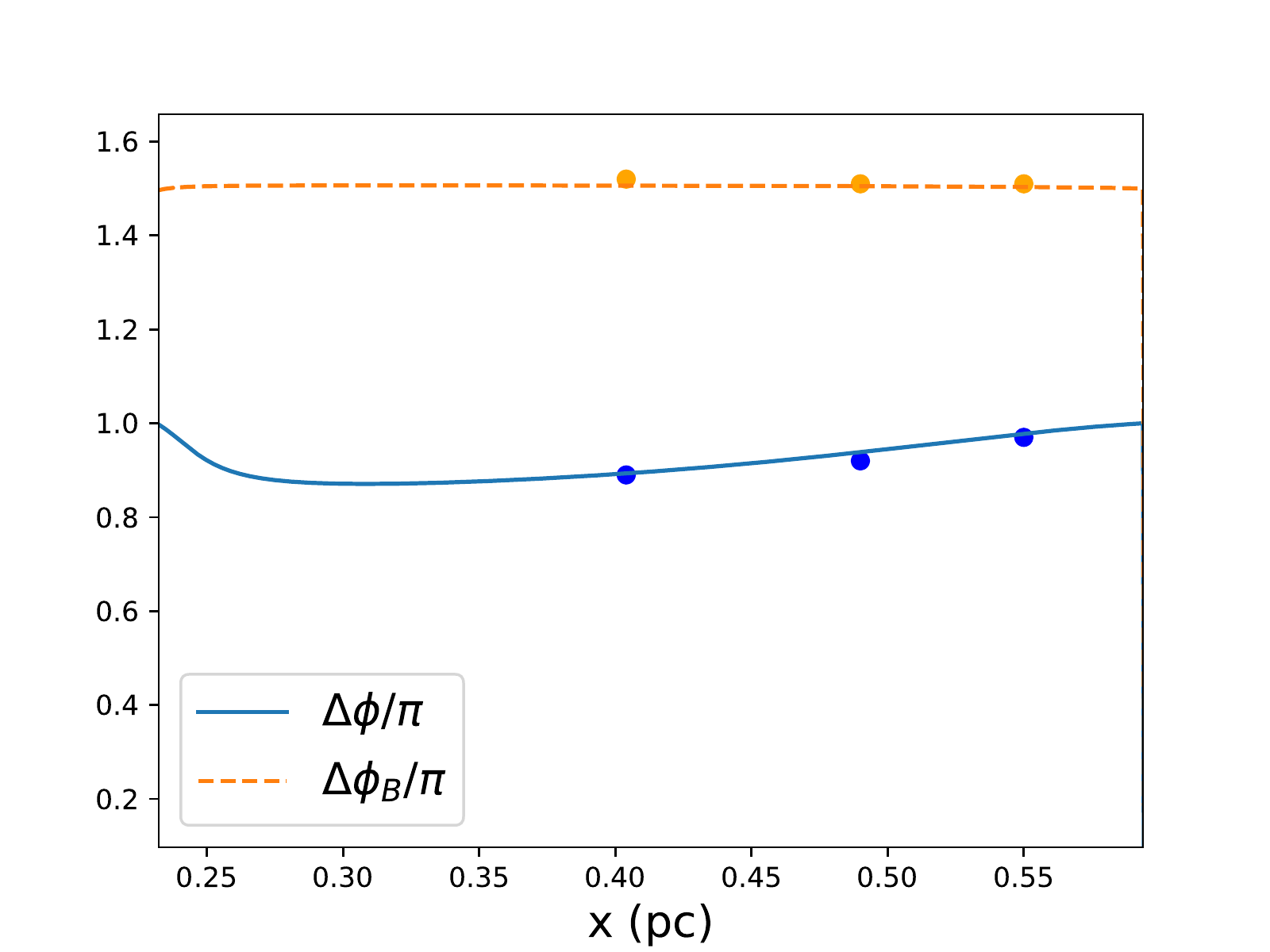} 
\end{tabular}
\caption{Comparison between the profiles of a few properties of the growing mode obtained from the MHD simulation (dots and crosses) and linear theory (curves) in model Fig3CO12. The simulation results correspond to the left three panels at three different locations shown in Figure~\ref{fig:perb_comp}. The curves from the linear theory are obtained from $\omega_{wave}=-v_0 k_0=-54$e$-12$ s$^{-1}$.
In the top left panel, the dots show the growth rate derived from Equation(\ref{eq:growthrate}) and the crosses present the growth rate obtained from Equation(\ref{eq:growthrate2}).}
\label{fig:linear_fig3_trend}
\end{figure}

\begin{figure}
\centering
\setlength{\tabcolsep}{0pt}
\renewcommand{\arraystretch}{0}
\begin{tabular}{cc}
\includegraphics[width=0.35\linewidth]{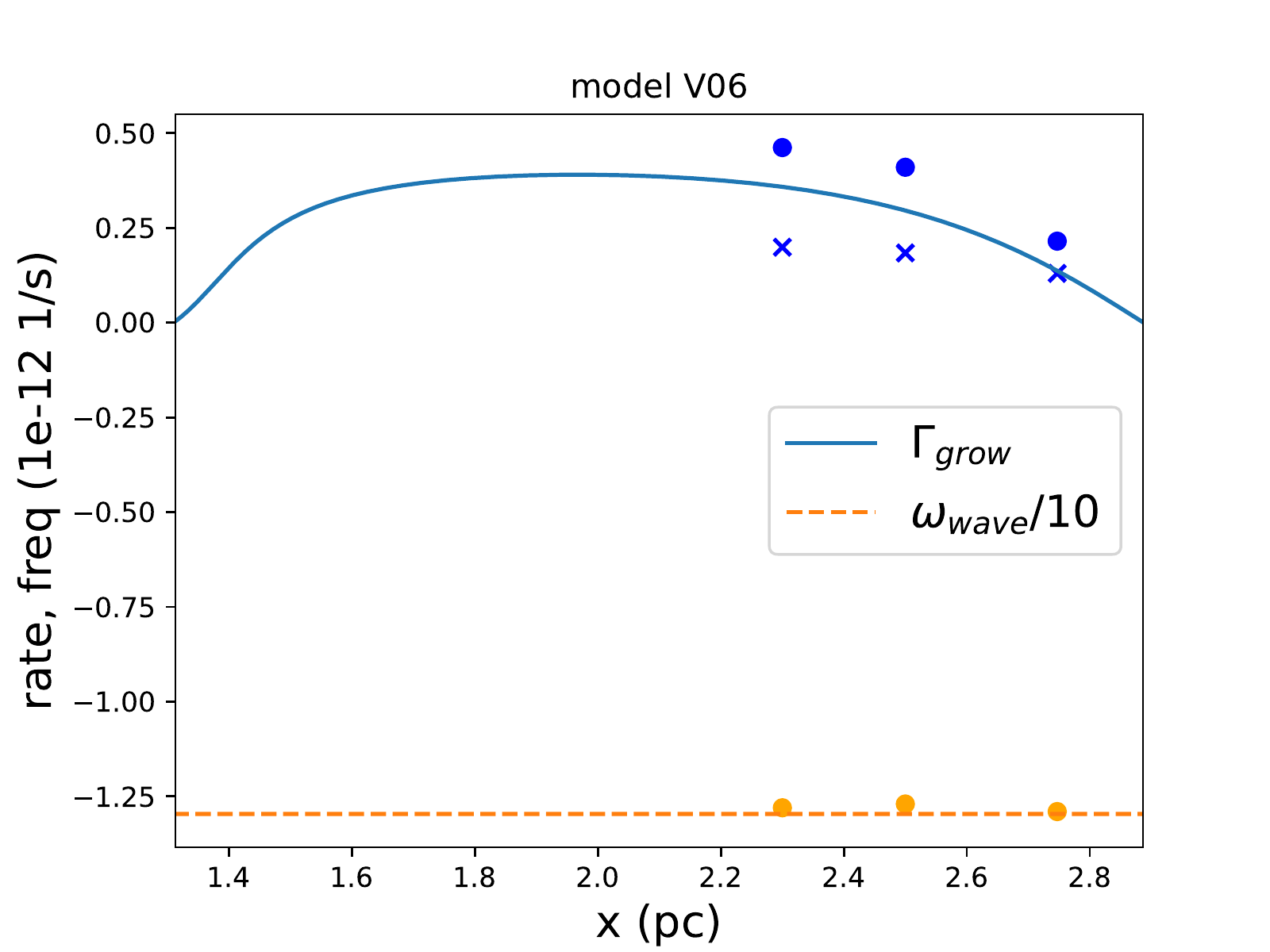} &
\includegraphics[width=0.35\linewidth]{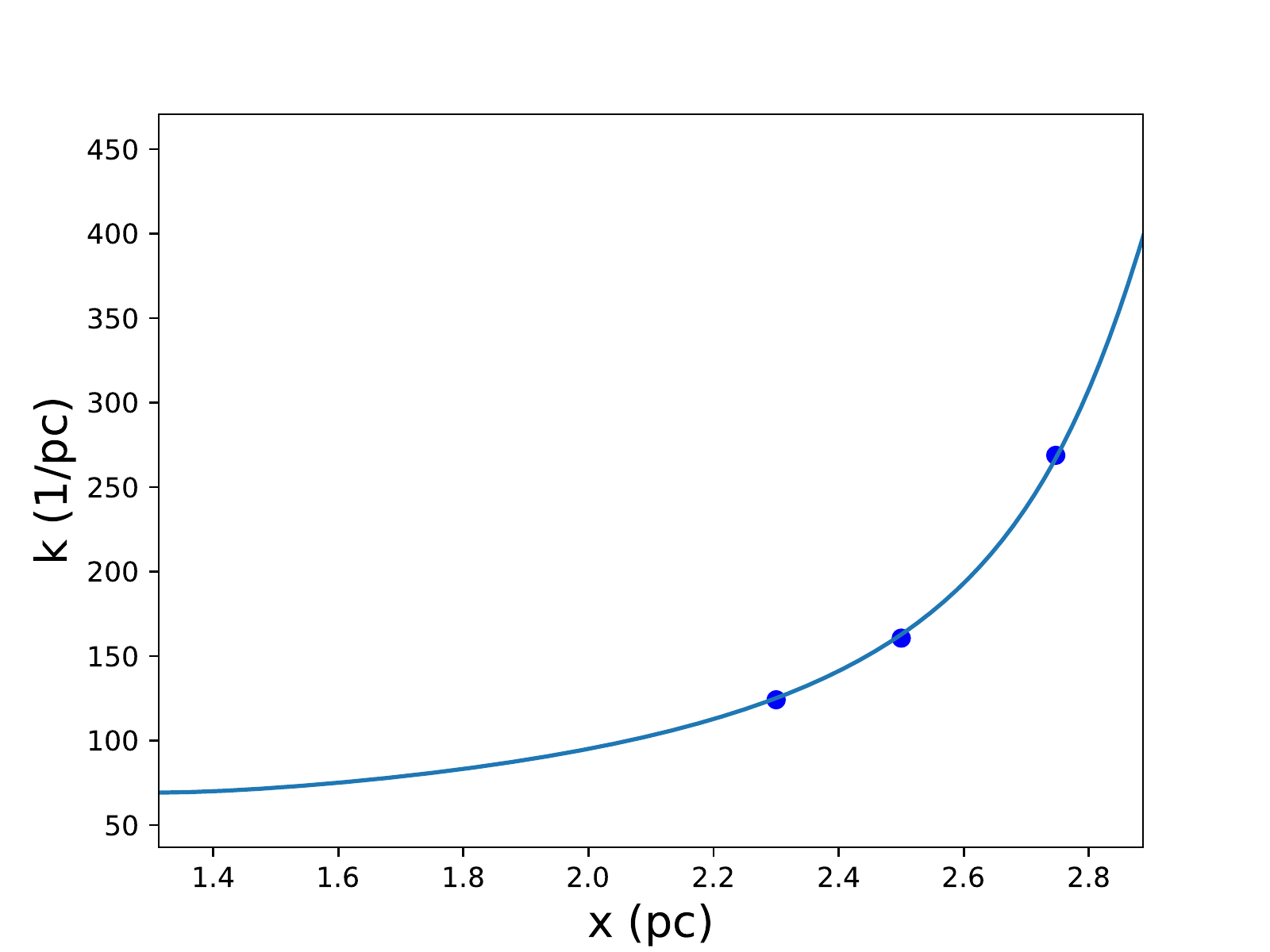} \\
\includegraphics[width=0.35\linewidth]{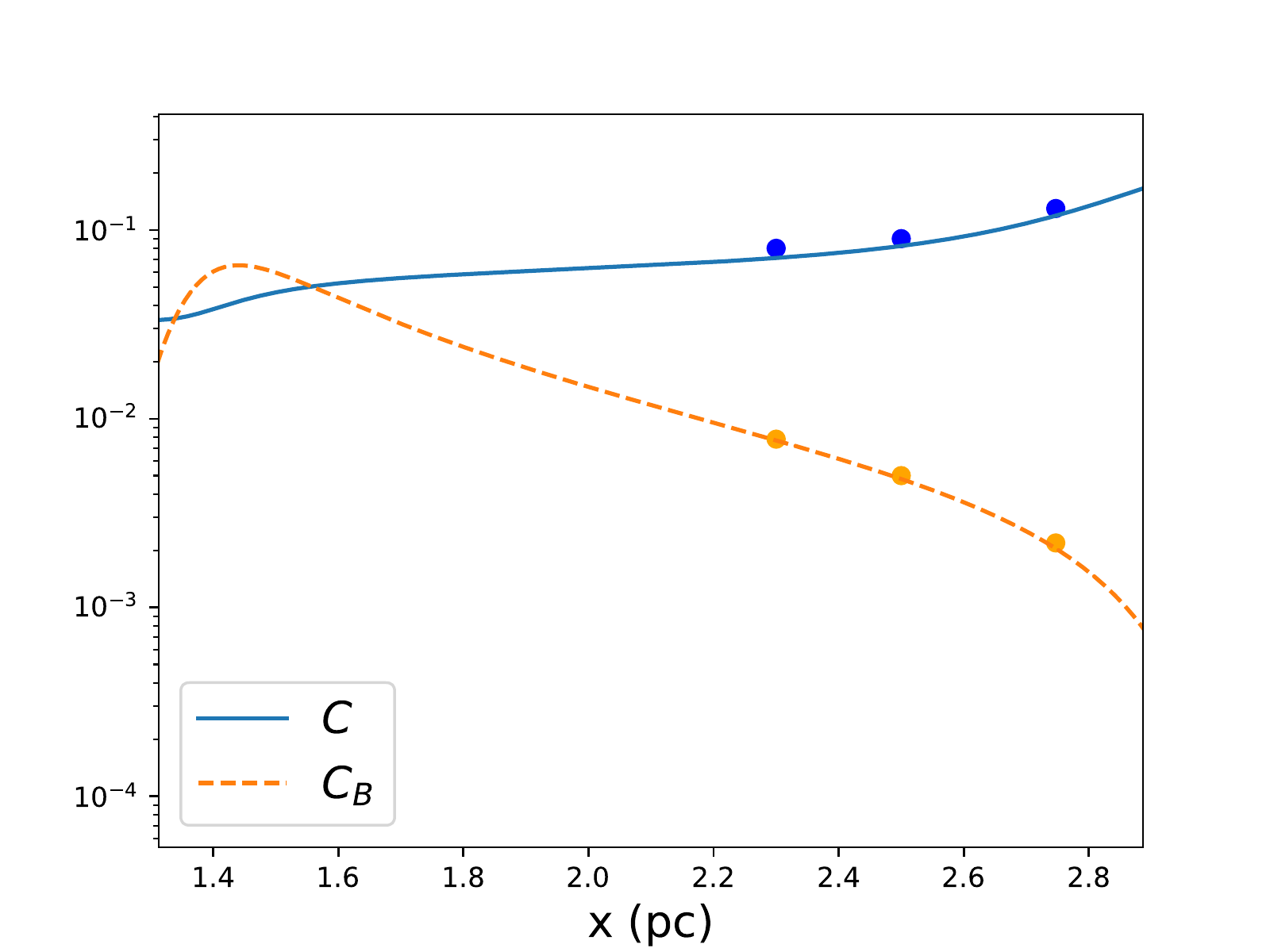}  &
\includegraphics[width=0.35\linewidth]{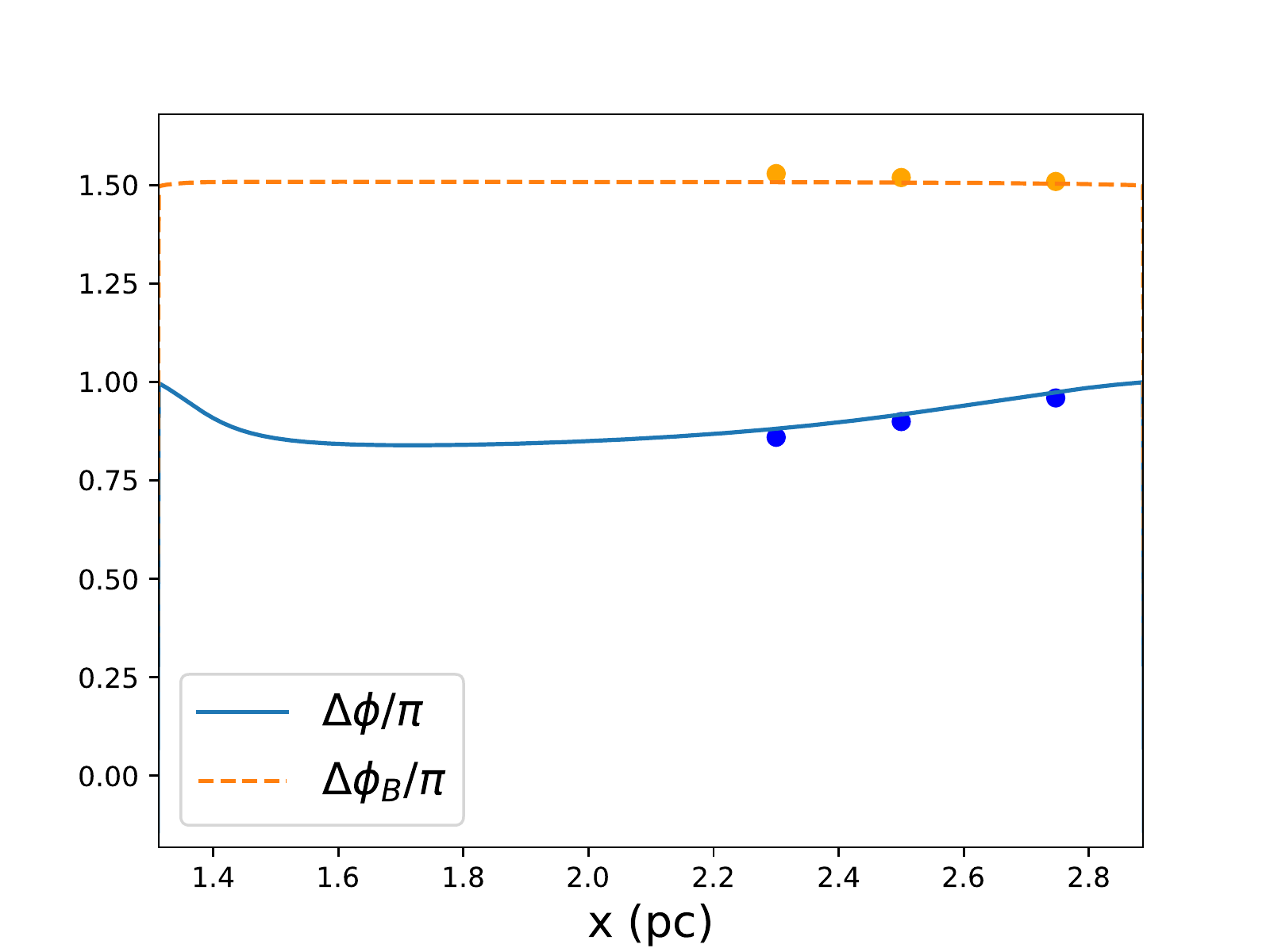} 
\end{tabular}
\caption{
Same as Figure~\ref{fig:linear_fig3_trend} but for model V06 with $\omega_{wave}=-v_0 k_0=-12.96$e$-12$ s$^{-1}$.
}
\label{fig:linear_V06_trend}
\end{figure}

Figure~\ref{fig:perb_comp} illustrates that  the simulated curves for perturbations almost match the analytical curves with little perceptible difference, strongly suggesting the presence of the drag instability in the simulation. Hence, we identify $C$, $C_B$, $\Delta \phi$, and $\Delta \phi_B$ from the simulated curves in Figure~\ref{fig:perb_comp} and present the numerical results to compare the analytical solutions in Figures~\ref{fig:linear_fig3_trend} and \ref{fig:linear_V06_trend} for models Fig3CO12 and V06, respectively. The amplitudes of $\delta \rho_n/\rho_n$ are much larger than those of $\delta v_n/V_n$ and $\delta B/B$, i.e., both $C$ and $C_B \ll 1$, as has been realized by \citet{GC20}. Additionally, $\delta v_n$ and $\delta B/B$ generally lead $\delta \rho_n$ by a phase of $\Delta \phi \sim 0.9\pi$ and $\Delta \phi_B \sim 1.5\pi$, respectively, which yields the instability.
Applying Equations(\ref{eq:growthrate})-(\ref{eq:growthrate2}), we then obtain $\omega_{wave}$ and $\Gamma_{grow}$ at the three different locations in each model. 

Figures~\ref{fig:linear_fig3_trend} and \ref{fig:linear_V06_trend}  show that in both models, the analytical and simulated growth rates have a similar trend along the shock flow, though they do not exactly match. The discrepancy is probably caused by the difference between the instantaneous growth rate of the predominant mode and the exact growth rate, i.e., the exact growth rate and wavenumber still slightly vary in the spatial range covering two wavelengths, where the single growth rate and $k$ of the predominant Fourier mode are obtained in our methodology. Nevertheless,
the curves for the analytical growth rate are sandwiched by the simulated
growth rates evaluated at the three locations using Equations(\ref{eq:growthrate}) and (\ref{eq:growthrate2}).
The wavenumber $k$, the relative amplitudes ($C$ and $C_B$), and phase differences  ($\Delta \phi$ and $\Delta \phi_B$) of perturbations from the simulated results follow the analytical curves reasonably well.
The top left panels of Figures~\ref{fig:linear_fig3_trend} and \ref{fig:linear_V06_trend} show that $\omega_{wave}$ is about 30-40$\times$ larger than $\Gamma_{grow}$ for the predominant mode extracted from the simulation, meaning that the instability propagates faster than its growth in accordance with the linear results.
Because $\omega_{wave}$ is dominated by the Doppler-shifted frequency $kV_n$ due to the fast downstream flow \citep{GC20}, $k$ increases with $x$ due to shock compression as shown in the figures, which has been observed in Figure~\ref{fig:pertD_inflow}. 
Further, $C$ slightly increases with $x$ but $C_B$ decreases with $x$. Overall, the simulated results agree with analytical results in terms of the mode profiles (i.e., the eigenvectors) shown in Figure~\ref{fig:perb_comp}  as well as the mode growth rates and frequencies (i.e., the eigenvalues) plotted in Figures \ref{fig:linear_fig3_trend} \& \ref{fig:linear_V06_trend}.
We confirm the presence of the drag instability in non-ideal MHD simulations for 1D isothermal C-shocks.

\subsection{Total growth across the shock width}
\label{subsec:growth}
Besides the confirmation of the linear theory for the drag instability,
numerical simulations enable us to directly evaluate the global total growth TG$_{sim}$ for an initial perturbation across the entire shock width, which cannot be achieved by the local linear theory. 
Recalling that the perturbations have the form $U \exp(\rmi kx+ \rmi \omega t) \equiv W$ in the local linear analysis, we may express the amplitude of the growing mode as follows:
\begin{equation}
    |W(x)|=\left(|U(x=x_1)|+ \int_{x_1}^x {d|U|\over dx}dx \right) \exp \left( \int_{x_1}^x \Gamma_{grow} {dx \over v_{ph}} \right),
    \label{eq:Ux}
\end{equation}
where $x_1$ is the location of the beginning of a C-shock and $|U(x_1)|=|W(x_1)|$. $|W(x)|/|W(x_1)|$ should equal TG$_{sim}$ when $x$ is the location of the shock end. However, $d|U|/dx$ is unknown
because $|U|$ is
arbitrary in a local linear analysis for normal modes.
To get a general sense of the total growth from the WKBJ analysis, \citet{GC20} set the amplitude of the unstable mode $|U|=1$ everywhere in the shock and thus $d|U|/dx=0$. The simplification allows the authors to estimate the total growth  TG$_{|U|=1}$  by integrating only the local growth of the mode over the entire shock width, i.e., the exponential term on the right-hand side of Equation(\ref{eq:Ux}). The total growth of the unstable mode at $x$ in a shock with $|U|=1$ is then given by
\begin{equation}
{\rm TG}_{|U|=1,x}=
\exp \left( \int_{x_1}^x \Gamma_{grow} {dx \over v_{ph}} \right).
\end{equation}
Therefore, the total growth across the entire shock, denoted by TG$_{|U|=1}$, is obtained by setting $x$ to be the location of the shock end in the above equation.

\begin{figure}
\plottwo{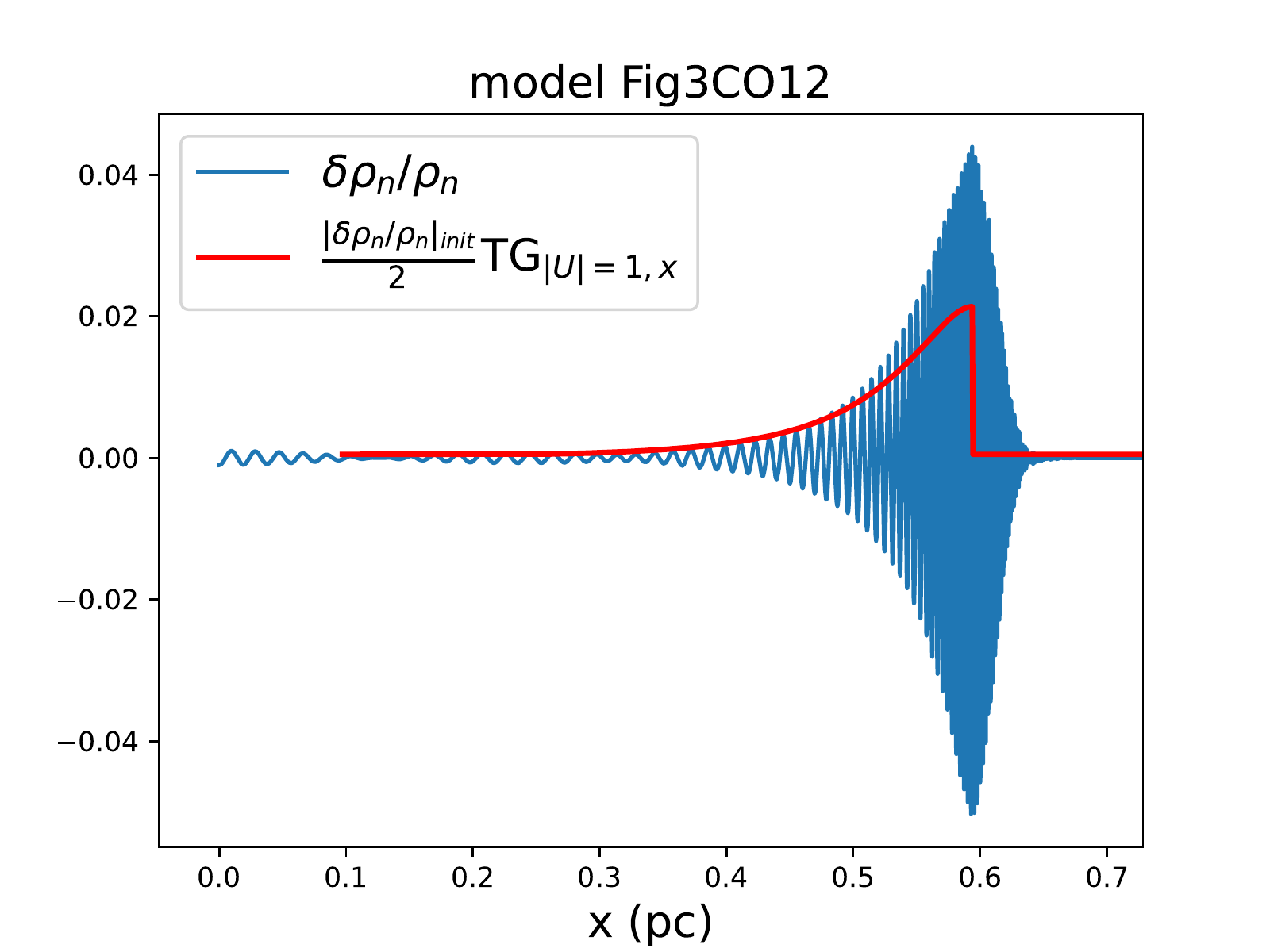}{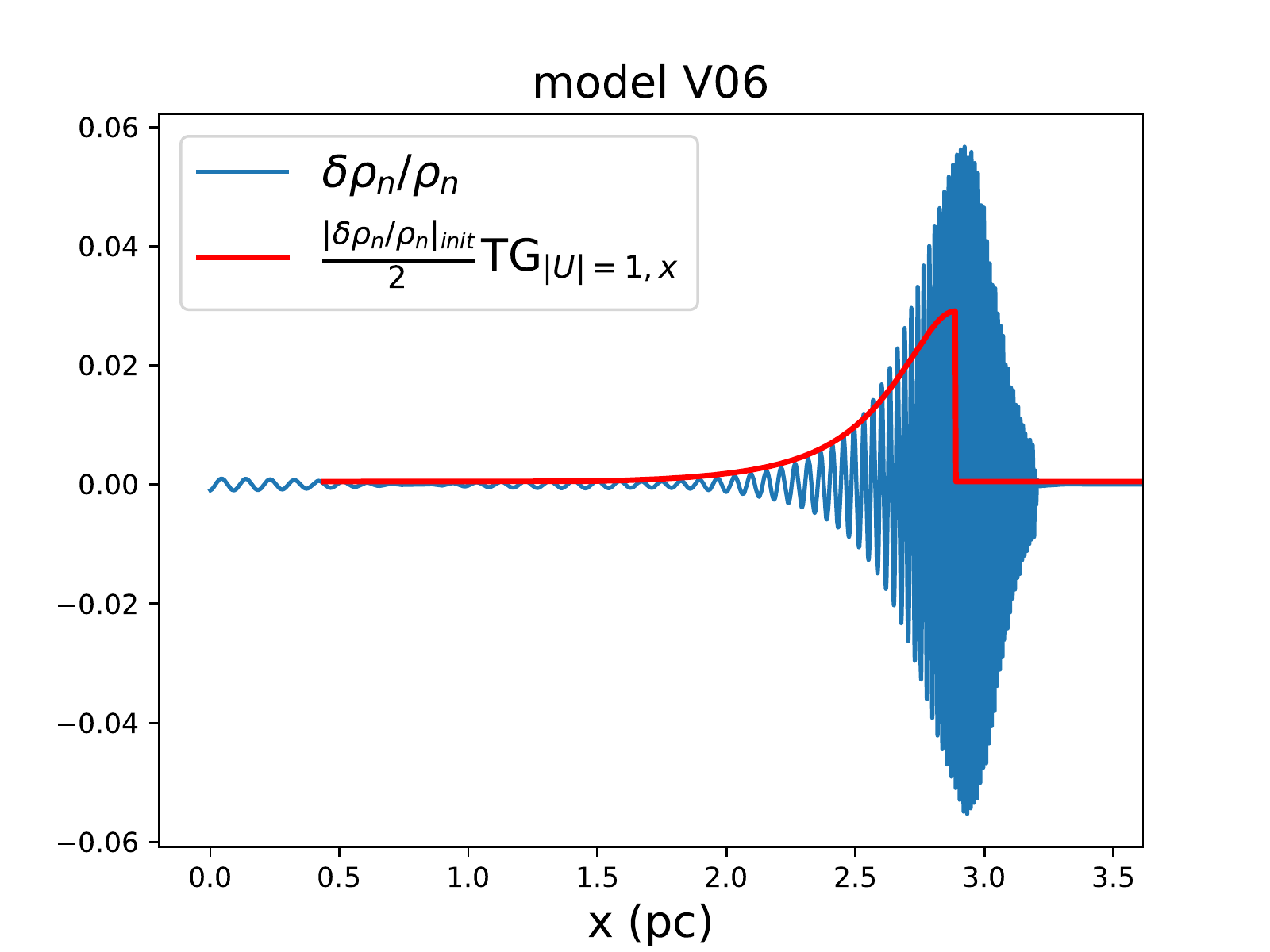}
\caption{Comparison between the simulated density profiles shown in the top panels of Figure~\ref{fig:pertD_inflow} and the profiles of the total growth ${|\delta \rho_n /\rho_n|_{init} \over 2}$TG$_{|U=1|,x}$ estimated based on the linear theory. Here $|\delta \rho_n /\rho_n|_{init}/2$ is the initial density perturbation of the unstable mode given by 0.001/2 (see the text).}
\label{fig:TG}
\end{figure}

\citet{GC20} used TG$_{|U|=1}$ as a proxy of TG$_{sim}$ to evaluate which shock model favors the growth of the drag instability in their parameter study. As shown above, $|U|$ should vary with $x$ in the shock and needs to smoothly connect to the initial condition at $x=0$. Figure~\ref{fig:TG} indicates that the profiles of simulated density perturbation $\delta \rho_n/\rho_n$ increase faster than 
TG$_{|U|=1,x}$, i.e,
the total growth from the beginning of the shock based on the linear theory with $|U|=1$.
The peak value of TG$_{|U|=1,x}$ in Figure~\ref{fig:TG} is TG$_{|U|=1}$. It is to be recalled from Section~\ref{sec:beat} that the amplitude of the unstable mode is made of only one-half of the initial density perturbation $|\delta \rho_n/\rho_n|_{init}=0.001$. The simulated density perturbation increases from 0.0005 to 0.044 in model Fig3CO12 and to 0.057 in model V06, which amount to TG$_{sim} \approx 88$ and 114 in model Fig3CO12 and V06, respectively. Comparatively,  TG$_{|U|=1}$ is only about 42.7 in model Fig3CO12 and  58.3 in model V06. Evidently, the simulated perturbations increase faster than those based on the simple estimate by a factor of two in the two models -- the local analysis performed at each $x$ cannot account for the growth contributed from $d|U|/dx$ in Equation(\ref{eq:Ux}). Additionally, the exact growth should be a little larger because the unstable mode has been slowly damped in the preshock region before it travels to the C-shock and starts growing.

\section{Nonlinear saturation}
\label{sec:nonlinear}
\begin{figure}
\centering
\setlength{\tabcolsep}{0pt}
\begin{tabular}{ccc}
\includegraphics[width=0.35\linewidth]{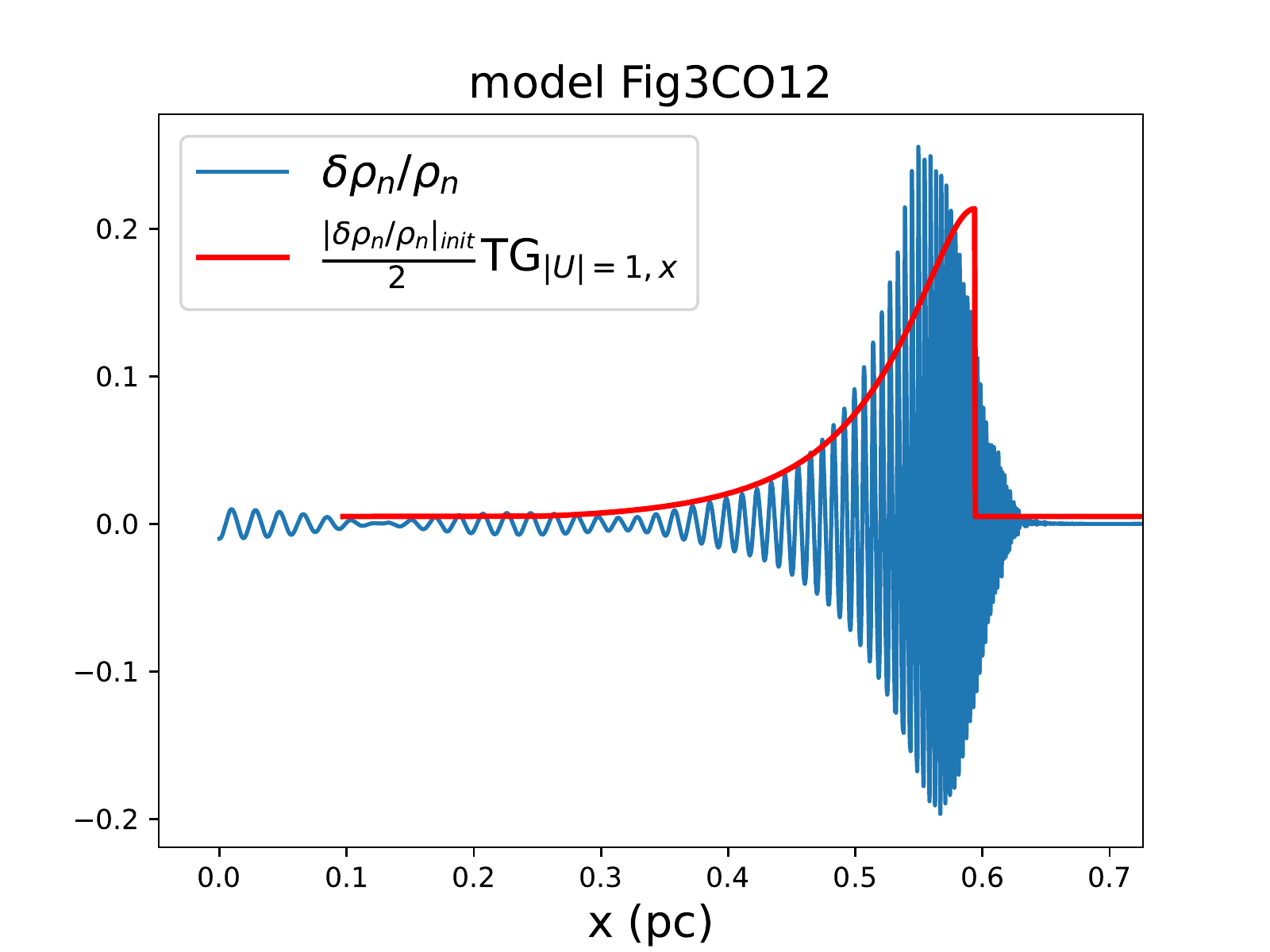} &
\includegraphics[width=0.35\linewidth]{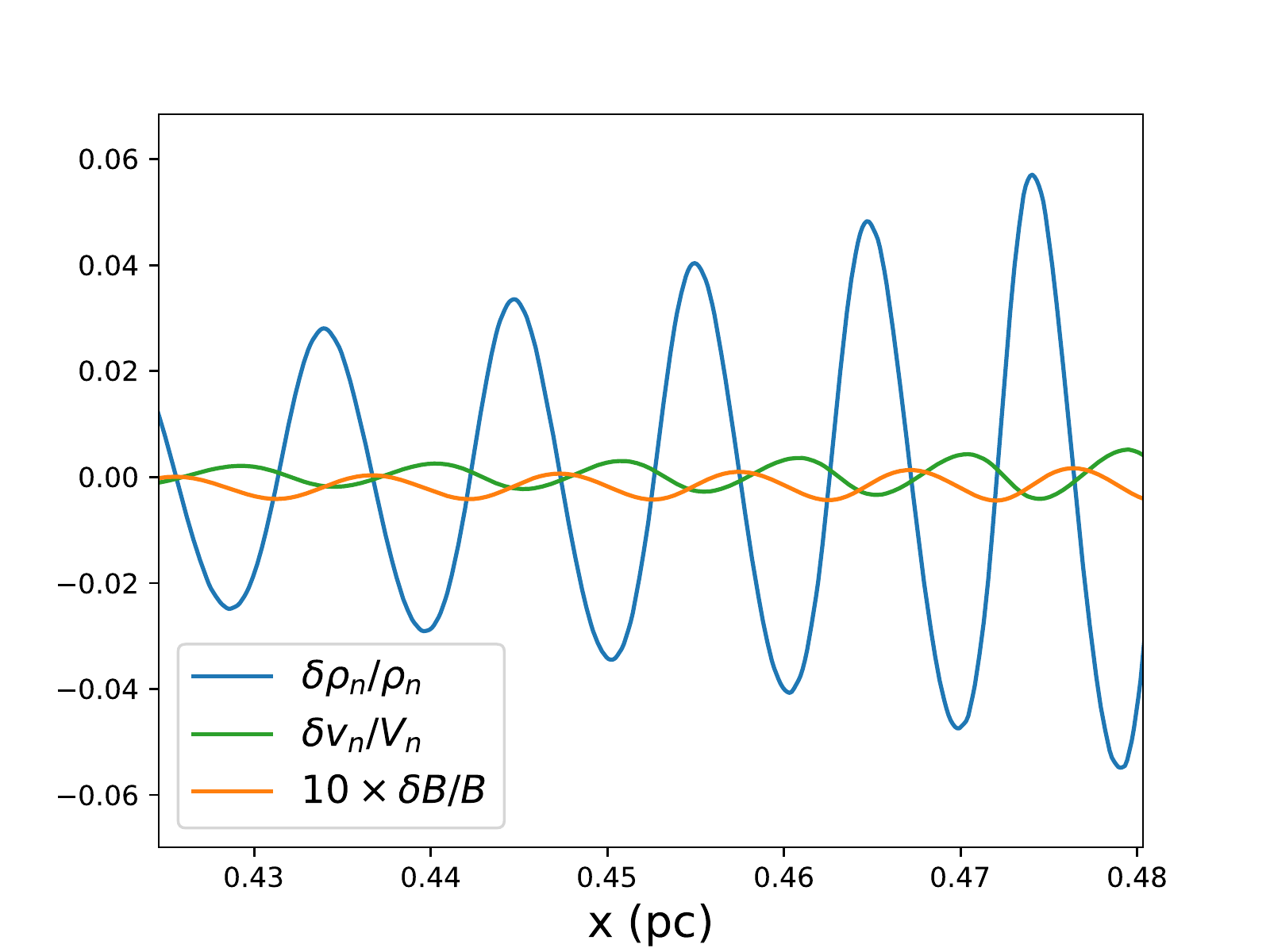}  &
\includegraphics[width=0.35\linewidth]{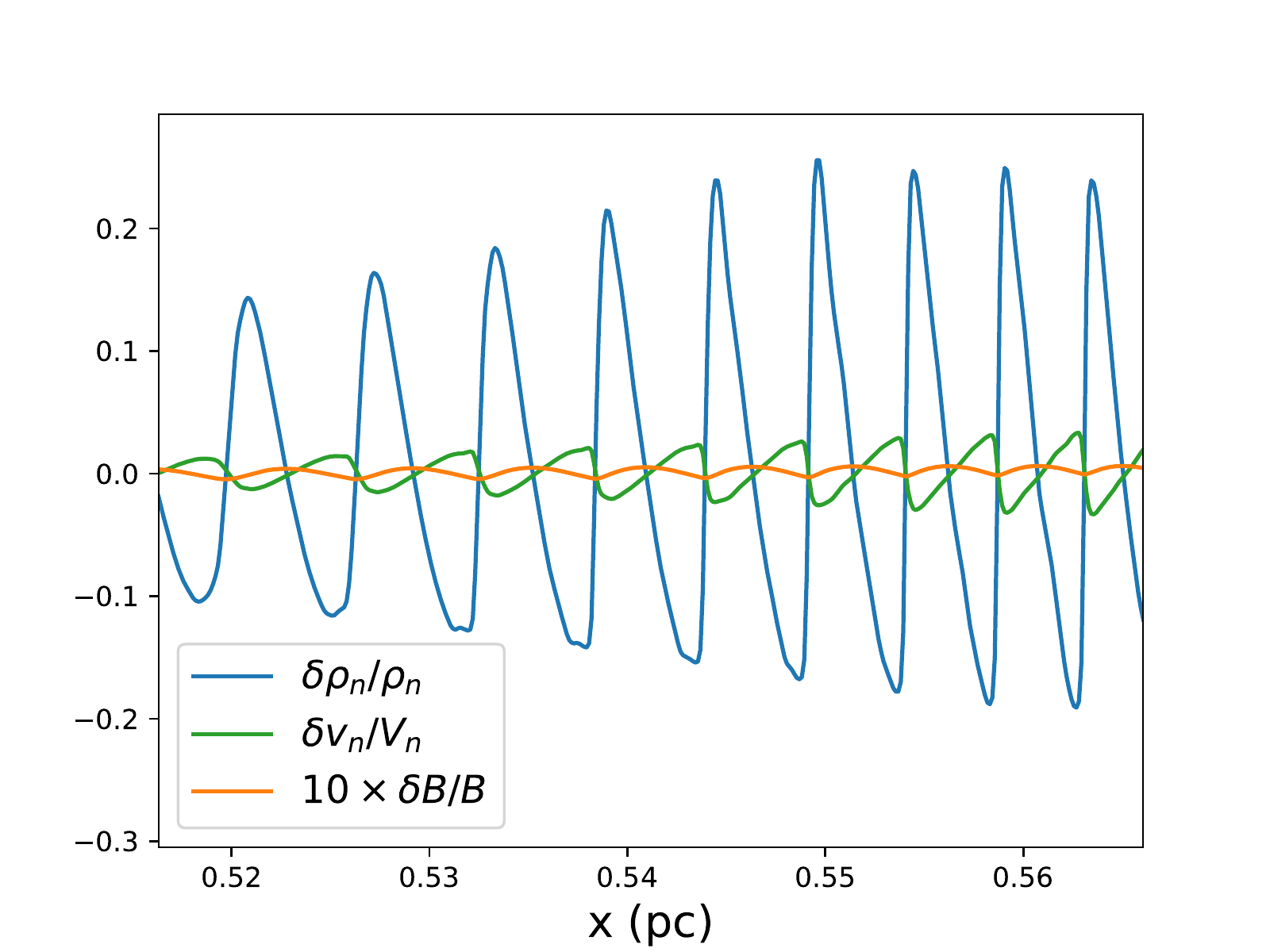}  \\
\includegraphics[width=0.35\linewidth]{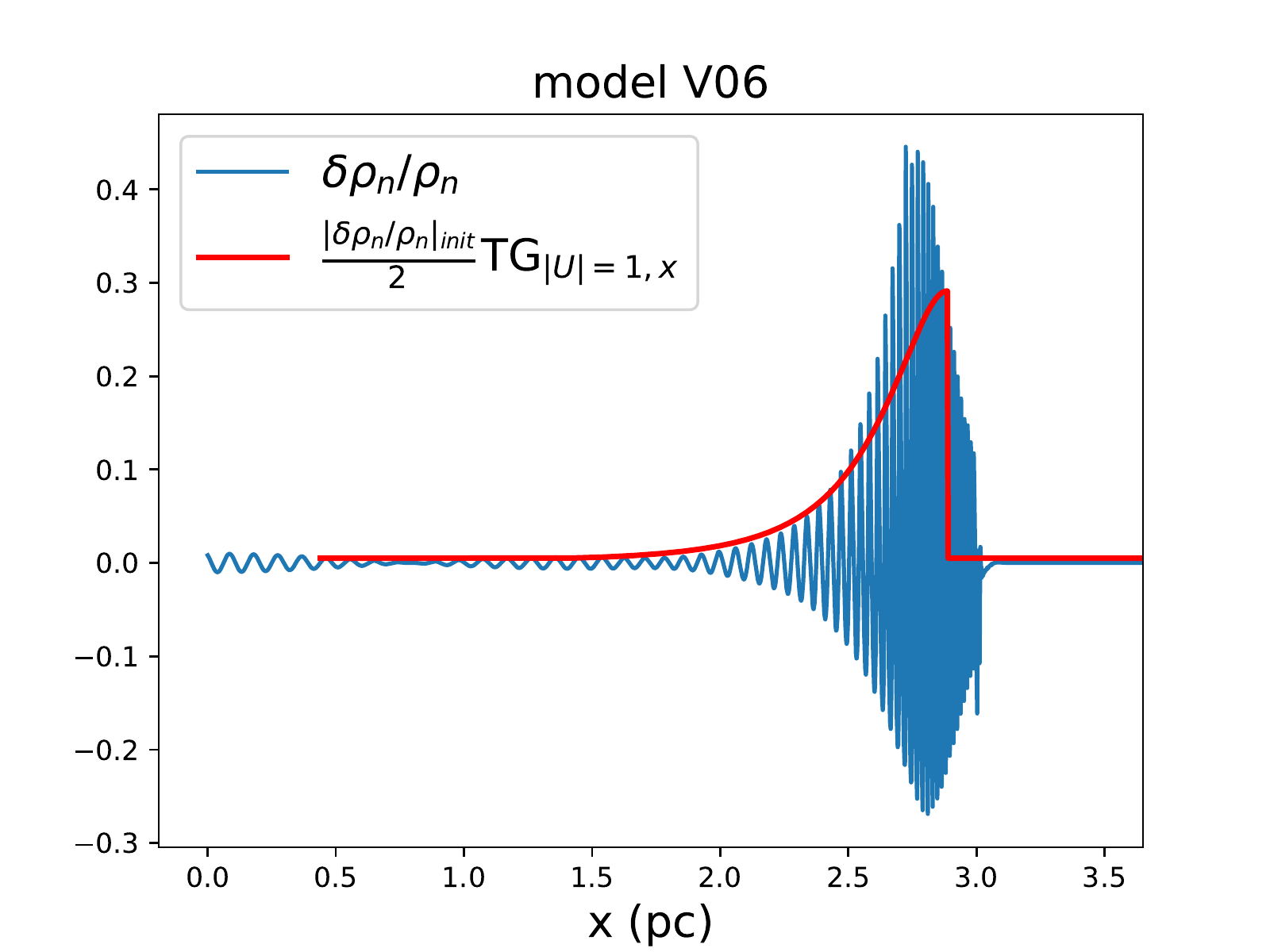} &
\includegraphics[width=0.35\linewidth]{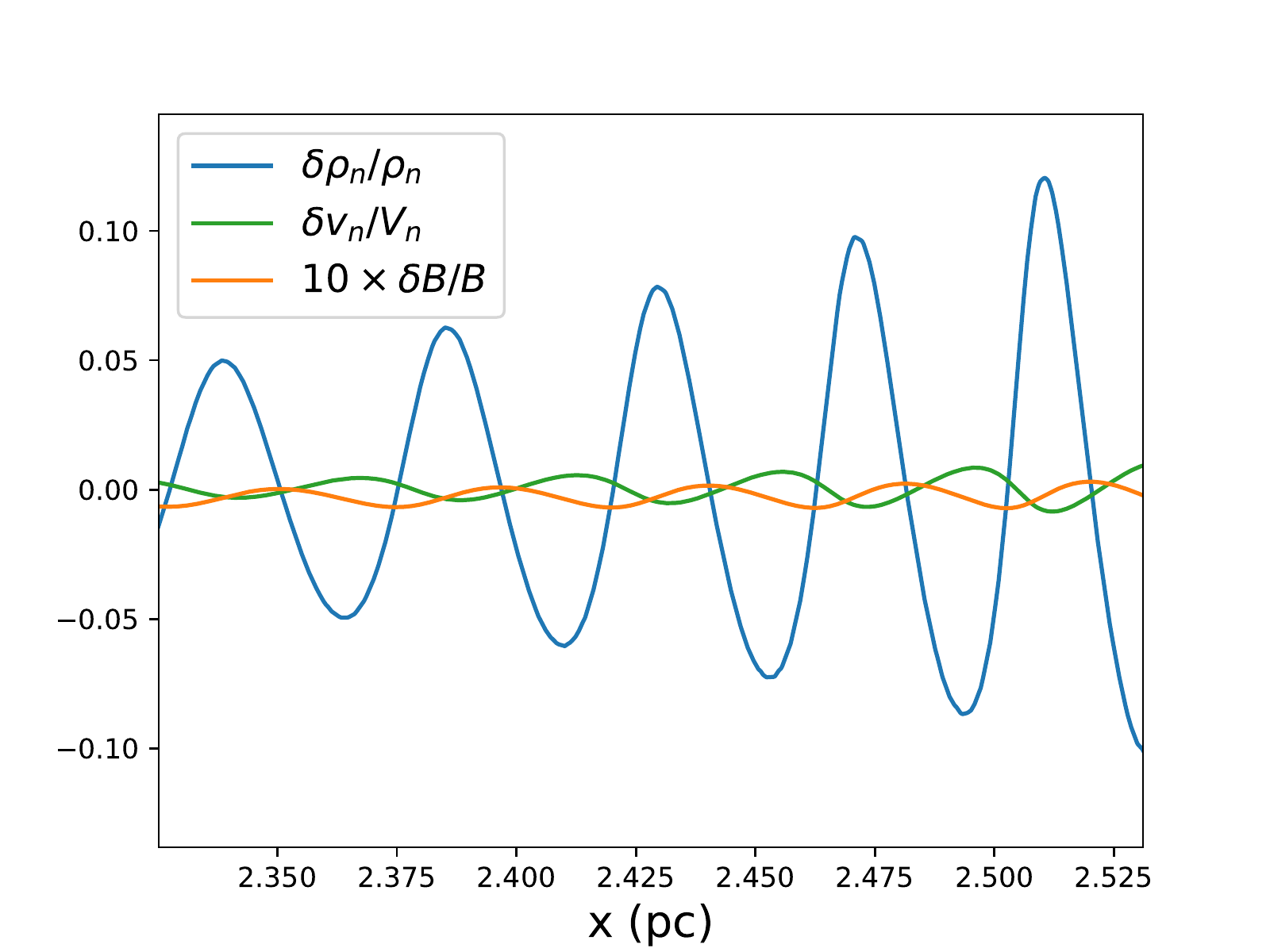} &
\includegraphics[width=0.35\linewidth]{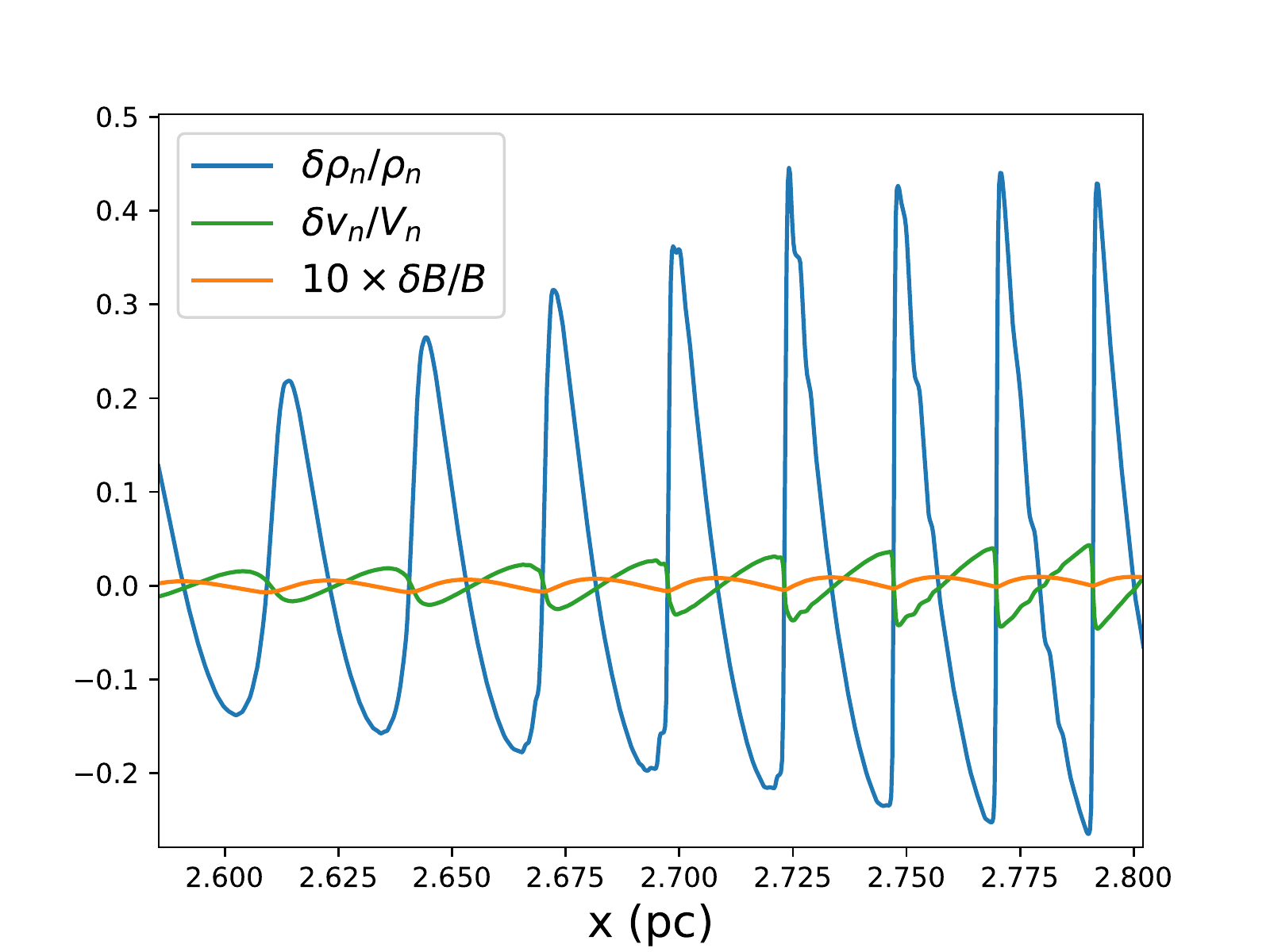}
\end{tabular}
\caption{Perturbation profile excited by $|\delta \rho_n/\rho_n|_{init}=0.01$ with  $\omega_{wave}=-v_0 k_0$ in models Fig3CO12 (top row) and V06 bottom row). Left panels: profile of the density perturbation in comparison with ${|\delta \rho_n /\rho_n|_{init} \over 2}$TG$_{|U=1|,x}$ from the linear theory. Middle panels: perturbations in the range around $x\approx 0.45$ pc (model Fig3CO12) and $\approx 2.43$ pc (model V06). Right panels: perturbations in the range around $x\approx 0.54$ pc (model Fig3CO12) and $\approx 2.7$ pc (model V06).
} 
\label{fig:Fig3_nonlinear}
\end{figure}

Apart from the global growth of propagating perturbations across the shock width, numerical simulations also provide information about the nonlinear behavior of the growing perturbation, which the linear theory is inadequate to describe. To examine the nonlinear effect, we impose a $10\times$ larger initial perturbation on the gas density 
$|\delta \rho_n/\rho_n|_{init}=0.01$ with the same frequency 
in model Fig3CO12.
The simulated profiles of perturbations are presented in Figure~\ref{fig:Fig3_nonlinear}.  The perturbations have propagated across
the entire shock width in both models. ${|\delta \rho_n/\rho_n|_{init} \over 2}$TG$_{|U|=1,x}$ from the linear theory with the initial value $0.01/2=0.005$ is plotted on top of the profile of the density perturbation in the left panel to gauge the growth. By comparing with Figure~\ref{fig:TG} (models with a smaller initial density perturbation)
where the perturbation amplitudes grow $\approx 2\times$ larger than ${|\delta \rho_n/\rho_n|_{init} \over 2}$TG$_{|U|=1,x}$,
we can see from Figure~\ref{fig:Fig3_nonlinear} that the growth of the large-amplitude perturbation is only slightly larger than ${|\delta \rho_n/\rho_n|_{init} \over 2}$TG$_{|U|=1,x}$. 
Additionally, the growing perturbation starts to decay before 
reaching the peak of TG$_{|U|=1,x}$, 
indicating that the growth of the unstable  mode is suppressed. 

In the case shown in Figure~\ref{fig:Fig3_nonlinear},
as the perturbations propagate into the shock region, they first maintain a nearly sinusoidal shape from the initial perturbation (middle panels of Figure~\ref{fig:Fig3_nonlinear}), but subsequently grow and evolve into more or less sawtooth waves with asymmetric amplitudes (right panels), 
i.e., waves are being steepened. 
The effect of wave steepening has also been observed in the middle panels of Figure~\ref{fig:pertD_inflow}, 
though with smaller initial perturbations the effect is relatively weak in those cases.
When investigating the perturbation evolution in 1D C-shock simulations, we see
the wave growth arises from the fact that $\delta v_n$ leads $\delta \rho_n$ by $\sim 0.9 \pi$ in the simulations (recall Figure~\ref{fig:linear_fig3_trend}), meaning that the $\delta v_n$ and $\delta \rho_n$ of the unstable mode are more or less out of phase. 
Consequently, the density trough (crest) is associated approximately with the velocity crest (trough), as shown in Figure 8. The amplitude of velocity perturbations has reached a non-negligible fraction of the phase velocity. Therefore, the velocity crest moves faster than the velocity trough, resulting in wave steepening.
When the amplitude is large, the effect of wave steepening is enhanced, leading to nonlinear dissipation, thus limiting the growth inferred from the linear theory. 
The right panels of Figure~\ref{fig:Fig3_nonlinear} imply that the nonlinear saturation of the drag instability has already occurred in the weak nonlinear regime where $|\delta v/v_{ph}| \sim |\delta v/V_n| \approx 0.03$ (model Fig3CO12) and $\approx 0.04$ (model V06), which correspond to $|\delta \rho_n/\rho_n|  \approx 0.25$ (model Fig3CO12) and $\approx 0.45$ (model V06) in the simulation. It is evident from the right panels of 
Figure~\ref{fig:Fig3_nonlinear} that the pressure scale height of the steep density perturbation is smaller than one wavelength, implying a strong pressure force in the nonlinear perturbations.
 
\section{Discussion}
\label{sec:discussion}
\subsection{Local Analysis}
\begin{figure}
    \plottwo{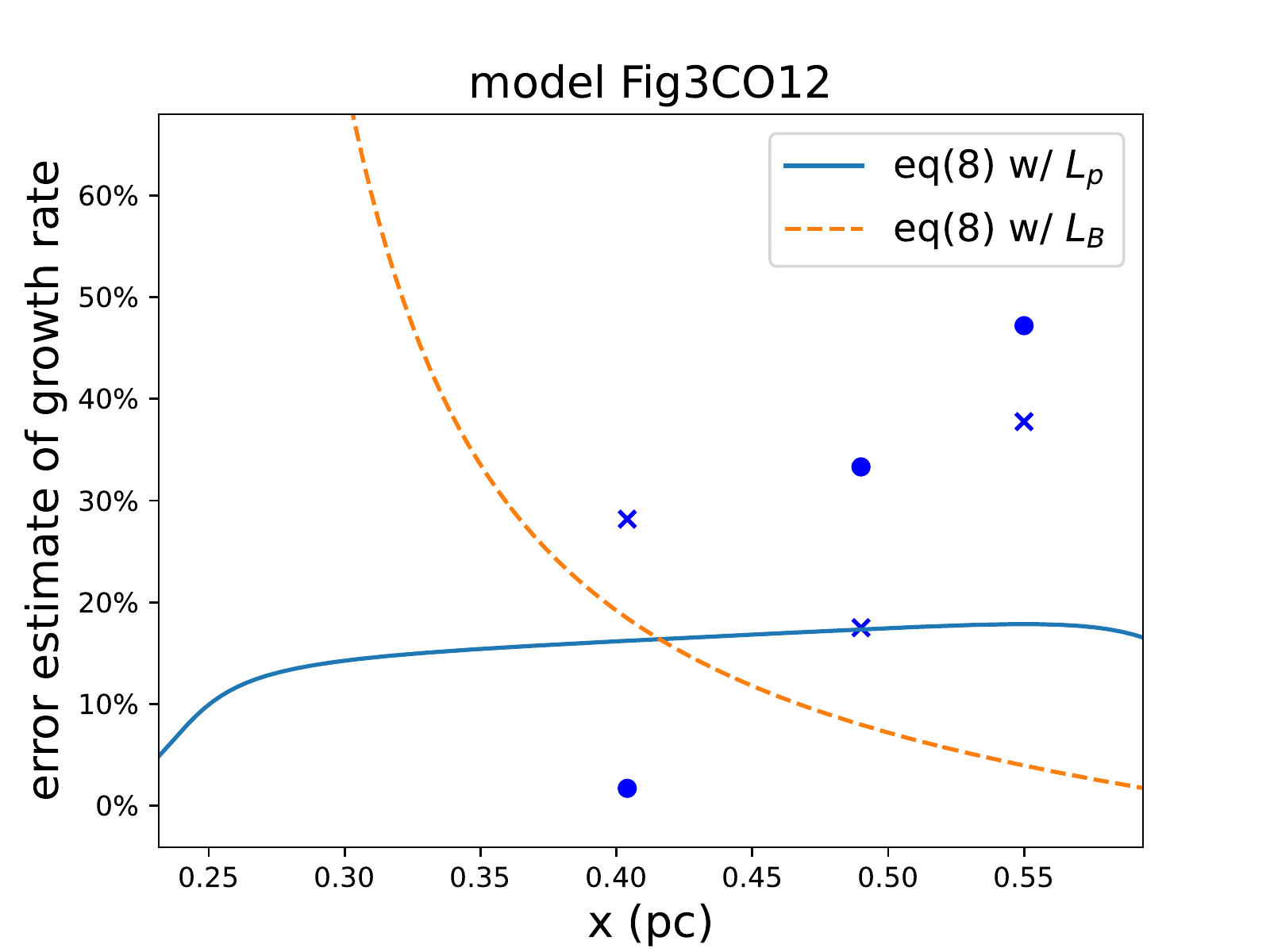}{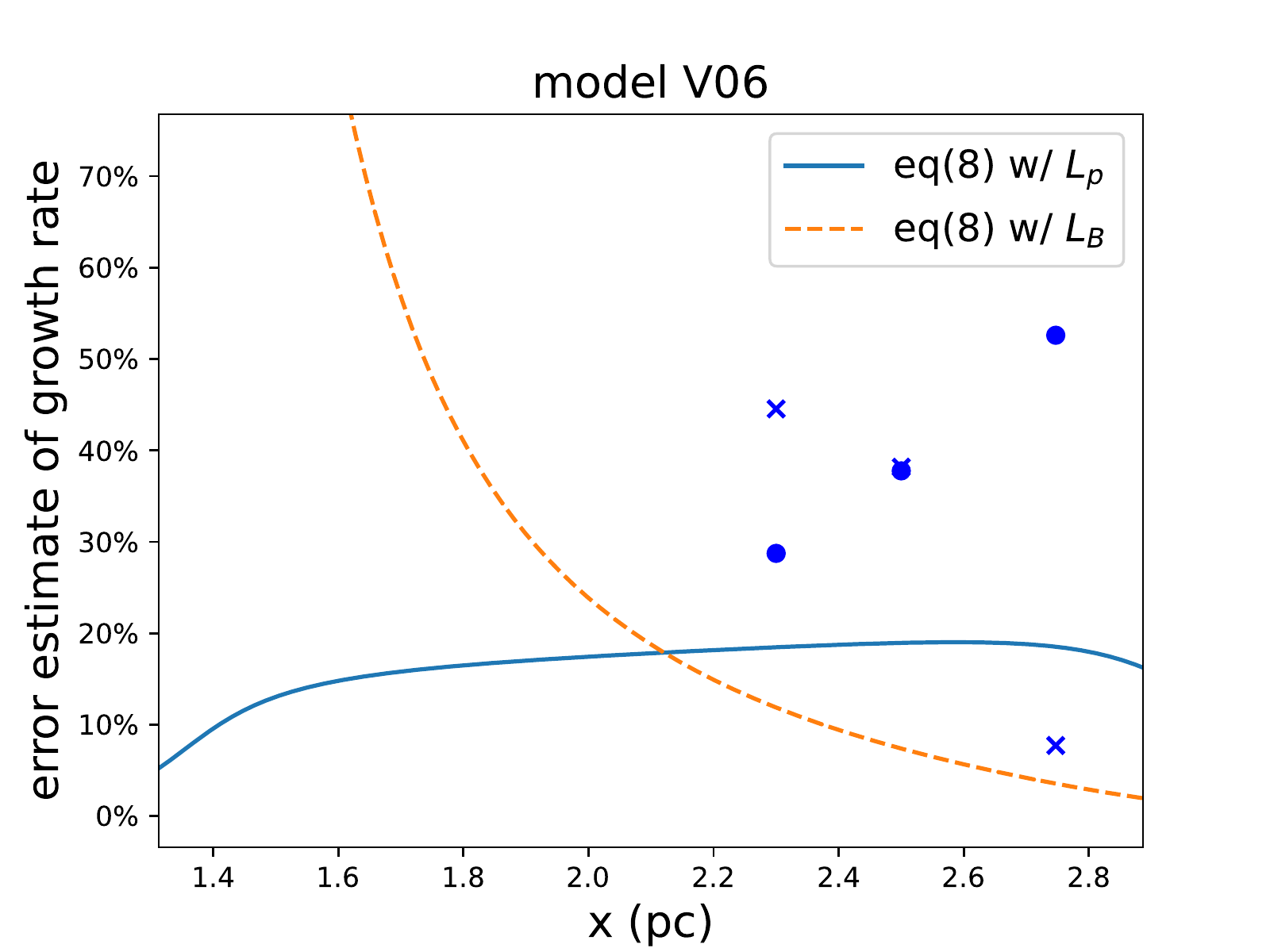}
    \caption{Error estimates of the growth rate based on  Equation(\ref{eq:Gamma}) to order of $k/L$ in models Fig3CO12 and V06. The results for $L=L_p$ (solid line) and $L=L_B$ (dashed line) are plotted. The dots and crosses correspond to those shown in the top left panel of Figures~\ref{fig:linear_fig3_trend} and \ref{fig:linear_V06_trend}. They indicate the percentage deviation of simulated growth rates from those derived in the complete linear analysis.}
    \label{fig:error}
\end{figure}

Although the simulated growth rate within the C-shock is generally consistent with the analytical growth rate, they do not exactly match in our study. It is probably because the exact local growth rate and wavenumber still slightly vary in the spatial range covering two wavelengths, where the single growth rate and wavenumber of the predominant Fourier mode are obtained by invoking the instantaneous growth rate. 
In such a case, the mismatch is expected to only be of order $1/(kL)$, where $L$ is the gradient length scale. The local analysis yields the dispersion relation described by Equation(\ref{eq:Gamma}). The analysis can approximately model the effect of a background gradient length scale $L$ by adding a small imaginary part, $\rmi/L$, to $k$. The dispersion relation then predicts a correspondingly small shift in $\omega$ in the complex plane. However, since $\omega$ is mostly real (i.e. $\omega \approx kV_n$), a small fractional shift of $\omega$ in the complex plane as $k$ is corrected to $k+\rmi/L$ is likely to change the imaginary part of $\omega$ by a significant fraction. The dispersion relation is then $\omega \sim kV_n + [(1+\rmi)/2] \sqrt{kV_d \gamma \rho_i}$, implying a correction to the imaginary part of $\omega$ of order $\rmi V_n/L$. Figure~\ref{fig:error} illustrates the error estimate $(V_n/L)/\Gamma$ for $L=$ the pressure scale length $L_p$ (solid curve) and $L=$ the magnetic field scale length $L_B$ (dashed curve). They are plotted to compare to the deviation percentage of simulated $\Gamma_{grow}$ from that obtained in the complete linear theory as shown in the top left panel of Figures~\ref{fig:linear_fig3_trend} \& \ref{fig:linear_V06_trend} (i.e. deviation of dots or crosses from the curve for $\Gamma_{grow}$ in the top left panel of Figures~\ref{fig:linear_fig3_trend} \& \ref{fig:linear_V06_trend}). We see that the deviation of simulated data from the linear results ($\sim 20$-$50$\%) is in general larger than the error estimate $(V_n/L)/\Gamma \sim$20\% in the second half of the shock regions. The effect of $k/L$ may contribute to part of the deviations. While the error estimate is made based on the dispersion relation Equation(\ref{eq:Gamma}), it is worth noting that the full linearized equations are more complicated than that. Namely, the result from the complete linear analysis does not exactly follow the dispersion relation due to the presence of other less dominant terms that are dropped to derive the simplified dispersion relation (refer to Section~\ref{sec:eqns}).

As described in Section~\ref{sec:method}, it can be difficult to directly compare the growth rates based on the local linear analysis to those from numerical simulations of the drag instability, which travels a few wavelengths in one growth timescale. Many of the difficulties in robustly connecting the simulations to the local linear analysis could be avoided by explicitly integrating the linearized equations globally through the steady C-shock background.
In contrast to local linear analysis, a global linear analysis preserves information on the causality of propagating waves. Therefore, it would be worthwhile to improve the comparison between the linear analysis and simulated results by performing a global linear analysis of the local, yet fast traveling, overstability.

\subsection{Nonlinear Saturation}
As has been noted by \citet{GC20}, the density perturbation dominates over velocity and magnetic field perturbations in the drag instability, which leads the authors to postulate that clump/core formation could be one of the nonlinear outcomes of the instability in C-shocks. 
However, our numerical simulations suggest that the growth of the instability in 1D isothermal C-shocks saturates in the weak nonlinear regime where $\delta \rho_n/\rho_n$ is only a fraction of unity. Because the corresponding $\delta v_n/v_{ph}$ is even smaller when the nonlinear saturation occurs, we conjecture that the steep pressure gradient of the predominant $\delta \rho_n/\rho_n$ would be important in saturating the growth of the drag instability. Analogous to the steepening of 1D ideal MHD waves, the pressure force resulting from the steepened density perturbation is expected to play a role in the nonlinear saturation of the drag instability in 1D C-shocks \citep{SI05,GS09}.
More numerical simulations for a parameter study with a large initial perturbation should be conducted to fully understand the nonlinear effects.

Larger perturbations can play a more important role in the thermal properties of the modes. The significance of thermodynamics for a perturbation may be characterized by $\delta v_n/c_s$, of which the maximum values are $\approx 0.2$-0.3 for the nonlinear cases presented in Figure~\ref{fig:Fig3_nonlinear}. Moreover,
the radiative cooling rate associated with a mode, $\Gamma_{rad}$, may be approximated as $\delta \rho_n c_s^2/\Lambda$, where $\Lambda$ is the cooling function. Adopting the expressions of $\Lambda$ from \citet{GL1978} due to molecular and atomic line emissions for $T\sim 10$ K and $n\sim $ a few $\times 10^3$ cm$^{-3}$, we have 
 $\Gamma_{rad} \sim \Gamma_{grow}$
in our nonlinear cases.
These all together imply that the adiabatic heating or cooling of the perturbations would be small but non-negligible. In our present study based on isothermal perturbations, heating effects (e.g., due to the ion-neutral drag, nonlinear dissipation) and aforementioned thermal processes are ignored. More self-consistent calculations incorporating thermochemical evolution would be worthwhile for future study of mode saturation.

\subsection{One-Fluid Approach}
\label{subsec:1fluid}
Although the numerical simulations for the drag instability are conducted for the one-fluid approach in this study, it is worth discussing the difference between the one-fluid approach and the more realistic two-fluid approach from the perspective of local linear theory. As described in Section~\ref{subsec:growth}, TG$_{|U|=1}=$42.7 and 58.3 in models Fig3CO12 and V06, respectively. However, \citet{GC20} identified the maximum-growth mode in the two-fluid approach and showed that the total growth of the maximum-growth mode, i.e. the maximum TG$_{|U|=1}$ by varying $\omega_{wave}$,\footnote{It is defined as the maximum total growth (MTG) in \citet{GC20}.} is only about 10 and 36 for models Fig3CO12 and V06, respectively. Apparently, the growth rate is overestimated in the one-fluid approach. Indeed, we apply the linear analysis in the two-fluid approach and find that TG$_{|U|=1}=$6.4 and 30.3 for model Fig3CO12 and 30.3 for model V06, which are smaller than those in the one-fluid approach by a factor of about 6.7 and 1.9, respectively. We now investigate the underlying physics.

\begin{figure}
\centering
\renewcommand{\arraystretch}{0}
\begin{tabular}{cc}
\includegraphics[width=0.35\linewidth]{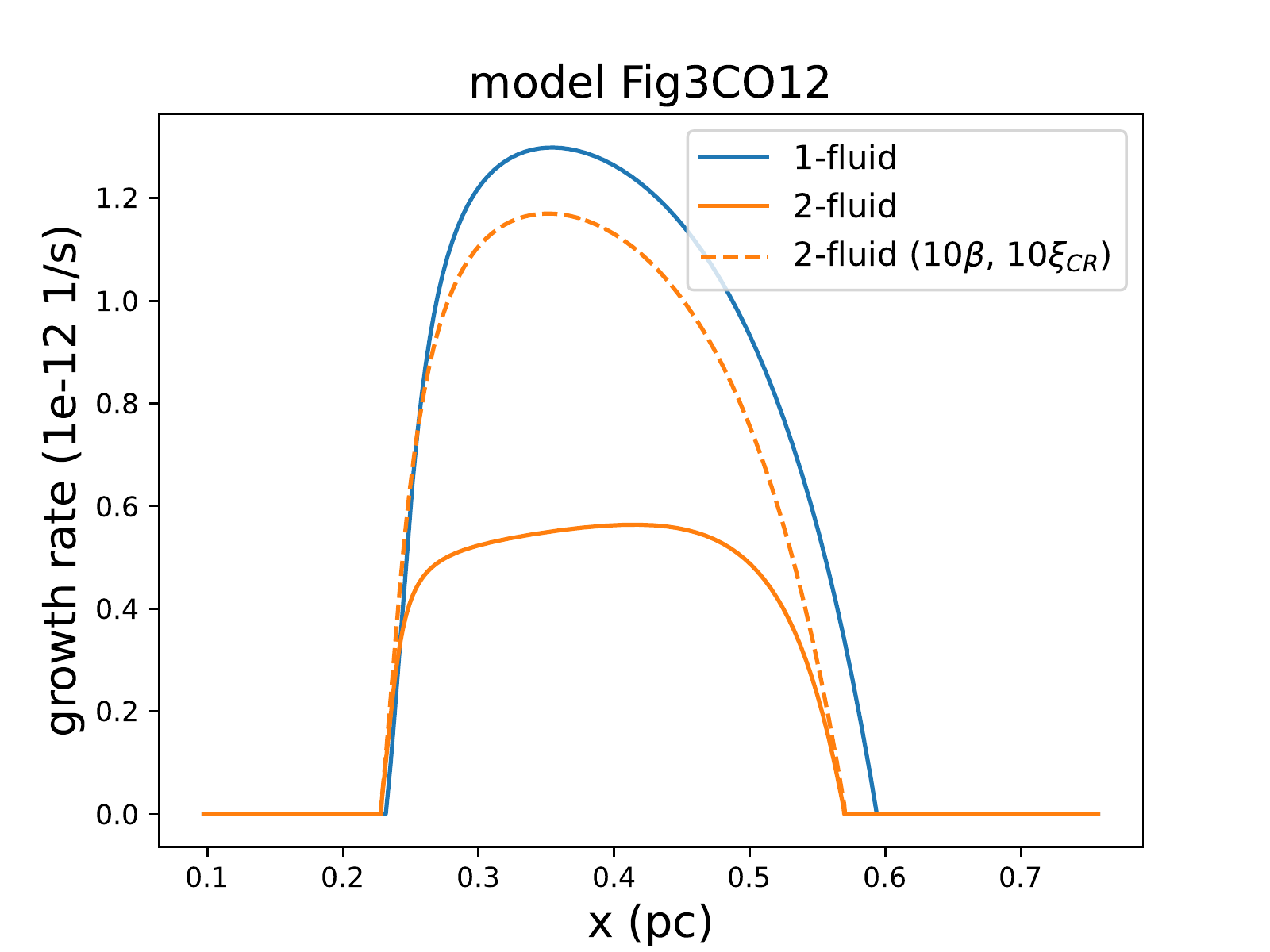} &
\includegraphics[width=0.35\linewidth]{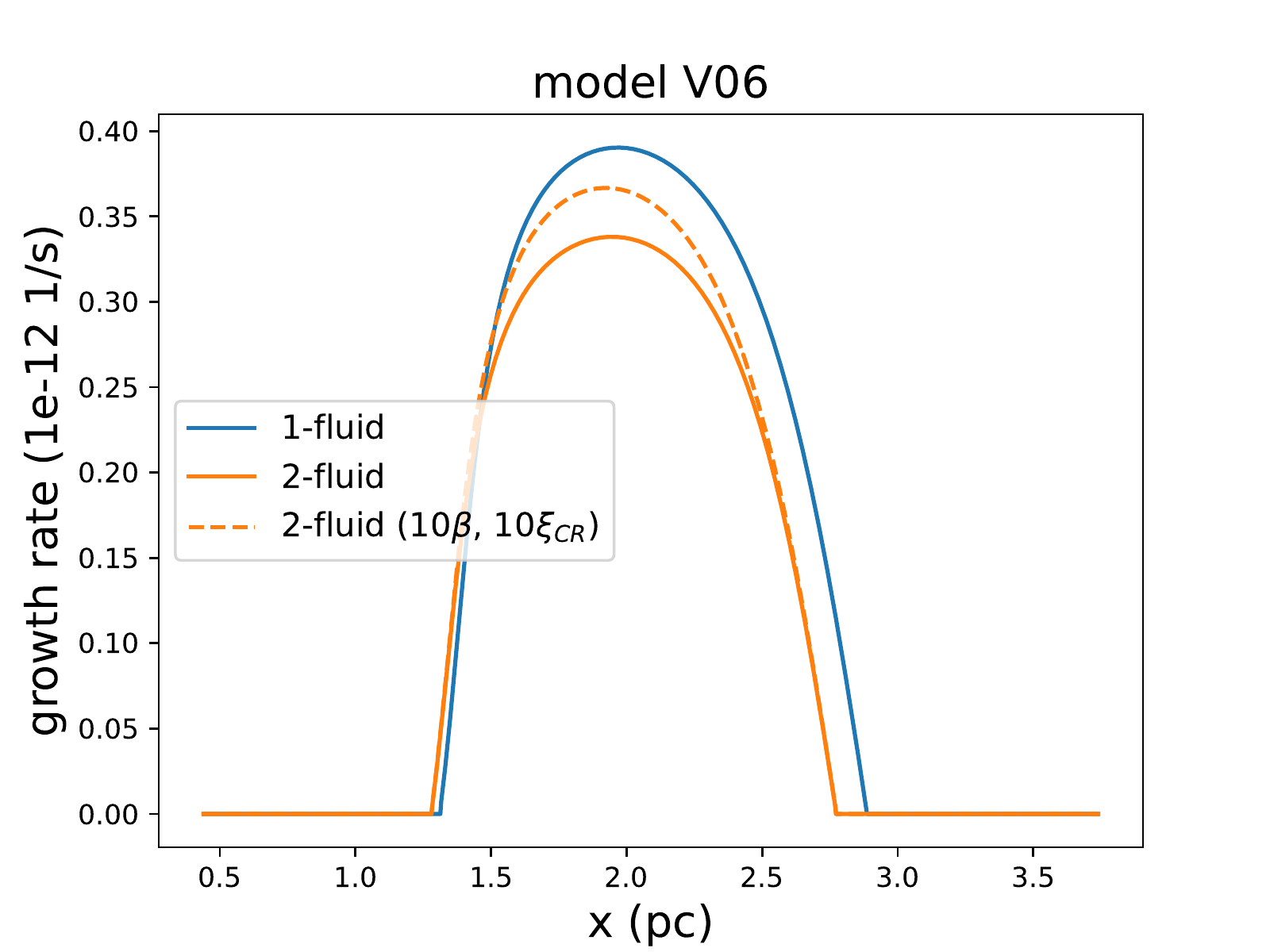} \\
\includegraphics[width=0.35\linewidth]{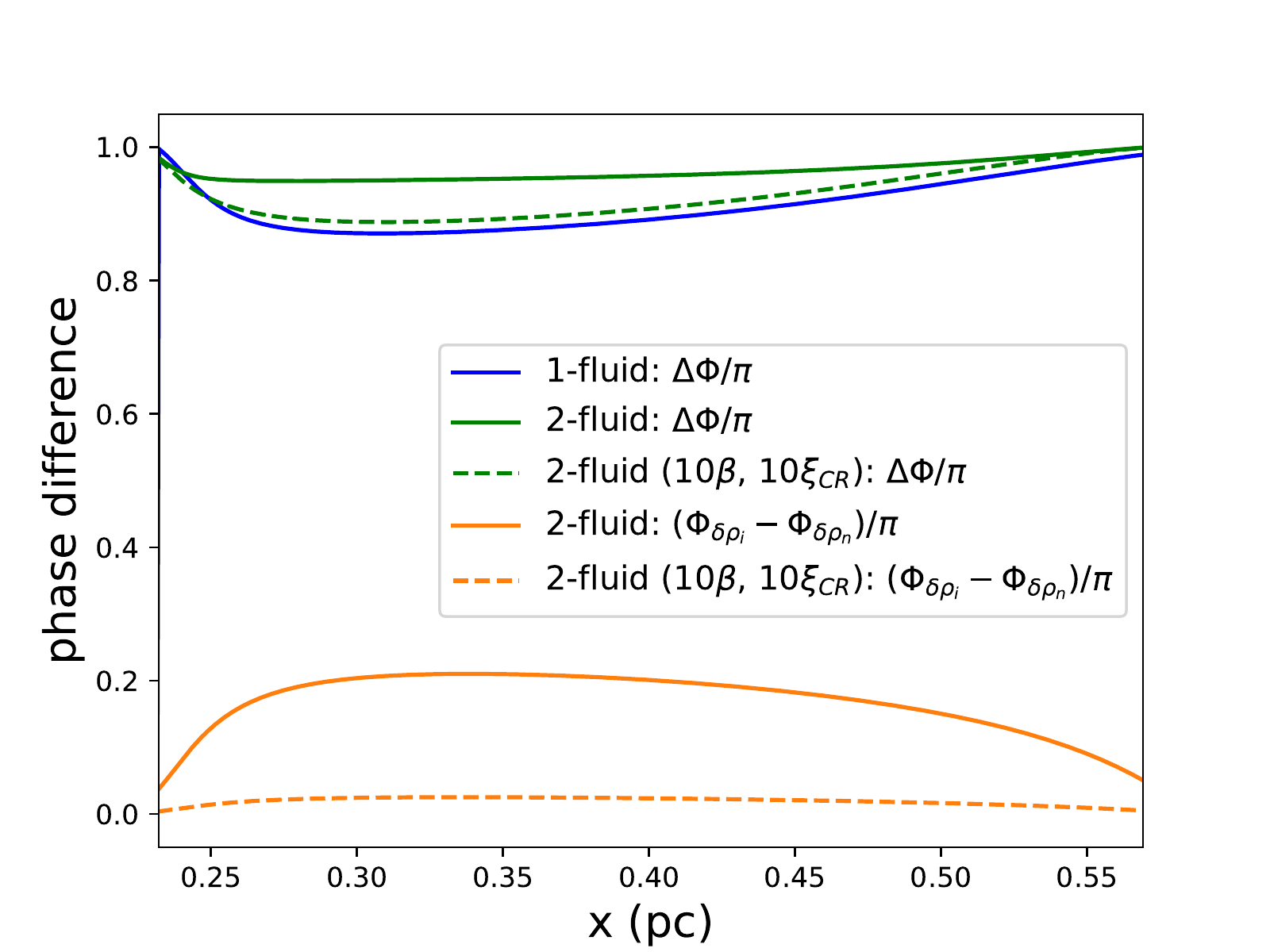}  &
\includegraphics[width=0.35\linewidth]{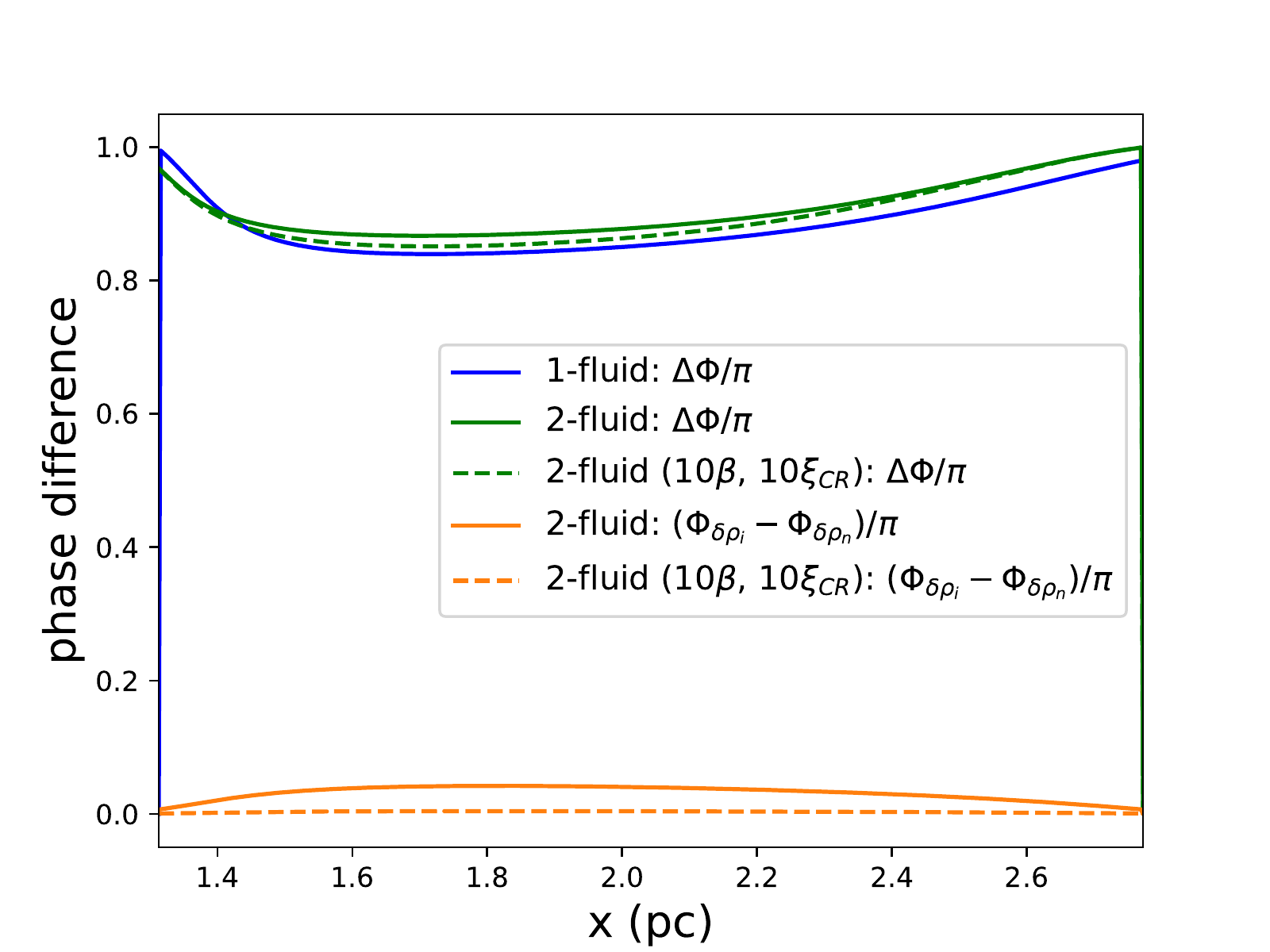} 
\end{tabular}
\caption{Comparison of the growth rate, $\Delta \phi$, and $\phi_{\delta \rho_i}-\phi_{\delta \rho_n}$ within the shock width between the 1- and 2-fluid linear analyses for model Fig3CO12 (left panels) and model V06 (right panels). As noted in the legends, the dashed lines illustrate the 2-fluid cases with ten times the ionization and recombination rates.}
\label{fig:1vs2fluid}
\end{figure}

Figure~\ref{fig:1vs2fluid} presents the comparison of the growth rate, $\Delta \phi$ ($=\phi_{\delta v_n}-\phi_{\delta \rho_n}$), and $\phi_{\delta \rho_i}-\phi_{\delta \rho_n}$ within the shock width between the one- and two-fluid linear approaches for model Fig3CO12 (left panels) and model V06 (right panels). The curves from the one-fluid theory have been demonstrated to agree with the numerical simulation in Figures~\ref{fig:linear_fig3_trend} and \ref{fig:linear_V06_trend}. However, the growth rate for the one-fluid approach is larger than that for the two-fluid approach, albeit still in the same order. Given the relatively high ionization and recombination rates in the problem, the primary approximation made for the one-fluid approach is that the ionization-recombination equilibrium is assumed to strictly hold for the density perturbations in the continuity equation for the ions such that $\delta \rho_i \propto \delta \rho_n^{1/2}$, which helps simplify the two-fluid model to the one-fluid model. Consequently, the ion and neutral density excesses are always in phase, maximizing the ion-neutral drag, thus promoting the growth due to the drag instability. To further demonstrate the physics more quantitatively, we increase both the ionization and recombination rate by the same factor of 10 and perform the linear analysis. In the process, the background state, which depends only on the ionization fraction, remains unchanged.
The dashed lines in Figure~\ref{fig:1vs2fluid}  show the results for the models with ten times the ionization and recombination rates.  It is evident that in both models, the growth rate in the two-fluid approach increases toward that in the one-fluid approach when both the ionization and recombination rates are enhanced. We can see from Figure~\ref{fig:1vs2fluid}  that $\delta \rho_i$ and $\delta \rho_n$ become closer in phase (i.e., $\phi_{\delta \rho_i}- \phi_{\delta \rho_n} \rightarrow 0$) due to the even faster processes of the cosmic ionization and ion-neutral recombination in the two-fluid approach. Therefore, the ionization equilibrium is almost attained to increase the phase difference $\Delta \phi$ and the growth rate closer to those in the one-fluid approach.

The importance of the ionization equilibrium in exciting the drag instability can be also realized by ignoring the ionization and recombination effects from the local linearized equations. In the two-fluid approach, the drag instability vanishes in a 1-D isothermal C-shock in the absence of ionization and recombination \citep{GC20}. In the one-fluid approach, the same physical process can be achieved by using
 $\delta \rho_i/\rho_i = \delta B/B$ (frozen-in field condition) instead of 
$\delta \rho_i/\rho_i = (1/2)\delta \rho_n/\rho_n$ (ionization equilibrium) in the linearized equations. Consequently, we find that the matrix element $(3/2)\rmi kV_d B/\rho_n$ in $C_n$ becomes $\rmi kV_d B/\rho_n$ in Equation(\ref{eq:linear_eq}). It can then be shown that there is no unstable eigenmode from Equation(\ref{eq:linear_eq}), with the frozen-in background state of 1D isothermal C-shocks where the ion compression ratio equals $r_B$ \citep{CO12}.

As $k$ increases in the two-fluid approach, the unstable mode is more akin to  an acoustic mode, and the ionization equilibrium becomes less valid. Hence, the drag instability is suppressed by the small-scale gas pressure perturbation along the shock direction \citep{GLV,GC20,Gu21}.  In contrast, the demise of the instability does not occur for a large $k$ in the one-fluid approach. As the initial $k$ of the perturbation increases in the incoming flow, we find that the total growth TG$_{|U=1|}$ increases to about 58.15 and 121.06 in models Fig3CO12 and V06, respectively, and then remains almost constant independent of the initial $k$. It arises because the thermal pressure term $k^2c_s^2$ becomes important in Equation(\ref{eq:complex_dispersion}) in the regime of a large $k$. Retaining this term in Equation(\ref{eq:dp2}) yields the simplified dispersion relation for the unstable mode to be $\Gamma \sim kc_s-\rmi (|V_d|/c_s)\gamma \rho_i/4$.
As in the two-fluid approach, the unstable mode of a large $k$ is transformed to an acoustic mode in the one-fluid analysis (i.e., Re$[\Gamma] \sim kc_s$), but with a growth rate $-$Im[$\Gamma$] independent of $k$. However, this growth is spurious because the ionization equilibrium breaks down in the limit of a large $k$, i.e., $kc_s > 2\beta \rho_i$, and thus the one-fluid approach fails. Therefore, while the one-fluid analysis is still reasonable to study the general properties of the drag instability for a moderate value of $k$, it is unable to hint at the maximum-growth modes as has been done by \citet{GC20} in their two-fluid analysis.

Adopting the two-fluid approach, \citet{Gu21} extended the linear analysis of the drag instability in 1D C-shocks to 2D perpendicular and oblique isothermal C-shocks. With one more dimension for fluid motion and magnetic field component, the author found that the properties of the unstable mode generally depend on the ratio of the transverse (i.e., normal to the shock flow) to longitudinal (i.e., parallel to the shock flow) wavenumber.  It follows that the growth rate of the instability is dynamically correlated with $\phi_{\delta \rho_i}- \phi_{\delta \rho_n}$ in different regimes of the wavenumber ratio, as illustrated in panel (a) in Figures~4 and 9 of  \citet{Gu21}. Nevertheless,
the 2D linear analysis shows that unless the shock is mildly oblique or perpendicular, the maximum growth of the drag instability is probably contributed by the transversely large-scale mode (i.e., almost 1D mode), suggesting that the numerical simulation in the one-fluid approach employed in this study can capture the drag instability in 2D oblique shocks, albeit with overestimated growth rates as obtained in this study. As for the drag instability in mildly oblique and perpendicular shocks,  \citet{Gu21} found a type of unstable modes, which almost co-moves with the shock and thus has sufficient time to grow without being limited by the shock width. Since the maximum-growth modes occur when the wavenumber ratio transitions from the transversely large-scale mode (i.e., ionization equilibrium is crucial) to  the transversely small-scale mode (i.e., slow modes become essential).  The numerical simulation based on the one-fluid approach due to ionization equilibrium may miss this piece of physics and hence fail to capture these modes.  More robust studies in the future would shed more 
light on this issue.

\subsection{Astrophysical Applications}
The linear theory, supported by the numerical simulations of this work, suggests that the drag instability imprints the density-banded substructure inside a C-shock in the typical environment of star-forming clouds. The maximum-growth modes have been suggested by \citet{GC20} in their two-fluid analysis. That is, for the predominant growing mode, $\omega_{wave} \approx -3\times 10^{-11}$ s$^{-1}$ with $k$ ranging from $\approx 190$-1300 pc$^{-1}$ across the shock for model Fig3CO12, and $\omega_{wave} \approx -2\times 10^{-11}$ s$^{-1}$ with $k$ ranging from $\approx 110$-610 pc$^{-1}$ throughout the shock for model V06. However, the density enhancement in a C-shock is still weaker than the density in the postshock region, which may imply that the density substructure is probably challenging to probe. For instance, there is no evidence of subcritical clouds in observations \citep{Crutcher12}, and indeed, \citet{Nakano98} demonstrated that a subcritical cloud would generally be quite unobservable because it would have too low a column-density contrast with its surroundings. 
Nevertheless, it is worth mentioning that \citet{HT19} recently proposed a new probe of the C-shock structure by detecting microwave emissions from fast spinning nanoparticles due to supersonic neutral gas-charged grain drift. In their work, the spin-up timescale of nanoparticles is inversely proportional to the neutral density. 
While the required resolution and sensitivity would be unrealistically high to resolve the density growth within the C-shock due to the drag instability, the new method could 
shed new light on observationally probing some plasma properties.

\section{Summary}
\label{sec:sum}

1D non-ideal MHD simulations with ambipolar diffusion are performed to investigate the presence of the drag instability in 1D isothermal C-shocks inferred from the linear theory of \citet{GC20}. While \citet{GC20} focused on a linear theory for both neutrals and ions, we limit this study to the one-fluid approach for the convenience of numerical computation. In the one-fluid approach, the continuity and momentum equations for the ions are reduced to the relation $\rho_i \propto \rho_n^{1/2}$ due to the ionization-recombination equilibrium. We first revisit the linearized equations in the one-fluid approach and confirm with \citet{GC20} that the drag instability still exists. We then set up 1D isothermal MHD simulations for two steady-state C-shock models with narrow (model Fig3CO12) and wide (model V06) shock widths. The density-only perturbation is excited at the inflow boundary of the preshock region, with the angular wave frequency given by $-v_0 k_0$. 

The perturbation generates a pair of density-dominated modes with equal amplitude but a slight difference in wavelength, leading to the wave beats in space as seen in the simulations (see the bottom panel of Figure~\ref{fig:pertD_inflow}). As the perturbations propagate into the C-shock region, one of the pairs grows within the C-shock width. To compare such growth with the local linear theory of the drag instability, we extract the predominant Fourier mode of perturbations over the range of two wavelengths at three locations within the C-shock in the simulations (Figure~\ref{fig:perb_comp}). We show that the profiles of the predominant mode agree with those derived from the linear theory with the same $k$. Additionally, the relative amplitudes and phase differences among the density, velocity, and magnetic-field perturbations of the predominant mode from simulations enable us to estimate the instantaneous growth rates and wave frequencies at the three locations, which are consistent with the linear results as well (see Figures~\ref{fig:linear_fig3_trend} \& \ref{fig:linear_V06_trend}). Therefore, we confirm the presence of the drag instability in the non-ideal MHD simulations.  
We also study the nonlinear outcome of the drag instability in a C-shock by exciting a large perturbation. We find that as the perturbation propagates into the shock, it grows but is subsequently saturated due to wave steepening (see Figure~\ref{fig:Fig3_nonlinear}).
The numerical simulations presented in this study demonstrate that the drag instability is a dynamically robust outcome of a 1D steady isothermal C-shock.

\acknowledgments

We thank Sheng-Yuan Liu for the helpful discussions. We also thank the anonymous referee for a constructive report, which improved the presentation and clarity of the paper. Some of the discussions on astrophysical applications were inspired by the referee report for \citet{Gu21}.
The numerical work was conducted on the high-performance computing facility at the Institute of Astronomy and Astrophysics in Academia Sinica (\url{https://hpc.tiara.sinica.edu.tw}) as well as the Research Computing at The University of Virginia (\url{https://rc.virginia.edu}).
P.-G.G. and E.S. acknowledge support from the Ministry of Science and Technology in Taiwan (MOST) through the grants 109-2112-M001-052 and 111-2112-M-001-037. 
C.-C.Y. acknowledges support from MOST through the grant 109-2115-M-030-004-MY2. M.-K.L. is supported by MOST (grants 107-2112-M-001-043-MY3, 110-2112-M-001-034-, 110-2124-M-002-012-, 111-2124-M-002-013-) and an Academia Sinica Career Development Award (AS-CDA-110-M06).
This work was performed under the auspices of the U.S.~Department of Energy (DOE) by Lawrence Livermore National Laboratory under Contract DE-AC52-07NA27344 (C.-Y.C). LLNL-JRNL-832183




\appendix
\section{Wave beats in the simulations}
\label{app}
The density-only perturbation at the left boundary of the simulations (i.e. $x=0$) is set up to follow
$\delta \rho_n/\rho_n=|\delta \rho_n/\rho_n|_{init} \exp(-\rmi \pi/2 + \rmi \omega_{wave} t)$. The perturbation can be expressed as the linear combination of the three modes $\propto \exp(\rmi \omega_{wave} t)$ illustrated in Figure~\ref{fig:Fig3_3modes} for model Fig3CO12. The same procedure is applied to model V06, though the corresponding modes are not shown. That is to say,
\begin{eqnarray}
\left[
\begin{array}{c}
0\\
\delta \rho_n/\rho_n \\
0
\end{array}
\right]
= \alpha_1 \left[
\begin{array}{c}
\delta B_1 /B\\
\delta \rho_{n,1}/\rho_n \\
\delta v_{n,1}/ V_{n}
\end{array}
\right]
+ \alpha_2  \left[
\begin{array}{c}
\delta B_2 /B\\
\delta \rho_{n,2}/\rho_n \\
\delta v_{n,2}/ V_{n}
\end{array}
\right]
+ \alpha_3  \left[
\begin{array}{c}
\delta B_3 /B\\
\delta \rho_{n,3}/\rho_n \\
\delta v_{n,3}/ V_{n}
\end{array}
\right],
\end{eqnarray}
where $\alpha_{1,2,3}$ are the projection coefficients of the initial perturbation on the left-hand side. Given the known modes, $\alpha_{1,2,3}$  can be solved. In model Fig3CO12, $\alpha_1 \approx 8.15 \times 10^{-4}\exp(\rmi 0.1\pi) |\delta \rho_n/\rho_n|_{init}$,   $\alpha_2 \approx 0.5\exp(\rmi 0.53\pi)|\delta \rho_n/\rho_n|_{init}$, and $\alpha_3 \approx 0.5\exp(\rmi 0.45\pi)|\delta \rho_n/\rho_n|_{init}$. In model V06, $\alpha_1 \approx 4.14 \times 10^{-4}\exp(-\rmi 0.69\pi)|\delta \rho_n/\rho_n|_{init}$,   $\alpha_2 \approx 0.5\exp(-\rmi 0.71\pi)|\delta \rho_n/\rho_n|_{init}$, and  $\alpha_3 \approx 0.5\exp(\rmi 0.11\pi)|\delta \rho_n/\rho_n|_{init}$. Unsurprisingly, $|\alpha_1| \ll |\alpha_2| \approx |\alpha_3|$ because the first mode is magnetically dominated and thus is not favorably excited by the density-only perturbation at the left boundary, whereas the second and third modes are density dominated and hence are primarily excited with equal amplitude.

The pair of the slowly decaying modes with almost equal amplitude but with a slight difference in wavelength is expected to generate wave beats in the preshock region as they propagate downstream. Specifically, ignoring the small imaginary part of angular frequencies, the linear superposition of the two traveling waves with the same amplitude and the mutual phase difference $\phi_{23}$ can be generally expressed as 
\begin{eqnarray}
{\rm Re} \left[ \delta \rho_n /\rho_n \right] &\propto &{\rm Re}[ \exp(\rmi k_2 x +\rmi \omega_2 t) + \exp(\rmi k_3 x + \rmi \omega_3 t + \phi_{23}) ]  \nonumber \\
&=& \left[ \cos \left( {k_2+k_3 \over 2} x + {\omega_2 + \omega_3 \over 2}t + {\phi_{23} \over 2} \right)  \right] \left[ 2\cos  \left( {k_2-k_3 \over 2} x + {\omega_2 - \omega_3 \over 2} t- {\phi_{23} \over 2}  \right) \right], 
\label{eq:beat}
\end{eqnarray}
where the first cosine factor describes the high-frequency traveling wave, modulated by the propagating envelope given by the second cosine factor. The condition that $\omega_2=\omega_3=\omega_{wave}$ leads to a complete cancellation of the two waves at the same locations during the evolution as seen in the simulation. In mode Fig3CO12 (V06), $\phi_{23}$ is contributed by the phase difference $-0.08\pi$ ($0.82\pi$) between the projection coefficients $\alpha_2$ and $\alpha_3$ as well as by the phase difference $0.14\pi$ ($-0.72\pi$) between $\delta \rho_{n,2}/\rho_n$ and $\delta \rho_{n,3}/\rho_n$. Hence, $\phi_{23}\approx 0.06\pi$ ($0.1\pi$). After being excited at the left boundary, the beat envelope first goes to zero at the location where the argument of the second cosine in Equation(\ref{eq:beat}) is $\pi/2$. Using $1/k_2=0.00288$ pc and $1/k_3=0.00312$ pc ($1/k_2=0.015$ pc and $1/k_3=0.0145$ pc), we have $x\approx 0.12$ pc ($x\approx 0.78$ pc) for complete destructive interference, in excellent agreement with the simulation shown in the lower panels of Figure~\ref{fig:pertD_inflow}.




\end{document}